\documentclass[twoside,11pt]{article}

\usepackage{jmlr2e}
\usepackage{enumerate}
\usepackage{amsmath,amsfonts,amsbsy}
\usepackage{multirow}
\usepackage{verbatim}
\usepackage{threeparttable}
\usepackage{booktabs}
\usepackage{bm}
\usepackage{soul}

\usepackage[usenames,dvipsnames]{xcolor}
\usepackage[boxed, vlined, lined, commentsnumbered]{algorithm2e}
\usepackage{url}
\usepackage{subcaption} % to create subfigures
\usepackage{chngcntr} % to change figure numbering for appendix

\newcommand{\calE}{\mathcal{E}}
\newcommand{\T}{\mathcal{T}}
\renewcommand{\H}{\mathcal{H}}
\renewcommand{\L}{\mathcal{L}}
\newcommand{\hbeta}{\widehat{\beta}}
\newcommand{\halpha}{\widehat{\alpha}}

\newcommand{\bbeta}{\bar{\beta}}
\newcommand{\balpha}{\bar{\alpha}}
\newcommand{\tbeta}{\tilde{\beta}}
\newcommand{\talpha}{\tilde{\alpha}}
\newcommand{\R}{\mathbb R}

\newcommand{\var}{\mathrm{var}}
\newcommand{\diag}{\mathrm{diag}}

\newcommand{\azero}{\mathbf{a}_0}
\newcommand{\bzero}{\mathbf{b}_0}
\newcommand{\bba}{\bar{\mathbf{a}}} % barbolda
\newcommand{\bbb}{\bar{\mathbf{b}}}
\newcommand{\ba}{\bar{a}}
\newcommand{\bb}{\bar{b}}

\newcommand{\ha}{\hat{a}}
\newcommand{\hb}{\hat{b}}

\newcommand{\tonetimesttwo}{\T_1 \times \T_2}
\newcommand{\inttonetimesttwo}{\int_{\T_1 \times \T_2}}

\newcommand{\zeronorm}[1]{\|#1\|_{0}}

\newcommand{\zerosnorm}[1]{\|#1\|_{0\alpha}}
\newcommand{\zerotnorm}[1]{\|#1\|_{0\beta}}
\newcommand{\zerosnormsq}[1]{\|#1\|_{0\alpha}^2}
\newcommand{\zerotnormsq}[1]{\|#1\|_{0\beta}^2}

\newcommand{\knorms}[1]{{\|#1\|_{K}^2}}

\newcommand{\konenormsq}[1]{{\|#1\|_{K_1}^2}}

\newcommand{\ktwonormsq}[1]{{\|#1\|_{K_2}^2}}

\newcommand{\ktwotnorm}[1]{{\|#1\|_{\tilde{K}_2}}}
\newcommand{\ktwotnormsq}[1]{{\|#1\|_{\tilde{K}_2}^2}}

\DeclareMathOperator*{\argmin}{arg\,min}

 \newcommand{\bes}{\begin{eqnarray*}}
 \newcommand{\ees}{\end{eqnarray*}}

\newcommand{\be}{\begin{equation}}
\newcommand{\ee}{\end{equation}}

\newcommand{\bea}{\begin{eqnarray}}
\newcommand{\eea}{\end{eqnarray}}
\newcommand{\beas}{\begin{eqnarray*}}
\newcommand{\eeas}{\end{eqnarray*}}

\newcommand{\benu}{\begin{enumerate}}
\newcommand{\eenu}{\end{enumerate}}
\newcommand{\bi}{\begin{itemize}}
\newcommand{\ei}{\end{itemize}}

\newcommand{\expec}{\mathbb{E}}
\newcommand{\expecs}{\mathbb{E}^*}
\newcommand{\prob}{\mathbb{P}}

\jmlrheading{1}{2000}{1-48}{4/00}{10/00}{meila00a}{}

\ShortHeadings{Functional Bilinear Regression with Two-way Functional Covariates}{Yang et al.}
\firstpageno{1}

\begin{document}
%\printlength\textwidth

\title{Optimal Functional Bilinear Regression with Two-dimensional Functional Covariates via Reproducing Kernel Hilbert Space}

\author{\name Dan Yang \email dyanghku@hku.hk \\
       \name Jianlong Shao \email sjlpku@hku.hk \\
       \name Haipeng Shen \email haipeng@hku.hk \\
       \addr %Innovation and Information Management, 
       Faculty of Business and Economics, 
       The University of Hong Kong
%       Fokfulam Road, Hong Kong
%      \addr Department of Statistics and Biostatistics\\
%       Rutgers University\\
%       Hill Center, Room 501, 110 Frelinghuysen Road, Piscataway, New Jersey 08854
      % \AND
      % \name Dong Wang \email dongwangunc@gmail.com \\
      % \addr Department of Statistics, 
      % Rutgers University
      %Piscataway, NJ 08854
    %   Hill Center, Room 501, 110 Frelinghuysen Road, Piscataway, New Jersey 08854
    %   \addr Department of Operations Research and Financial Engineering\\
    %   Princeton University
    %   \AND
%       \name Haipeng Shen \\
%       \addr Innovation and Information Management, Faculty of Business and Economics\\
%       University of Hong Kong\\
%       Fokfulam Road, Hong Kong
%       \AND
%       \name Haipeng Shen \email haipeng@hku.hk \\
%       \addr Innovation and Information Management, Faculty of Business and Economics\\
%       University of Hong Kong\\
%       Fokfulam Road, Hong Kong
       \AND
       \name Hongtu Zhu  \email htzhu@email.unc.edu\\
       \addr Department of Biostatistics,
       University of North Carolina at Chapel Hill
%       Chapel Hill, NC 27599 
%       \AND for the Alzheimer's Disease Neuroimaging Initiative
       }

% \editor{Kevin Murphy and Bernhard Sch{\"o}lkopf \dyR{carl, check!}}
\editor{}%\carlB{Francis Bach and David Blei} \dyR{carl, check!}}
\maketitle

\begin{abstract}%   <- trailing '%' for backward compatibility of .sty file
Traditional functional linear regression usually takes a one-dimensional functional predictor as input and estimates the continuous coefficient function. Modern applications often generate two-dimensional covariates, which become matrices when observed at grid points. To avoid the inefficiency of the classical method involving estimation of a two-dimensional coefficient function, we propose a functional bilinear regression model, and introduce an innovative three-term penalty to impose roughness penalty in the estimation. The proposed estimator exhibits minimax optimal property for prediction under the framework of reproducing kernel Hilbert space.  An iterative generalized cross-validation approach is developed to choose tuning parameters, which significantly improves the computational efficiency over the traditional cross-validation approach. The statistical and computational advantages of the proposed method over existing methods are further demonstrated via simulated experiments, the Canadian weather data, and a biochemical long-range infrared light detection and ranging data. 
\end{abstract}


%%%%%%%%%%%%%%%%%%%%%%%%%%%%%%%%%%%%%%%%%%%%%%%%%%%%%%%%%%%%%%%%%%%%%%%

% Carl: Reduce the space between formula and text
\setlength{\abovedisplayskip}{.5pt}
\setlength{\belowdisplayskip}{.5pt}
% Reduce the space between figure and text
\setlength{\abovecaptionskip}{0pt}
\setlength{\belowcaptionskip}{0pt}

\vspace{-1em}
\section{Introduction}
The functional linear regression (FLR) is a powerful approach for predicting a scalar response from a one-dimensional functional predictor. It was first introduced by \citet{Ramsay+1991a}, and has been widely used in functional data analysis since then \citep{Ramsay+2002a,Ramsay+2005a,wang2016review,reiss2017methods}. Consider a scalar response $Y$ and a square integrable random function $X(\cdot)$ with mean $0$ defined over the domain $\T$, the FLR model adopts the following form,
\begin{equation}
\label{eqn:1dflr}
Y = \mu_0 + \int_{\T} X(t)\beta_0(t)~dt + \epsilon ,
\end{equation}
where $\mu_0$ is the intercept, $\beta_0(\cdot)$ is the unknown coefficient function, and $\epsilon$ is zero-mean noise.

The predictor $X(\cdot)$ in Model \eqref{eqn:1dflr} is often one-dimensional, and is therefore sometimes referred to as one-way functional input. However, in recent years, it has been increasingly common to collect data in a two-way fashion. To be more specific, the random predictor $X(\cdot)$ in these cases is a bivariate function defined in the domain $\T_1 \times \T_2$. 
%\textcolor{red}{Replace with one of the two from Section 5? For instance, the functional magnetic resonance imaging (fMRI) data is the blood-oxygen-level dependent contrast observed for the whole brain over time, where the spatial and temporal dimensions correspond to the two-way functional input.  The two-dimensional functional input can be used to predict the severity of the attention deficit hyperactivity disorder.}
%, for instance, spatially and temporally, in cases where multiple time series were observed at different locations. 
For instance, the well-known Canadian weather data traditionally uses one-dimensional function of daily temperature to predict precipitation. In Section \ref{sec:realdata}, it is shown that the hour-of-the-day and day-of-the-year information constitutes two-dimensional functional predictor, which predicts precipitation more accurately and reveals more meteorological phenomena. 
Besides meteorology, such two-way functional data are now frequently encountered in fields like finance, economics, social science, neuroimaging, and so on. See Section \ref{sec:rd lidar} in Appendix for another real data example, where the two domains correspond to wavelength and range, respectively. 

To generalize Model~\eqref{eqn:1dflr} in order to deal with two-way covariate, we propose the following functional bilinear regression (FBLR) model,
\begin{equation}
\label{eqn:2dflr}
Y = \mu_0 + \int_{\tonetimesttwo} \alpha_0(s)X(s,t)\beta_0(t)~dsdt + \epsilon.
\end{equation}
Here, $Y$, $\mu_0$ and $\epsilon$ are again the scalar response, intercept, and error terms, respectively, $X(\cdot,\cdot)$ is a square integrable bivariate zero-mean function defined on $\T_1 \times \T_2$, and $\alpha_0(\cdot)$ in the domain $\T_1$ and $\beta_0(\cdot)$ on $\T_2$ are two unknown coefficient functions. 
%For ease of presentation, we will further assume $\mu_0=0$ since its estimate can be easily obtained. 
The error term $\epsilon$ is assumed to have mean zero and finite variance. We study the random design case where $X(\cdot,\cdot)$ is a stochastic process and independent of $\epsilon$. The goal of the FBLR is to recover the two coefficient functions from $n$ independent and identically distributed training sample $\{x_i(\cdot,\cdot), y_i\}_{i=1}^n$ and make predictions for testing data. 

Compared with Model \eqref{eqn:1dflr}, the two coefficient functions in Model \eqref{eqn:2dflr} preserves the two-way functional structural information through a bilinear form, $\alpha_0(s) X(s,t) \beta_0(t)$. In the literature, this type of bilinear/multilinear combination is becoming increasingly common when dealing with two-way/multi-way/tensor data \citep[e.g.,][]{Dyrholm+2007a, zhou2013tensor, bi2018multilayer, bi2021tensors, chen2022factor}.%\citep[e.g.,][]{Dyrholm+2007a, Li+2010a, Hung+2013a, zhou2013tensor, Zhao+2014a, bi2018multilayer, bi2021tensors, chen2022factor}.

A naive and straightforward approach to deal with the two-way functional covariate is to convert the two-dimensional predictor into one-dimensional through stacking the data along one direction, then followed by implementing the traditional FLR Model \eqref{eqn:1dflr}. However, this conversion would destroy the two-way functional structure of the covariate, and the resulting one-way predictor is typically no longer a smooth function, which violates the underlying assumption of FLR. Therefore, adopting Model \eqref{eqn:1dflr} is not a good choice for a two-dimensional functional predictor. More discussions on the vectorization methods to make the resulting long vector smoother can be found in Appendix \ref{sec:exsim-FLR-vec}. But even though the resulting long vector is smoother, applying FLR on it still leads to worse performance than keeping the two-dimensional structure.

Note that $\alpha_0(\cdot)$ and $\beta_0(\cdot)$ are only identifiable up to a scalar, but their product $\alpha_0(\cdot)\beta_0(\cdot)$ is identifiable. That is, for any $c\ne 0$, $\alpha_0(\cdot)/c$ and $c \beta_0(\cdot)$ will lead to an equivalent model. For our primary focus on  prediction accuracy, the un-identifiability is not a concern: as equivalent models will lead to identical predictions. To make the coefficient functions completely identifiable, one can adopt three choices of common practices to control the scaling issue: (1) assume $\|\alpha_0\|=\|\beta_0\|=1$ and introduce one extra scaling scalar parameter; (2) assume either $\|\alpha_0\|=1$ or $\|\beta_0\|=1$ and absorb the scalar into the other; (3) assume $\|\alpha_0\|=\|\beta_0\|$. To further make the sign identifiable, one could assume the integral $\int_{\T_1}\alpha_0(s)ds$ to be positive and adjust the signs of $\alpha_0$ and $\beta_0$ accordingly, or one could use domain expertise to determine the signs.
Please refer to Section \ref{ssec:obj} for detailed discussion of the impact of this un-identifiability property on the choice of the penalty involved.

A few articles have examined the problem of regression with two-way predictor, sometimes referred to scalar-on-image regression, as summarized in \citet{happ2018impact}. \citet{Reiss+2010a} studied this problem with 2D images as predictors by regressing on the principal component (PC) scores obtained from two-dimensional principal component analysis (PCA) of the observed images. 
% where radial cubic B-splines or thin plate penalty were used \citep{Saranli+1998a,Green+1994a}. 
\citet{Sangalli+2009a} proposed to penalize with the integral of the square of the Laplacian of the two-dimensional coefficient function. \citet{Guillas+2010a} approached the problem through fixed bivariate spline. \citet{Wang+2014b} and \citet{Reiss+2015a} transformed the multi-dimensional functional problem into estimation of wavelet coefficient via wavelet transformation. %\citet{kang2018scalar} investigated this problem from a Bayesian perspective. 
These aforementioned papers considered the following model
\begin{equation}
\label{eqn:2dflr-other}
Y = \mu_0 + \int_{\tonetimesttwo} X(s,t)\beta_0(s,t)~dsdt + \epsilon,
\end{equation}
and focused on the estimation of the coefficient function $\beta_0(\cdot,\cdot)$. 
%Few have studied the theoretical properties of the proposed estimator.
%, except for \citet{balasubramanian2022unified}, which used an infinite-dimensional Gaussian Stein's identity to relax some assumptions typically required.  
% keep
% In the literature, the regression with two-way predictor is sometimes also referred to as the scalar-on-image regression. Among the works on scalar-on-image regression, some utilized the functional property as mentioned above while others did not, such as \citet{ZL14} penalized the loss function by the nuclear norm of the coefficient matrix instead of the coefficient function and \citet{wang2017generalized} added the penalty of the total variation of the coefficient matrix.

The key difference between Model \eqref{eqn:2dflr} and Model \eqref{eqn:2dflr-other} is that our Model \eqref{eqn:2dflr} adopts two one-dimensional coefficient functions compared to 
a single two-dimensional coefficient function in Model \eqref{eqn:2dflr-other}. Model \eqref{eqn:2dflr} is a special case of Model \eqref{eqn:2dflr-other} with restriction. The primary motivation for proposing this seemingly restrictive Model \eqref{eqn:2dflr} is that estimation of Model \eqref{eqn:2dflr} can lead to estimation the following model,
\begin{equation}
\label{eqn:2dflr-2}
Y = \mu_0 + \sum_{r=1}^{R}\int_{\tonetimesttwo} \alpha_0^{[r]}(s)X(s,t)\beta_0^{[r]}(t)~dsdt + \epsilon.
\end{equation}
Model \eqref{eqn:2dflr-2} is also a special case of Model \eqref{eqn:2dflr-other} with extra assumption that the true two-dimensional function is approximated by the summation of a few terms of the products of two one-dimensional functions, i.e., $\beta_{0}(s,t)=\sum_{r=1}^{R}\alpha_{0}^{[r]}(s)\beta_{0}^{[r]}(t)$. 

If Model \eqref{eqn:2dflr} can be estimated via some approach, then Model \eqref{eqn:2dflr-2} can be estimated in an iterative fashion via deflation: apply the approach to the original data $\{x_i(\cdot,\cdot), y_i\}_{i=1}^n$, obtain the estimate $\hat\alpha_{0}^{[1]}(s)\hat\beta_{0}^{[1]}(t)$ and retain the residuals $\{e_i\}_{i=1}^n$; re-apply the approach to the predictors and residuals $\{x_i(\cdot,\cdot), e_i\}_{i=1}^n$, obtain the estimate $\hat\alpha_{0}^{[2]}(s)\hat\beta_{0}^{[2]}(t)$ and retain the updated residuals; and repeat. Such deflation approach is commonly adopted in the literature of PCA. 

There are a few reasons why Model \eqref{eqn:2dflr-2} is preferred over Model \eqref{eqn:2dflr-other} for two-dimensional functional regression. 
The first reason comes from the necessity of low-rank assumption. Suppose the two-dimensional functional predictors are observed on a dense grid of equidistant points $(s_i,t_j)$ for $i=1,\ldots,m_1$, $j=1,\ldots,m_2$ and denote the observed data matrix as $x_{ij} = X(s_i,t_j)$, correspondingly $b_{ij}=\beta_0(s_i,t_j)$. Then Model \eqref{eqn:2dflr-other} becomes $Y = \mu_0 + \sum_{ij}x_{ij}b_{ij} + \epsilon = \mu_0 + \langle \mathbf{X}, \mathbf{B}\rangle + \epsilon$. Such a scalar-on-matrix regression is one special type of tensor regression. For tensor regression, since the coefficient tensor/matrix, has multiple modes and high dimensions, it is necessary to assume the coefficient has low-rank structure \citep{zhou2013tensor,lock2018tensor,raskutti2019convex,chen2019non}.
%, which resembles the spirit of reduced rank regression. Therefore, it is natural to assume that 
Suppose $\mathbf{B}$ has low-rank singular value decomposition (SVD) $\mathbf{B} = \mathbf{U}\mathbf{D}\mathbf{V}^T = \sum_{r=1}^Rd_r\mathbf{u}_r\mathbf{v}_r$, where $\mathbf{u}_r,\mathbf{v}_r$ are the left and right singular vectors of unit length. Then on the observed grid, the low-rank model is $Y = \mu_0 + \sum_{r=1}^R\sum_{ij}x_{ij}d_r u_{ir}v_{jr} + \epsilon$. 
Bringing this approximation from the discrete case back to the continuous case, it boils down to $Y = \mu_0 + \sum_{r=1}^{R}\int_{\tonetimesttwo} d_r u_r(s)X(s,t)v_r(t)~dsdt + \epsilon$. Note that the constants $d_r$ can be absorbed into $u_r(\cdot)$ or $v_r(\cdot)$, which becomes Model \eqref{eqn:2dflr-2}. 

Second, adopting a framework of reproducing kernel Hilbert space (RKHS) to solve Model \eqref{eqn:2dflr-other} will eventually lead to Model \eqref{eqn:2dflr-2} as well. To solve Model \eqref{eqn:2dflr-other} via RKHS, a four-dimensional kernel needs to be specified. To the best of our knowledge, there is no literature that study the theories of Model \eqref{eqn:2dflr-other} via RKHS. 
One natural choice for the four-dimensional kernel is the kernel associated with tensor product RKHSs in the context of smoothing spline \citep{gu2013smoothing}. Model \eqref{eqn:2dflr-other} with the tensor product kernel will be referred to as FLR+TPK from now on. By the representer theorem, derivations in Section \ref{sec:sim} show that the solution of FLR+TPK will be of the form  $\sum_{jk}c_{jk}\phi_j^1(s)\phi_k^2(t) + \ldots$, where $\ldots$ represents some less important terms, and $\phi_j^1(s)$ and $\phi_k^2(t)$ are the basis of the two kernels from two domains. Plugging the representation back into objective function and solving for coefficient matrix $\mathbf{C}=(c_{jk})$ is again a scalar-on-matrix regression problem. Extending such a framework to even higher dimensional problem is scalar-on-tensor regression. It is well known in the tensor regression literature that it is imperative to assume low-rankness of $\mathbf{C}$, which is {\it exactly} equivalent to assuming Model (4) after simplifications.

The previous two reasons are rooted in the modeling perspective, the next two reasons show the theoretical and numerical advantages of Model \eqref{eqn:2dflr-2} over Model \eqref{eqn:2dflr-other}. 
Third, our theory shows that the two-dimensional FBLR has the same convergence rate as one-dimensional FLR. There is no literature that studies Model \eqref{eqn:2dflr-other} via RKHS, but our conjecture is that the convergence rate should have another factor of ``2'' in the shoulder of the rate, just as in the non-parametric regression, which is slower than ours. 

Fourth, extensive simulation studies and two real data examples demonstrate the superior statistical and computational performances of FBLR over FLR+TPK under all three two-dimensional models mentioned above, even when the data are generated according to Model \eqref{eqn:2dflr-other}. Furthermore, visualization and interpretation of the two one-dimensional functions from Model \eqref{eqn:2dflr-2} are more straightforward and meaningful than the two-dimensional function from Model \eqref{eqn:2dflr-other}; see the real data applications for the details. Moreover, for many applications, the two domains are very different. In FLR+TPK, although two distinct kernels can be adopted, but only one hyperparameter can be used. However, it is very likely that the levels of smoothness in the two domains are very different. Such difference requires not only two kernels, but also two hyperparameters, which FBLR can accommodate but FLR+TPK cannot.

Lastly, the advantage of Model \eqref{eqn:2dflr-2} over Model \eqref{eqn:2dflr-other} is magnified when extension to even higher dimension $d$ is made. Model \eqref{eqn:2dflr-2} can be extended straightforwardly via multilinear form and Model \eqref{eqn:2dflr-other} can be extended via multivariate integral. However, when Model \eqref{eqn:2dflr-other} is extended, issues like infeasible computation, curse of dimensionality, and slow theoretical convergence rate will inevitably occur. These issues were mentioned in the two-dimensional functional PCA literature as well \citep{chen2017modelling}.
On the other hand, when Model \eqref{eqn:2dflr-2} is extended, we expect the same computational cost (by a factor of $d$, not exponentially in $d$) and identical theoretical convergence to remain. 

Although the extension from linear to bilinear seems natural, the generalization is in fact non-trivial. Due to the interplay of the two coefficient functions in a product form, several challenges arise from the design of the penalty, the development of the algorithm, and the theoretical analysis. Therefore, the five-fold main contributions are elaborated below.

First, there are mainly two categories of approaches for one-way FLR, including functional PCA regression (FPCR) \citep[e.g., ][]{Ramsay+2002a,Ramsay+2005a,CH06} %\citep[e.g., ][]{Ramsay+2002a,Ramsay+2005a,CH06,HH07}
and smoothness penalization under the framework of RKHS \citep[e.g., ][]{Yuan+2010b, cai2012minimax, balasubramanian2022unified}. 
~\citet{cai2012minimax} showed that the penalty approach has an advantage over FPCR, because the penalty approach does not require the alignment of the reproducing kernel and the covariance kernel of the predictor, while the FPCR approach does. 
%In other words, when the PCs obtained from PCA of the covariates are not the basis functions for the coefficient function, FPCR does not perform well.
Hence, we take the penalty approach for the FBLR problem when extending from 1D to 2D. Our key innovation is the proposal of a three-term penalty that involves the Hilbert norm based on the reproducing kernel and the norm associated with the covariance kernel. This three-term penalty enjoys the invariant property and successfully separates the effects from the smoothness levels of the two coefficient functions $\alpha(\cdot)$ and $\beta(\cdot)$, which leads to a minimax rate optimal solution.

Second, upon the proposal of the penalty function, an iterative algorithm is developed to optimize the objective function because of the bi-convexity of the function. The main idea for the block descent algorithm is to reduce the two-way problem into a one-way FLR problem in each iteration, which can then be solved by the representer theorem \citep{Wahba+1990a}, with slight complications since the one-way problem involves updated 1D stochastic processes and reproducing kernels. Furthermore, there are two tuning parameters in the penalty corresponding to the different degrees of smoothness of the two coefficient functions. Naive cross validation (CV) over two-dimensional grid is outrageously time-consuming. A novel iterative generalized cross validation (iGCV) approach is proposed whose computational time is only slightly more than the iterative FBLR algorithm with {\it fixed} tuning parameters, while achieving similar performance as the computationally expensive CV.

Third, one interesting finding is that the minimax convergence rate for the FBLR is identical to that of the FLR if we assume the domains, kernels and covariances are the same for the two dimensions of the two-way functional data. Unlike the FLR problem, whose solution is explicit and can be directly analyzed theoretically, the FBLR problem does not have an explicit solution since the two coefficient functions interact with each other. Therefore, the techniques from \citet{cai2012minimax} cannot be applied. To prove the minimax property, we need to combine 2D linearization, two-dimensional G\^ateaux derivatives with sophisticated expressions, block matrix inversion, and RKHS, among others.

Fourth, this article is the first attempt to study the theoretical properties of scalar-on-matrix functional regression with low rank assumption, where the two-dimensional coefficient function has low rank of products of one-dimensional coefficient functions. Such low rank assumption is helpful to extend to functional tensor predictor of even higher order in the future. Tensor regression has become increasingly important recently. There are different combinations of response types and predictor types motivated by various applications, such as tensor-on-vector regression {\citep[e.g.,][]{sun2017store,zhou2021partially}}%\citep[e.g.,][]{li2017parsimonious,sun2017store,zhou2021partially}
, scalar-on-tensor regression {\citep[e.g.,][]{zhou2013tensor,liu2019generalized}}%\citep[e.g.,][]{zhou2013tensor,li2015spatial,liu2019generalized}
, and tensor-on-tensor regression {\citep[e.g.,][]{hoff2015multilinear,chen2021autoregressive}}%\citep[e.g.,][]{hoff2015multilinear,raskutti2019convex,chen2021autoregressive}
. Most of the existing work focus on the low tensor rank assumption on the coefficient tensor with or without sparsity assumption. To the best of our knowledge, work on functional matrix or functional tensor predictor is very rare. 

Lastly, we apply FBLR to two real data examples. One is the Canadian weather data, a well-known example in the functional data analysis (FDA) literature. The goal is to predict precipitation at different weather stations with temperature information. The existing FDA studies \citep[e.g.,][]{Ramsay+1991a,ramsay2005principal, cai2012minimax} focus on 1D PCA in Model \eqref{eqn:1dflr}, where each observation is a vector of 365 daily average temperatures. Besides daily variation, we introduce 2D predictors with the second domain reflecting the hourly temperature variation in Model \eqref{eqn:2dflr-other}. The extra domain not only boosts the prediction accuracy, but also echos some meteorological phenomena. The other real data example is the Light Detection and Ranging (LIDAR) data from the biochemistry field \citep{xun2013parameter}. The data generating process naturally produces smooth data and call for FDA method. For both datasets, FBLR and its iterative variant have higher prediction accuracy and better interpretability compared to existing 2D FLR and 1D FLR methods.

The rest of the paper is organized as follows. A brief review of FLR and the methodology for FBLR are provided in Section \ref{sec:method}. The optimal theoretical property of the proposed method is discussed in Section \ref{sec:theory}. Simulation and the Canadian data analysis are presented in Sections \ref{sec:sim} and \ref{sec:realdata}. Section \ref{sec:dis} concludes with discussion. All proofs, more simulation results, and the biochemical data application are provided in the supplementary materials.

%%%%%%%%%%%%%%%%%%%%%%%%%%%%%%%%%%%%%%%%%%%%%%%%%%%%%%%%%%%%%%%%%%%%%%%
\vspace{-.5em}
\section{Methodology}
\label{sec:method}

\subsection{Notation and definitions}
\label{sec:notation}

Suppose that $\T$ is a compact set. 
%An RKHS is a Hilbert space $\H$ of functions on $\T$ which is equipped with a reproducing kernel $K:\T\times\T \mapsto \R$ and the reproducing kernel is real, symmetric and nonnegative definite. 
We denote by $\H(K)$ an RKHS associated with the reproducing kernel $K$, $\langle \cdot, \cdot \rangle_K$ the associated inner product, and $\|\cdot\|_K$ the induced norm. Then, we have $K(s,\cdot) \in \H(K)$ for all $s \in \T$ and $f(t) = \langle K(t,\cdot),f \rangle_{K}$ for all $f \in \H(K)$. We refer the readers to \citet{Wahba+1990a}, \citet{gu2013smoothing} and references therein for more details. Let $K_1(\cdot,\cdot): \T_1 \times \T_1 \mapsto \R$ and $K_2(\cdot,\cdot): \T_2 \times \T_2 \mapsto \R$ be two reproducing kernels, and $\H(K_1)$ and $\H(K_2)$ be the corresponding RKHS's. The coefficients $\alpha_0(\cdot)$ and $\beta_0(\cdot)$ of Model \eqref{eqn:2dflr} reside in $\H(K_1)$ and $\H(K_2)$, respectively.

The covariance function of the mean zero bivariate random function $X(\cdot,\cdot)$ plays another important role in developing both methodology and theory of FBLR. We define it as $C(s_1,t_1,s_2,t_2) = \expec[X(s_1,t_1)X(s_2,t_2)]$, for any  $s_1, s_2 \in \T_1$ and $t_1, t_2 \in \T_2 $.
As mentioned earlier, the two dimensions of $X(\cdot,\cdot)$ usually correspond to different domains, and hence the covariance can be reasonably assumed to have a decomposable or separable structure, that is,
$C(s_1,t_1,s_2,t_2) = C_\alpha(s_1,s_2)C_\beta(t_1,t_2)$,
where $C_\alpha(\cdot,\cdot)$ and $C_\beta(\cdot,\cdot)$ are two real bivariate functions that characterize the covariance structures along the first and second dimensions respectively. We note that this type of decomposable covariance structure has been widely used in the literature recently when dealing with two-way data %\citep[e.g.,][]{Werner+2008a, Zhou+2014b, FH14, VH15, hafner2020estimation, aston2017tests,chen2017modelling,sun2018knowledge,chen2021autoregressive,drton2021existence,chen2023testing}. 
\citep[e.g.,][]{Zhou+2014b, VH15, hafner2020estimation, aston2017tests,chen2021autoregressive,chen2023testing}. 
Similar to the reproducing kernels $K_1$ and $K_2$, the two covariance functions $C_\alpha$ and $C_\beta$ are also symmetric and nonnegative definite. The subscripts $\alpha$ and $\beta$ will be used frequently to differentiate between the two dimensions.

Lastly, for any $f \in \H(K_1)$ and $g \in \H(K_2)$, we define two semi-norms as follows,
\begin{equation}
  \begin{split}
    \label{eqn:zerosnorm}
    \zerosnorm{f} &= \left(\int_{\T_1 \times T_1} f(s_1) C_\alpha(s_1,s_2) f(s_2) ds_1 ds_2 \right)^{1/2}, \\
    \zerotnorm{g} &= \left( \int_{\T_2 \times T_2} g(t_1) C_\beta(t_1,t_2) g(t_2) dt_1 dt_2 \right)^{1/2}.
  \end{split}
\end{equation}
They will appear in the penalty term of the regularization approach and play a crucial role in the development of the asymptotic theory. It can be verified that the variance of the integral of the bilinear form in Model \eqref{eqn:2dflr} is equal to the square of the product of the two norms defined above, 
\begin{equation}
\label{eqn:var-zeronorm}
  \expec \left( \inttonetimesttwo f(s) X(s,t) g(t) ds dt \right)^2 =
  \zerosnormsq{f} \zerotnormsq{g}.
\end{equation}
%which makes it unsurprising to appear in the penalty term that will be defined momentarily in Section \ref{ssec:obj}.
Throughout this paper, following \citet{Yuan+2010b}, we assume that $\zerosnormsq{f}\ne 0$ holds for any $f\ne 0$ that belongs to the null space of $K_1$, and similarly $\zerotnormsq{f}\ne 0$ holds for any $f\ne 0$ that belongs to the null space of $K_2$. This assumption is necessary to ensure that even if the estimation of the coefficient functions is constrained to the null spaces of $K_1$ and $K_2$, the objective function to be proposed in Section \ref{ssec:obj} can still be uniquely optimized.

%--------------------------------------------------------------------
%--------------------------------------------------------------------
%--------------------------------------------------------------------
%--------------------------------------------------------------------
%--------------------------------------------------------------------
\subsection{Review of the smoothness regularization approach for one-way FLR}

We provide a brief review of the smoothness regularization approach for the FLR model \eqref{eqn:1dflr} in this section to facilitate discussion of FBLR. For more details, please see \citet{Yuan+2010b}, \citet{cai2012minimax}, and references therein.

Consider the reproducing kernel $K(\cdot,\cdot)$ and the corresponding RKHS $\H(K)$. Assume that the coefficient function in Model \eqref{eqn:1dflr} belongs to $\H(K)$. The smoothness regularized estimator can be obtained via minimizing the following objective with loss and penalty,
\be
\label{eqn:onewayobj}
\hbeta = \argmin_{\beta \in \H(K)} \ell_{n\lambda}(\beta) 
\stackrel{def}{=} \argmin_{\beta \in \H(K)} \left\{\ell_n(\beta) + \lambda J(\beta)\right\},
\ee
where $\ell_n(\beta) = n^{-1} \sum_{i=1}^n \left(y_i - \int_\T x_i(t) \beta(t)dt \right)^2$ is the normalized residual sum of squares measuring the goodness-of-fit, $J(\beta)=\knorms{\beta}$ is the squared RKHS norm measuring the smoothness, and $\lambda$ is a tuning parameter that balances the trade-off between them.

Although the optimization in \eqref{eqn:onewayobj} is taken over an infinite-dimensional space, it can be solved by the representer theorem, i.e., Theorem 1 in \citet{Yuan+2010b}. This representer theorem is a generalization of the well-known representer lemma for smoothing splines \citep{Wahba+1990a}. The optimizer of \eqref{eqn:onewayobj} then has the following expression,
\begin{equation}
\label{eqn:onewaycconstants}
%\hbeta(t) = \sum_{i=1}^n c_i \int_\T K(s,t) \beta(s) ds,
\hbeta(t) = \sum_{i=1}^n c_i \int_\T K(s,t) x_i(s) ds,
\end{equation}
where the unknown scalars $c_1, c_2, \ldots, c_n\in\R$ can be readily computed once \eqref{eqn:onewaycconstants} is plugged back into \eqref{eqn:onewayobj}, which leads to a quadratic function of $c_1,...,c_n$. Please refer to Section 2 of \citet{Yuan+2010b} for the details on the explicit expression and implementation.

%--------------------------------------------------------------------
%--------------------------------------------------------------------
%--------------------------------------------------------------------
%--------------------------------------------------------------------
%--------------------------------------------------------------------
\subsection{Objective function of two-way FBLR}
\label{ssec:obj}

The smoothness regularization approach of the one-way FLR can be extended to the two-way FBLR. Assume that the coefficient function $\alpha_0$ resides in $\H(K_1)$ and $\beta_0$ in $\H(K_2)$, it is natural to estimate them by minimizing an objective function, 
\begin{equation}
\label{eqn:twowayobj}
(\halpha,\hbeta) = \argmin_{\alpha \in \H(K_1),\beta \in \H(K_2)}\ell_{n\lambda}(\alpha,\beta)
\stackrel{def}{=} \argmin_{\alpha \in \H(K_1),\beta \in \H(K_2)} \left\{\ell_n(\alpha,\beta) + J(\alpha,\beta)\right\}.
\end{equation}
This is a direct analogy of \eqref{eqn:onewayobj}. $\ell_n(\alpha,\beta)=n^{-1} \sum_{i=1}^n \left( y_i - \int_{\tonetimesttwo} \alpha(s) x_i(s,t) \beta(t)dsdt \right)^2$, the first part, is again the data fidelity term. However, the second part of the objective function \eqref{eqn:twowayobj}, which is the penalty $J(\alpha,\beta)$, cannot be trivially extended from the one-way case and will be discussed in details below.

For the choice of $J(\alpha,\beta)$, we point out four properties that one wishes to consider. (1) As mentioned in the introduction, the functions $\alpha$ and $\beta$ are only identifiable up to a scalar. %since $\alpha(s)X(s,t)\beta(t) = [c\alpha(s)]X(s,t)[c^{-1}\beta(t)]$ for any non-zero constant $c$. 
We intentionally do not enforce any identifiability constraint so that the design of the penalty could be more convenient. Since $\alpha(s)\beta(t)$ is scale-invariant, the penalty $J(\alpha,\beta)$ should be scale-invariant as well; see \citet{Huang+2009a} for similar requirement on two-way functional SVD. (2) The loss term $\ell_n(\alpha,\beta)$ in \eqref{eqn:twowayobj} is bi-quadratic in $(\alpha, \beta)$. Hence, the penalty part should ideally be bi-quadratic as well. (3) The norms $\konenormsq{\alpha}$ and $\ktwonormsq{\beta}$ should encourage smoothness. (4) The two functions $\alpha$ and $\beta$ are from two domains and can have quite different levels of smoothness, which requires two tuning parameters $\lambda_\alpha, \lambda_\beta$ to control their smoothness respectively in the penalization. 

These considerations suggest three potential candidates for the penalty $J(\alpha,\beta)$:
\begin{eqnarray*}
J(\alpha,\beta)=
\left\{\begin{array}{l}
\mbox{candidate 1: }\lambda_\alpha\lambda_\beta \konenormsq{\alpha}\ktwonormsq{\beta},\\
\mbox{candidate 2: } \lambda_\alpha\konenormsq{\alpha}\zerotnormsq{\beta} + \lambda_\beta\zerosnormsq{\alpha}\ktwonormsq{\beta},\\
\mbox{candidate 3: } \lambda_\alpha\konenormsq{\alpha}\zerotnormsq{\beta} + \lambda_\beta\zerosnormsq{\alpha}\ktwonormsq{\beta}
+\lambda_\alpha\lambda_\beta \konenormsq{\alpha}\ktwonormsq{\beta} ,
\end{array}\right.
\end{eqnarray*}
where the norms $\zerosnorm{\cdot}$ and $\zerotnorm{\cdot}$ involving covariance structure of the input are in \eqref{eqn:zerosnorm}.

A careful study of these three candidates reveals the following insights. Candidate 1 is simply the product of two one-way penalties, $\lambda_\alpha\konenormsq{\alpha}$ and $\lambda_\beta\ktwonormsq{\beta}$, which is scale-invariant and bi-quadratic. However, it is deficient because it cannot specialize to one-way penalty of FLR by setting one of $\lambda_\alpha$ and $\lambda_\beta$ to be zero when it is desirable to only penalize one dimension. Candidates 2 and 3 are both scale-invariant, bi-quadratic, and can both specialize to a form of one-way FLR. Nevertheless, Candidate 3 is the optimal choice since it ensures that the smoothness levels of $\alpha$ and $\beta$ are detached as shown below, whereas Candidate 2 entangles the smoothness levels of $\alpha$ and $\beta$, which is undesirable.

The advantage of Candidate 3 can be seen as follows. After completing the squares of the data fidelity $\ell_{n}$, the objective function $\ell_{n\lambda}$ with Candidate 3 has three types of terms as functions of $\alpha$ and $\beta$: bi-quadratic, bi-linear $n^{-1} \sum_{i=1}^n y_i\int \alpha(s) x_i(s,t) \beta(t)dsdt $, and constant $n^{-1} \sum_{i=1}^n y_i^2$. When Candidate 3 is adopted, all the bi-quadratic terms become $$n^{-1} \sum_{i=1}^n \left(\int \alpha(s) x_i(s,t) \beta(t)dsdt \right)^2+\lambda_\alpha\konenormsq{\alpha}\zerotnormsq{\beta}
+ \lambda_\beta\zerosnormsq{\alpha}\ktwonormsq{\beta}
+ \lambda_\alpha\lambda_\beta \konenormsq{\alpha}\ktwonormsq{\beta}.$$ Because of \eqref{eqn:var-zeronorm}, the population version of the bi-quadratic terms then becomes $$\zerosnormsq{\alpha}\zerotnormsq{\beta}
+ \lambda_\alpha\konenormsq{\alpha}\zerotnormsq{\beta}
+ \lambda_\beta\zerosnormsq{\alpha}\ktwonormsq{\beta}
+ \lambda_\alpha\lambda_\beta \konenormsq{\alpha}\ktwonormsq{\beta}.$$ 
These four bi-quadratic terms can be written as the product of terms related to $\alpha$ and $\beta$ separately as $$\left(\zerosnormsq{\alpha}+\lambda_\alpha\konenormsq{\alpha}\right)
\left(\zerotnormsq{\beta}+\lambda_\beta\ktwonormsq{\beta}\right).$$ 

Here, the bi-quadratic terms of $\alpha$ and $\beta$ are completely decoupled, which has the following benefits. 
Imagine that $\alpha$ is known, then optimizing the objective function with respect to $\beta$ will degenerate to a one-way FLR problem, where the quantities in front of $\zerotnormsq{\beta}$ and $\lambda_\beta\ktwonormsq{\beta}$ are always proportional to each other no matter what value $\alpha$ takes, so that the level of smoothness of $\alpha$ does not affect the optimization over $\beta$ or the smoothness of $\beta$.

However, if Candidate 2 is chosen, the bi-quadratic term cannot be written as a product of terms related to $\alpha$ and $\beta$ separately, and the quantities in front of $\zerotnormsq{\beta}$ and $\lambda_\beta\ktwonormsq{\beta}$ are $\lambda_\alpha\konenormsq{\alpha}$ and $\zerosnormsq{\alpha}$ respectively. This implies that when $\alpha$ is smoother, so that $\lambda_\alpha\konenormsq{\alpha}$ is smaller, $\lambda_\beta\ktwonormsq{\beta}$ will have a larger impact and $\beta$ will also tend to be smoother. 
In summary, Candidate 2 will make the levels of smoothness of the two coefficient functions depend on each other.

This decoupling property of Candidate 3 is not only necessary to separate the smoothness of two coefficient functions in the objective function, but also crucial in Theorem \ref{thm:upper} and its proof, where the measurements of the smoothness of $\alpha$ or $\beta$ appear separately. Otherwise, some kind of measurement of the joint level of smoothness will be necessary to understand the theoretical property.

Note that the phenomenon of decoupling only occurs because of the choice of the norm $\zerosnormsq{\alpha}, \zerotnormsq{\beta}$ defined in \eqref{eqn:zerosnorm}, which appears in Candidate 3. Adopting other norms to replace \eqref{eqn:zerosnorm} in Candidate 3 will not simultaneously satisfy the invariant, bi-quadratic, specialization to one-way FLR, and decoupling requirements. Therefore, we will use Candidate 3 from now on.

We comment that other three-term penalties have been used to address the problem of SVD of two-way functional data \citep{Huang+2009a} and the problem of bivariate smoothing \citep{xiao2013fast}. However, the penalty in \citet{Huang+2009a} only involves the standard $l_2$ norm of a vector and the norm of second order differences, which is a discrete version of $\left(\int(f'')^2\right)$, and the penalty in \citet{xiao2013fast} involves spline basis and differencing matrix,
while our penalty involves the relatively more complicated norm defined in \eqref{eqn:zerosnorm} and the Hilbert norm in a more general framework.

Due to the un-identifiability issue of Model \eqref{eqn:2dflr} and the design of the penalty in the objective function \eqref{eqn:twowayobj}, the optimization does not have unique solution. Specifically, given $(\halpha,\hbeta)$ as the optimizer of \eqref{eqn:twowayobj}, for any nonzero constant $c$, $(\halpha/c,c\hbeta)$ is also the optimizer. We consider these as an equivalent set of solutions by varying $c$ since they will lead to identical model fit and prediction. As mentioned in the introduction right after Model \eqref{eqn:2dflr}, there are various ways to make the final optimization solution unique if needed, which can be appended to the algorithm in Section \ref{ssec:algorithm-FBLR}.

\subsection{Optimization algorithm of two-way FBLR}
\label{ssec:algorithm-FBLR}

Recall that the objective function is defined as follows,
% \begin{eqnarray}
% \label{eqn:twoway-FLR}
%   \ell_{n \lambda}(\alpha,\beta) &=&
%   n^{-1} \sum_{i=1}^n \left( y_i - \int_{\tonetimesttwo} \alpha(s)
%   x_i(s,t) \beta(t)dsdt \right)^2 \nonumber \\
%   && + \lambda_\alpha\konenormsq{\alpha}\zerotnormsq{\beta} + \lambda_\beta\zerosnormsq{\alpha}\ktwonormsq{\beta}
% +\lambda_\alpha\lambda_\beta \konenormsq{\alpha}\ktwonormsq{\beta}.
% \end{eqnarray}
The bi-quadratic property of the function naturally calls for the block-descent algorithm to optimize the function iteratively. Given a starting point $\alpha^{(0)}$, one can iterate between minimizing one of $\alpha$ and $\beta$ while holding the other fixed until convergence. For the rest of this section, the focus will be on the discussion of how to update $\beta$ given $\alpha$ at each iteration. The updating rule for $\alpha$ given $\beta$ can be obtained analogously.

For any integer $k \geq 1$, suppose the estimation of $\alpha$ in the $(k-1)$th step is $\alpha^{(k-1)}$. Denote a new 1D random input function
\begin{equation}
  \label{eqn:x-tilde}
  \tilde{x}_i(t) = \int_{\T_1} \alpha^{(k-1)}(s)x_i(s,t)ds,
\end{equation}
and for any $f \in \H(K_2)$, define,
\begin{equation}
  \label{eqn:K-tilde}
  \ktwotnormsq{f} =
  \left(\lambda_\alpha\konenormsq{\alpha^{(k-1)}}\right)\zerotnormsq{f}
  +\left(\lambda_\beta \zerosnormsq{\alpha^{(k-1)}}
  +\lambda_\alpha\lambda_\beta\konenormsq{\alpha^{(k-1)}}\right)
  \ktwonormsq{f},
\end{equation}
where, given $\alpha^{(k-1)}$, the terms $\lambda_\alpha \konenormsq{\alpha^{(k-1)}}$ and $\lambda_\beta \zerosnormsq{\alpha^{(k-1)}}+\lambda_\alpha\lambda_\beta\konenormsq{\alpha^{(k-1)}}$ are both known constants. 
%Throughout this paper, we assume that $\zerosnormsq{f}\ne 0$ and $\zerotnormsq{f}\ne0$ hold for any $f$ that belongs to the null space of $K_1$ or $K_2$ and $f\ne 0$. This assumption is necessary to guarantee that the minimizer of $\expec\ell_{n \lambda}(\alpha,\beta)$ is unique even when the optimization is constrained to the null space of $K_1$ or $K_2$. 
Recall the assumption on the relationship between $C_\beta$ and $K_2$, which we make at the end of Section \ref{sec:notation}. It is seen that $\ktwotnorm{\cdot}$ is a norm since $\ktwotnormsq{f}=0$ if and only if $f=0$ and $\ktwotnormsq{f}$ is quadratic. We further let $\tilde{K}_2(\cdot, \cdot)$ be the reproducing kernel associated with the norm $\ktwotnorm{\cdot}$.

Given $\alpha^{(k-1)}$, a new one-dimensional predictor \eqref{eqn:x-tilde} and a new kernel \eqref{eqn:K-tilde}, the objective function $\ell_{n \lambda} (\alpha, \beta)$ for the $k$-th step becomes a functional of $\beta$ alone, denoted by $\ell_{n\lambda}(\beta;\alpha^{(k-1)})$, and can be re-expressed in a compact form,
\begin{equation}
\label{eqn:twotooneobjfcn}
\ell_{n\lambda}(\beta;\alpha^{(k-1)})= n^{-1/2} \sum_{i=1}^n \left(y_i -
\int_{\T_2} \tilde{x}_i(t)\beta(t) dt \right)^2 + 1 \times
\|\beta\|_{\tilde{K}_2}^2.
\end{equation}
The above objective function \eqref{eqn:twotooneobjfcn} is the same as the 1D FLR objective function \eqref{eqn:onewayobj} with inputs $\{(\tilde{x}_i(\cdot),y_i)\}_{i=1}^n$, kernel $\tilde{K}_2(\cdot,\cdot)$, and tuning parameter $\lambda=1$. In other words, by fixing $\alpha^{(k-1)}$, the FBLR problem degenerates to an FLR problem with respect to $\beta$. Hence, the intermediate $\beta^{(k)}$ can be obtained from the 1D FLR via the representer theorem (\ref{eqn:onewaycconstants}). 

To summarize, the complete approach to obtain the estimators of FBLR is schematically presented in Algorithm \ref{algo:iter} with given initialization, known covariance structure of the predictor, and fixed tuning parameters. Some details of the initialization, covariance structure, and tuning parameter selection are provided below.

%%%%%%%%%%%%%%%%%%%%%%%%%%%%%%%%%%%%%%%%%%%%%%%%%%%%%%%%%%%%%%%%%%%%%%%
\vspace{-1em}
\begin{algorithm}[!ht]
\SetAlgoLined
\KwIn{}
1. The observations $(x_i(\cdot,\cdot),y_i), i=1,...,n$; initial estimator $\alpha^{(0)}$\;
2. Penalty parameters $\lambda_\alpha$ and $\lambda_\beta$; reproducing kernels $K_1$ and $K_2$\;
3. Pre-specified norms $\zerosnorm{\cdot}$ and $\zerotnorm{\cdot}$ as in \eqref{eqn:zerosnorm}\;
{\Repeat{Convergence}{
\nl To obtain $\beta^{(k)}$ while fixing $\alpha^{(k-1)}$ \label{algo:iter-step-1}\\
    \Begin{
        \nl Compute $\tilde{x}_i(\cdot)$ according to (\ref{eqn:x-tilde})\\
        \nl Evaluate $\|\alpha^{(k-1)}\|_{K_1}^2$ and $\zerosnormsq{\alpha^{(k-1)}}$\\
        \nl Derive the reproducing kernel $\tilde{K}(\cdot,\cdot)$ associated with the norm defined in (\ref{eqn:K-tilde})\\
        \nl Solve \eqref{eqn:twotooneobjfcn} for $\beta^{(k)}$ via the one-way FLR approach
    }
\nl To obtain $\alpha^{(k)}$ while fixing $\beta^{(k)}$ \label{algo:iter-step-2}\\
    \Begin{
        \nl Switch the role of $\alpha$ and $\beta$ in Steps 2-5
      }}
\KwOut{Estimators $\halpha$ and $\hbeta$}
}
\caption{The smoothness regularization approach to the FBLR problem \eqref{eqn:2dflr}.}
\label{algo:iter}
\end{algorithm}

\vspace{-1em}
\paragraph{Initialization.} 
%\textcolor{red}{remove the following: It is commonly understood that different starting points will lead to multiple stationary solutions rather than the global optimizer due to the bi-convexity of the objective function \eqref{eqn:twowayobj}. In practice, it is recommended and frequently adopted to try multiple random starting points and choose the one with minimum value of the objective function after convergence. The theoretical justification of such practice can be found, for instance, for the tensor decomposition problem, in \citet{AGJ14}.} Another possibility is 
Based on the understanding that if the initial point is close to the truth, then local contraction property will make the convergent point to be the global optimal one with appealing theoretical property. Hence it is often true that one only needs a consistent estimator to begin with and the iterative procedure will produce an optimal solution. 
There are two possible choices for initialization. The first one is to regress on the 2D functional PCs \citep{chen2017modelling}, because these estimated PCs have desirable theoretical properties. The second one is to initialize our algorithm with the estimator obtained from Ridge regression after naive vectorization of the two-way covariates. We recommend the latter one for two reasons: 
(i) when the reproducing kernel and the covariance kernel align,
the former one is computationally much more expensive than the latter while leading to almost identical results; (ii) when these two kernel are misaligned, the former one is inconsistent. These facts are revealed further in the simulation.

\paragraph{Estimation of the covariance function.} To implement Algorithm \ref{algo:iter}, the input of $\zerosnorm{\cdot}$ and $\zerotnorm{\cdot}$ as in \eqref{eqn:zerosnorm} is required and it depends upon the separable covariance structures $C_\alpha$ and $C_\beta$ of the covariate. However, for most applications, the two covariance functions are unknown. Hence, we adopt an iterative algorithm introduced in \citet{Werner+2008a} to estimate them. It basically fixes one of $C_\alpha$ and $C_\beta$ and estimates the other. %Other non-iterative algorithm to estimate such covariance structure also exists, but may have inferior estimation accuracy.

\paragraph{Tuning parameter selection.} The selection of tuning parameters plays a crucial role in determining the eventual performance of the algorithm. The most straightforward way is to use cross validation. However, since there are two tuning parameters, $\lambda_\alpha$ and $\lambda_\beta$, the search grid will be two dimensional and the computational cost will be extremely high. So we propose to use the following iGCV approach. As described in Algorithm \ref{algo:iter}, $\alpha$ or $\beta$ is updated via one-way FLR given the other. Fixing one of them, say $\beta$ and $\lambda_\beta$, the selection of $\lambda_\alpha$ can be done via generalized cross validation 
and the $\alpha$ can be updated once $\lambda_\alpha$ is chosen, and vice versa. The iGCV algorithm terminates when the choices of $\lambda_\alpha$ and $\lambda_\beta$ in the current iteration remain the same as the previous one, and the distances of $\alpha$ and $\beta$ between the current iteration and the previous one are less than some pre-determined tolerance level. We emphasize that once the iGCV algorithm stops, the tuning parameters $\lambda_\alpha$ and $\lambda_\beta$ are selected, and the estimation of $\alpha$ and $\beta$ is completed as well. Hence there is no need to perform another round of iteration for the finally chosen tuning parameters.

%%%%%%%%%%%%%%%%%%%%%%%%%%%%%%%%%%%%%%%%%%%%%%%%%%%%%%%%%%%%%%%%%%%%%%%
\section{Theoretical results: optimal rate of convergence}
\label{sec:theory}

\subsection{Preliminary}
\label{ssec:theory-preliminary}

In this section, we introduce some notations and assumptions on the reproducing kernels and covariances that will be used in the development of the minimax rate of convergence.

From the methodology point of view, it is implicitly assumed that the two domains, the covariance structures of the two dimensions, and the levels of smoothness of the two coefficient functions may be different and require different penalties thereby. For notational simplicity, throughout the theoretical section, we assume that they are the same, that is, $\T_1 = \T_2 = \T, ~K_1=K_2=K, ~C_\alpha=C_\beta=C$ and $\lambda_\alpha=\lambda_\beta=\lambda$. 
%But the discussions and theorems in this section are still applicable with slight adjustment when the two domains and their corresponding properties are different.
It follows that the $\zerosnorm{\cdot}$ and $\zerotnorm{\cdot}$ norms are identical, and hence we write them as $\zeronorm{\cdot}$. The theoretical results and proofs follow similar logic for the distinct version, with more complicated notation. In particular, the theorems below will hold with the slower rate produced by the two dimensions. See Section \ref{sec:theory-distinct} in the Appendix for the statement of the theorems and a brief sketch of the proofs for the version with distinct domains.

%% To avoid ambiguity, we restate the model assumptions and optimization criterion. The covariance operator of the predictor $X(\cdot,\cdot)$ can be rewritten as
%% \begin{equation*}
%% C(s_1,t_1,s_2,t_2) = \expec[X(s_1,t_1)X(s_2,t_2)] = C(s_1,s_2)C(t_1,t_2).
%% \end{equation*}
%% The $\zerosnorm{\cdot}$ and $\zerotnorm{\cdot}$ norms are identical and defined as
%% \begin{equation}
%% \label{eqn:zeronorm}
%% \zeronorm{f} = \left(\intttimest f(s) C(s,t) f(t) ds dt\right)^{1/2}.
%% \end{equation}
%% The objective function \eqref{eqn:twoway-FLR} now becomes
%% \begin{eqnarray*}
%% \ell_{n \lambda}(\alpha, \beta) =
%% n^{-1} \sum_{i=1}^n\left(y_i-\int_{\T\times\T}\alpha(s) x_i(s,t) \beta(t)  dsdt\right)^2 + \lambda\|\alpha\|_0^2\|\beta\|_K^2+\lambda\|\beta\|_0^2\|\alpha\|_K^2+\lambda^2\|\alpha\|_K^2\|\beta\|_K^2.
%% \end{eqnarray*}

The eigen-structures of the kernel $K$ and the covariance $C$, and their alignment jointly determine the performance of FLR \citep{Yuan+2010b, cai2012minimax}. Similarly, they play an important role in the theoretical property of FBLR as well. Suppose that both the reproducing kernel $K$ and the covariance $C$ are continuous and square integrable. By Mercer's Theorem, $K$ and $C$ have the
following spectral decompositions,
$ K(s,t) = \sum_{k=1}^\infty s_k^K \phi_k^K(s)\phi_k^K(t)$ and $  C(s,t) = \sum_{k=1}^\infty s_k^C \phi_k^C(s)\phi_k^C(t)$,
where $s_1^K\ge s_2^K \ge \cdots \geq 0 $ and $s_1^C\ge s_2^C \ge \cdots \ge 0 $ are the eigenvalues of $K$ and $C$ in descending order, and $\{\phi_1^K,\phi_2^K,...\}$ and $\{\phi_1^C,\phi_2^C, \ldots\}$ are the corresponding orthonormal eigenfunctions of $K$ and $C$, respectively.

Given the norms $\|\cdot\|_0$ and $\|\cdot\|_K$, for any function $f \in \H(K)$, define a new norm $\|\cdot\|_R$ that combines the two as $\|f\|_R^2 = \|f\|_0^2 + \|f\|_K^2$.
Note that $\|\cdot\|_R$ is indeed a norm as discussed earlier in Section \ref{ssec:algorithm-FBLR}, since $\|f\|_0^2\ne 0$ holds for any $\|f\|_K^2=0$ and $f\ne 0$. 
Let $R$ be the corresponding kernel associated with the $\|\cdot\|_R$ norm. Since $R$ is also continuous and square integrable, it follows from Mercer's Theorem that $R$ has the following spectral decomposition, $R(s,t) = \sum_{k=1}^\infty s_k^R \phi_k^R(s)\phi_k^R(t)$,
where $s_1^R\ge s_2^R \ge \cdots \geq 0$ and $\{\phi_k^R:k=1,2,\ldots\}$ are eigenvalues and eigenfunctions respectively.

In general, for a square integrable, symmetric, and non-negative definite function $R:\T \times \T \mapsto \R$ (similarly for $K$ and $C$), its corresponding linear operator can be defined as $\L_R(f)(\cdot) =\int_\T R(t,\cdot)f(t)dt$.
It follows from the definitions of the eigenvalues and eigenfunctions of the spectral decomposition of $R$ that
$\L_R(\phi_k^R) = s_k^R\phi_k^R$, for $k=1,2,\ldots$.
Define the square root of the linear operator as
$\L_{R^{1/2}}(\phi_k^R) = (s_k^R)^{1/2}\phi_k^R$, for $k=1,2,\ldots$.
Now consider a new linear operator $\L_T = \L_{R^{1/2}CR^{1/2}}$, that is $\L_T(f)=\L_{R^{1/2}}(\L_C(\L_{R^{1/2}}(f)))$. Since $\L_T$ is a bounded linear operator, there exist eigenvalues %$\{s_k^T:k=1,2,\ldots\}$ 
$\{s_1^T, s_2^T,\ldots\}$ in descending order and the corresponding eigenfunctions %$\{\phi_k^T:k=1,2,\ldots\}$
$\{\phi_1^T, \phi_2^T,\ldots\}$, such that 
$\L_T(\phi_k^T) = s_k^T\phi_k^T$, for $k=1,2,\ldots$.

Define $\omega_k = (s_k^{T})^{-1/2}\L_{R^{1/2}}(\phi_k^T)$ and
$\gamma_k = (1/s_k^T-1)^{-1}$. 
% Then it is easy to validate the following identities,
% \begin{equation}
% \label{eqn:kernelidentities}
% \langle\omega_j,\omega_k\rangle_{R} = \delta_{jk} (1/\gamma_k+1),~~
% \langle C\omega_j,\omega_k\rangle_{\L_2} = \delta_{jk},~~ \mbox{and }
% \langle \omega_j,\omega_k\rangle_{K} = \delta_{jk} /\gamma_k.
% \end{equation}
The functions $\{\omega_k:k=1,2,\ldots\}$ are essential in the proof, since we will expand all of the functions of interest onto these basis functions. The decay rate of $\gamma_k$ plays a prominent role in the convergence rate.

We will impose the following conditions:\\
{\bf Condition 1:} the values $\gamma_k$ satisfy the decay rate,
\begin{equation}
  \label{eq:gamma-order}
  \gamma_k\asymp k^{-2r},
\end{equation}
for some constant $0<r<\infty$.\\
{\bf Condition 2:} for any two functions $f, g \in \L_2(\T)$, we further assume that the
following fourth moment condition holds,
\begin{equation}
\label{eq:4th-moment-condition}
\expec \left(\int_{\T\times\T} f(s)X(s,t)g(t)dsdt \right)^4 \le M
\left( \expec \left( \int_{\T\times\T} f(s)X(s,t)g(t)dsdt \right)^2 \right)^2,
\end{equation}
for some constant $M>0$.

Note that Condition 2 is satisfied with $M=3$ when the process $X$ is assumed to be normal and is therefore weaker than the Gaussian assumption.

\begin{remark}[On Condition 1]
\label{rmk:condition}
There are a few facts which can facilitate the understanding of this condition on the decay rate of $\gamma_k$. First, in the literature of nonparametric statistics, it is known that when Sobolev space is studied, the kernel $K$ has eigenvalues decaying as $s_k^K\asymp k^{-2r_K} \mbox{ for some } r_K>1/2$ \citep{MW81}. Second, if the covariance $C$ satisfies the Sacks-Ylvisaker conditions of order $r_C-1$, then its eigenvalues decay as $s_k^C \asymp k^{-2r_C}$ \citep{SY66, SY68, SY70}. Third, if the kernel $K$ and the covariance $C$ share the same set of ordered eigenfunctions $\phi_k^K=\phi_k^C$ for all $k$, that is, under the scenario with perfect alignment, Proposition 4 in \citet{Yuan+2010b} shows that $\gamma_k=s_k^Cs_k^K$. This implies that when perfect alignment happens and eigenvalues of $C$ and $K$ decay with the parameters $r_C$ and $r_K$, then $\gamma_k$ decays with parameter $r=r_C+r_K$. 
Fourth, even if the assumption of perfect alignment is violated, $\gamma_k\asymp s_k^Cs_k^K$ holds in other situations such as Sobolev space $\H$ and $C$ with Sacks-Ylvisaker condition (see Theorem 5 in \citet{Yuan+2010b}), or when $K$ and $C$ are commutable. Lastly, the decay rate of $\gamma_k$ depends on not only the decay rates of $s_k^K$ and $s_k^C$, but also the alignment of the eigenfunctions of $K$ and $C$. For instance, when $\phi_k^K=\phi_{k^2}^C$, and $s_k^C \asymp k^{-2r_C}, s_k^K\asymp k^{-2r_K}$, we have $\gamma_k\asymp k^{-(4r_C+2r_K)}\asymp k^{-2r}$, where $r=2r_C+r_K$. Eventually, $r$ will show up in the minimax upper and lower bounds.
\end{remark}

\subsection{Optimal rate of convergence}
In this section, we study the asymptotic properties of the FBLR and provide justification of the methodology proposed in Section \ref{sec:method}. We first establish a minimax upper bound for the smoothness regularization estimator in Theorem \ref{thm:upper} and then derive a minimax lower bound for all possible estimators in Theorem \ref{thm:lower}. The upper bound matches the lower bound and hence our proposed smoothness regularization estimator is rate optimal.

We assess the accuracy of the estimators by the excess prediction risk. Suppose $(X^*,Y^*)$ has the same distribution as $(X,Y)$ and is independent of the training data $\{(x_i(\cdot,\cdot),y_i)\}_{i=1}^n$. By taking expectation only with respect to $(X^*,Y^*)$, denoted by $\expecs(\cdot)$, the excess prediction risk of the estimates $\halpha,\hbeta$ over  the true coefficient functions $\alpha_0,\beta_0$ is defined as follows,
\vspace{-.5em} 
{\small
\begin{eqnarray*}
  \calE (\halpha,\hbeta;\alpha_0,\beta_0) &=&
  \expecs \left( Y^*-\int_{\T\times\T}\halpha(s)X^*(s,t)\hbeta(t)dsdt\right)^2
 -\expecs \left(
 Y^*-\int_{\T\times\T}\alpha_0(s)X^*(s,t)\beta_0(t)dsdt\right)^2\\ [-0.3em]
&=&\expecs \left( \int_{\T\times\T}X^*(s,t) \left( \halpha(s)\hbeta(t)-\alpha_0(s)\beta_0(t) \right)dsdt \right)^2.
%% &=&\int_{\T^4} C(s_1,s_2) C(t_1,t_2)
%% \left( \halpha(s_1)\hbeta(t_1)-\alpha_0(s_1)\beta_0(t_1) \right)
%% \left( \halpha(s_2)\hbeta(t_2)-\alpha_0(s_2)\beta_0(t_2) \right)ds_1ds_2dt_1dt_2.
\end{eqnarray*}
}
Theorem \ref{thm:upper} states the upper bound for the excess prediction
risk of the smoothness regularized estimator with a properly chosen tuning parameter $\lambda$.
\begin{theorem}
  \label{thm:upper}
  Under Conditions 1-2, the smoothness regularization estimators $(\halpha,\hbeta)$ defined in \eqref{eqn:twowayobj} with Candidate 3 as the penalty and %$\lambda=O(n^{-2(r_c+r_k)/(2(r_c+r_k)+1)})$
  $\lambda=O(n^{-2r/(2r+1)})$ satisfies
  \begin{equation*}
    \lim_{A\rightarrow\infty}\lim_{n\rightarrow\infty}\sup_{\alpha_0 \in
      \H(K),\; \beta_0 \in \H(K)}
    \prob \left(\mathcal{E}(\halpha,\hbeta;\alpha_0,\beta_0)\ge
    A%n^{-\frac{2(r_c+r_k)}{2(r_c+r_k)+1}} 
    n^{-\frac{2r}{2r+1}}
    \right)=0.
  \end{equation*}
\end{theorem}

\vspace{-1em}
\begin{remark}[On Theorem \ref{thm:upper}]
\label{rmk:thm}
By assuming the same domains, kernels, and covariance structures along the two dimensions of the 2D functional covariates, the convergence rate is determined jointly by the joint properties of the covariance $C$ and the kernel $K$ and behaves like most of the non-parametric statistical problems. Furthermore, this result demonstrates that faster decay rate of the eigenvalues of the covariance $C$ will lead to faster convergence rate of the estimator. Interestingly, this upper bound recovers the same convergence rate of the one-way FLR. We can relax these assumptions to allow different domains, kernels and/or covariance structures. The proof is essentially the same but with more complicated notation. The convergence rate will be the maximum of the convergence rates from the two domains %$\max\{n^{-\frac{2(r_{c\alpha}+r_{k\alpha})}{2(r_{c\alpha}+r_{k\alpha})+1}},n^{-\frac{2(r_{c\beta}+r_{k\beta})}{2(r_{c\beta}+r_{k\beta})+1}}\}$, 
$\max\{n^{-\frac{2r_\alpha}{2r_\alpha+1}},n^{-\frac{2r_\beta}{2r_\beta+1}}\}$, 
where the subscripts $_\alpha$ and $_\beta$ differentiate the discrepancies between the two domains corresponding to the $\alpha$ and $\beta$ dimensions, respectively.
\end{remark}

\vspace{-1em}
%\textcolor{red}{Remark 4 right after Remark 1?} NO, it follows after the theorem, which shows the rate.
\begin{remark}[On the impact of alignment between kernel and covariance]
\label{rmk:align}
Since the convergence rate depends on $r$, defined in \eqref{eq:gamma-order}, in a form of $n^{-\frac{2r}{2r+1}}$, Theorem \ref{thm:upper} suggests that, the larger the $r$, the faster the convergence. Remark \ref{rmk:condition} discussed the relationship between $r$ and $r_C$, $r_K$, which measures the decay of the covariance and kernel, respectively. When the perfect alignment occurs, $r$ takes its largest value of $r_C+r_K$, achieving the fastest convergence. When misalignment occurs, $r$ is smaller, which leads to slower convergence. 
\end{remark}

Theorem \ref{thm:lower} provides the lower bound for the excess prediction risk over all possible estimators, which is achieved by our estimator.
\begin{theorem}
  \label{thm:lower}
  Under the same assumptions as in Theorem \ref{thm:upper}, for any estimate $(\talpha,\tbeta)$ based on the observations $\{(x_i(\cdot,\cdot),y_i): i=1,2,\ldots,n\}$, we have the following lower bound,
  \begin{equation*}
    \lim_{a\rightarrow 0}\lim_{n\rightarrow\infty}\inf_{\talpha,\tbeta}\sup_{\alpha_0 \in
      \H(K),\; \beta_0 \in \H(K)} \prob\left(\mathcal{E}(\talpha,\tbeta;\alpha_0,\beta_0)\ge
    a%n^{-\frac{2(r_c+r_k)}{2(r_c+r_k)+1}}
    n^{-\frac{2r}{2r+1}}
    \right)=1.
  \end{equation*}
\end{theorem}

The upper bound in Theorem \ref{thm:upper} and the lower bound in Theorem \ref{thm:lower} match each other. So we establish the rate optimality of our proposed smoothness regularization estimator.

\section{Simulations}
\label{sec:sim}

\subsection{Simulation settings}
\label{ssec:sim-setting}

We now demonstrate numeric properties of our proposed methodology, in comparison with a few existing methods. In particular, we consider three model settings as discussed in the introduction: our proposed functional bilinear Model \eqref{eqn:2dflr}, the broader low rank functional bilinear Model \eqref{eqn:2dflr-2}, and the broadest Model ~\eqref{eqn:2dflr-other} for robustness verification. The last setting is unfavorable to our methodology as the true coefficient function is two-dimensional instead of the product of two 1D functions. However, it will be shown that our method still performs well with the deflation approach mentioned in Section~1, right after introducing Model \eqref{eqn:2dflr-2}.

We consider three streams of existing methods. The first stream is the Bayesian approach with Gaussian Markov random field (GMRF) priors \citep{happ2018impact}. The second one is what we refer to as 2D-FPCR, which is based upon the regression of the scalar response on the estimated 2D functional PCs (2D-FPCA). \citet{chen2012modeling,park2015longitudinal,chen2017modelling} all studied the problem of 2D-FPCA and we will adopt the last one because of its nice properties. \citet{chen2017modelling} described two versions of 2D-FPCA to estimate the PCs, % for iid 2D observations $x_i(\cdot,\cdot), i=1,\ldots, n$, 
namely, product FPCA and marginal FPCA, %depending on two different assumptions of Karhunen–Lo\`eve representation of the covariance operator. These two versions 
which lead to two estimators of $\beta_0(\cdot,\cdot)$ in Model \eqref{eqn:2dflr-other} respectively, referred to as PFPCR and MFPCR accordingly. Implementation of FPCR requires the knowledge of the number of PCs. In what follows, we will compare various possibilities for the unknown ranks.\footnote{For GMRF, the implementation is available on the webpage \href{https://github.com/ClaraHapp/SOIR}{https://github.com/ClaraHapp/SOIR}. For the calculation of 2D-FPCA, we use the \texttt{ProductFPCA} and \texttt{MarginalFPCA} functions in the \texttt{PACE} package at \href{https://www.stat.ucdavis.edu/PACE/}{https://www.stat.ucdavis.edu/PACE/}.}

The third stream targets at the estimation of $\beta_0(\cdot,\cdot)$ in Model \eqref{eqn:2dflr-other} directly via an RKHS framework. Such a task can be accomplished via solving \eqref{eqn:onewayobj} while considering a four-dimensional reproducing kernel $K(\cdot,\cdot,\cdot,\cdot)$. To the best of our knowledge, there is no literature that specifically studies the theoretical properties for solving the problem of estimating Model \eqref{eqn:2dflr-other} via adopting a four-dimensional kernel in \eqref{eqn:onewayobj}. Although \citet{sun2018optimal,sang2022nonlinear} considered a four-dimensional kernel, they were for the problem with one-dimensional functional input, one-dimensional functional output, and two-dimensional coefficient function. 

A natural choice for the four-dimensional kernel is associated with the tensor product RKHS. Suppose for one-dimensional domain, $K(\cdot,\cdot)=K_0(\cdot,\cdot)+K_1(\cdot,\cdot)$, where the $K_0(\cdot,\cdot)$ and $K_1(\cdot,\cdot)$ correspond to the null space and its orthogonal complement, respectively, of some penalty. For two-dimensional domain, consider the two marginal reproducing kernels $K^1(\cdot,\cdot)=K_0^1(\cdot,\cdot)+K_1^1(\cdot,\cdot)$ and $K^2(\cdot,\cdot)=K_0^2(\cdot,\cdot)+K_1^2(\cdot,\cdot)$. Note that whenever we discuss FLR+TPK in this section, the superscript denotes the domain and the subscript denotes either null or orthogonal complement, which is slightly different from the notations in Sections \ref{sec:method}-\ref{sec:theory}. The four-dimensional kernels corresponding to the null space and the orthogonal complement of the tensor product space satisfy that $$K_0((s_1,t_1),(s_2,t_2)) = K_0^1(s_1,s_2) K_0^2(t_1,t_2),\quad {\rm and},$$  
\be
K_1((s_1,t_1),(s_2,t_2)) = K_0^1(s_1,s_2) K_1^2(t_1,t_2) + 
K_1^1(s_1,s_2) K_0^2(t_1,t_2) + 
K_1^1(s_1,s_2) K_1^2(t_1,t_2).\nonumber
\ee
%, where the reproducing kernel $K \left((\cdot,\cdot),(\cdot,\cdot)\right)$: $(\T_1 \times \T_2) \times (\T_1 \times \T_2) \mapsto \R$, is formed by the product of reproducing kernels on the two marginal domains $K_1(\cdot,\cdot)$: $\T_1 \times \T_1 \mapsto \R$ and $K_2(\cdot,\cdot)$: $\T_2 \times \T_2 \mapsto \R$, i.e., $K((s_1,t_1),(s_2,t_2)) = K_1(s_1,s_2) K_2(t_1,t_2)$.
We refer to this approach as FLR+TPK, since it is functional linear regression with tensor product kernel. For more details on the tensor product RKHS, see Chapter 2 of \citet{gu2013smoothing}. Note that FLR+TPK only has one tuning parameter that controls the overall smoothness of $\beta(\cdot,\cdot)$, unlike the two tuning parameters in our method that control the smoothness of $\alpha(\cdot)$ and $\beta(\cdot)$ separately.

%\carlB{We also consider a specific form of FLR model under RKHS framework with a tensor product kernel and we refer it as FLR+TPK. Specifically, The associated RKHS $\mathcal{H}(K)$ has a tensor decomposition, $\mathcal{H}(K) = \mathcal{H}_1(K_1) \otimes \mathcal{H}_2(K_2)$, where $\mathcal{H}_1(K_1)$ and $\mathcal{H}_2(K_2)$ are the respective RKHSs associated with $K_1(.,.)$ and $K_2(.,.)$. Notably, while the tensor product RKHS framework has been used in function-on-function regression \citep[e.g][]{balasubramanian2022unified,sun2018optimal,sang2022nonlinear}, it has not yet been applied, to the best of our knowledge, in scalar-on-matrix function regression. }

We consider several other far less competitive competitors in the online supplementary materials (Appendix \ref{sec:exsim}). The first one is FLR of one-dimensional input function after vectorization. %, with either CV or GCV to select tuning parameter. 
The second one is a naive implementation of Ridge after plain vectorization without considering the smoothness.
The third one is bilinear regression with no penalty, which is a special case of FBLR when $\lambda_\alpha=\lambda_\beta=0$ and will be called BLR. A quick summary of the message is that all these methods are much worse than FBLR, because they either do not keep matrix/tensor structure or do not take advantage of the smoothness.

% We generate data from six different settings: four settings from Model \eqref{eqn:2dflr}, in favor of our methods, one setting from Model \eqref{eqn:2dflr-2}, and one more from Model \eqref{eqn:2dflr-other}, in favor of the other methods. \textcolor{red}{Below you said first frou from Model 2 and last two from Model 3. Inconsistent?} 
%The detailed setup of the six settings is as follows. 
Under all six settings, we adopt the same covariance structures from \cite{cai2012minimax},  given by
\be
\label{eq:cov-200}
       C_{\alpha}(s,t) = C_{\beta}(s,t) = \sum_{i = 1}^{200} 2 i^{-2r_c} \textrm{cos}(i \pi s)\cos(i \pi t),
\ee
where $r_c$ controls the smoothness of the function. The parameter $r_c$ appears implicitly in the upper bound of Theorem \ref{thm:upper} and drives the convergence rate, according to Remark \ref{rmk:condition}. Four choices of $r_c$ are considered: 1, 1.5, 2, and 2.5.

\begin{figure}[!ht]
	\centerline{\includegraphics[width=\textwidth]{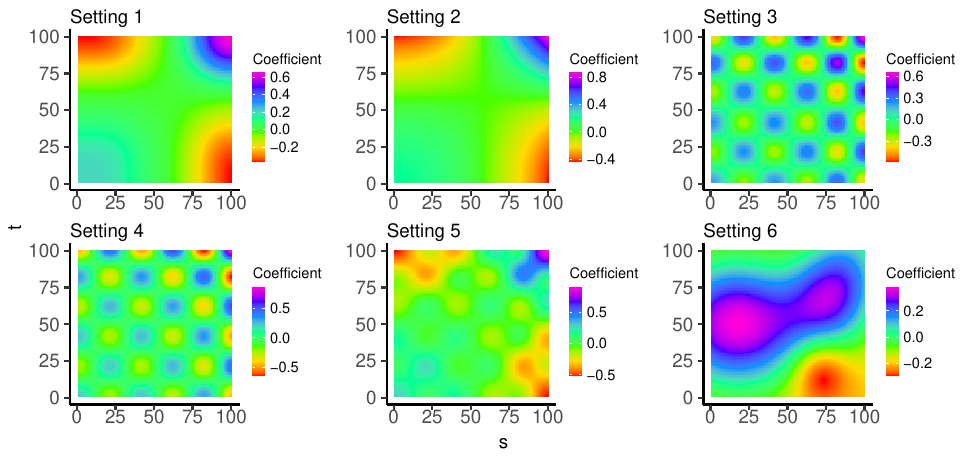}}
	\caption{Settings 1-6: Heat-map plots of the true coefficient functions $\beta_0(s,t)$.}
	\label{fig:r-beta0}
\end{figure}

Under all settings, the coefficient functions are set up differently for different purposes. The heat-map plots of $\alpha_{0}(s)\beta_{0}(t)$ in Model \eqref{eqn:2dflr} for the first four settings, the heat-map plot of $\alpha_{0}^{[1]}(s)\beta_{0}^{[1]}(t)+\alpha_{0}^{[2]}(s)\beta_{0}^{[2]}(t)$ in Model \eqref{eqn:2dflr-2} for Setting 5, and the heat-map plot of $\beta_0(s,t)$ in Model \eqref{eqn:2dflr-other} for Setting 6 are given in Figure \ref{fig:r-beta0}. 

Specifically, for the first four settings, the coefficient functions are given by
\begin{equation}
    \label{eq:slope_1to4}
    \alpha_0(t) = \beta_0(t) = 4\sqrt{2}\sum_{i=1}^{n_{\textrm{eig}}}(-1)^i i^{-2} \cos \big((i + k_{\textrm{mis}}) \pi t\big).
\end{equation}
Here, $n_{\textrm{eig}}$ controls the number of eigenfunctions in the coefficient function and will be set to be 4 and 200 later. Furthermore, $k_{\textrm{mis}}$ controls the degree of misalignment between the covariance eigen structure in \eqref{eq:cov-200} and the leading basis functions in the coefficient function \eqref{eq:slope_1to4}. 
Because the leading eigenfunctions in \eqref{eq:cov-200} are ordered from the first to the last as $i=1,2,\ldots, 200$ in $\cos(i\pi t)$, while the first leading basis function in \eqref{eq:slope_1to4} is $\cos \big((1 + k_{\textrm{mis}}) \pi t\big)$. When $k_{\textrm{mis}}=0$, there is no misalignment. As $k_{\textrm{mis}}$ increases, the misalignment becomes more severe. Settings 1-4 consider combinations of $n_{\textrm{eig}}\in\{4,200\}$ and $k_{\textrm{mis}}\in \{0,4\}$, with detailed configurations in Table \ref{tab:cases1to4}.

\begin{table}[ht]
    \centering
    \begin{tabular}{c|cc|c|c}
    \hline
    Setting &  $n_{\textrm{eig}}$ & $k_{\textrm{mis}}$ &
    indices $i$ of basis function $\cos(i\pi t)$ 
    & indices $i$ of basis function $\cos(i\pi t)$ \\
    &&&in the covariances $C_{\alpha}$ and $C_{\beta}$& in the coefficient $\alpha_0(\cdot),\beta_0(\cdot)$\\
    \hline
    1 & 4 & 0 &  $1,2, \ldots, 200$ &  $1,2, 3, 4 $\\
    \hline
    2 & 200 & 0 & $1,2, \ldots, 200$& $1,2, \ldots, 200$\\
    \hline
    3 & 4 & 4 &  $1,2, \ldots, 200$& $5,6, 7, 8$\\
    \hline
    4 & 200  &  4 & $1,2, \ldots, 200$& $5,6, \ldots, 204$\\
    \hline
    \end{tabular}
    \caption{Settings 1-4: The configurations of the covariances and coefficient functions. %The number of eigenfunctions in the coefficient functions $n_{\textrm{eig}}$ takes value 4 or 200. The degree of misalignment $k_{\textrm{mis}}$ takes value 0 or 4. All of these four settings share the same covariance structure, whose indices of eigenfunctions include $1,2,\ldots,200$.
    }
    \label{tab:cases1to4}
\end{table} 

Settings 5-6 are based on Model \eqref{eqn:2dflr-other}, where the true coefficient function $\beta_{0}(s,t)$ is indeed two-dimensional, instead of the product of two 1D functions as our model assumes. Setting 5 considers a two-dimensional coefficient function, as the sum of two terms, where each term is a product of two 1D functions, $\beta_{0}(s,t)=\alpha_{0}^{[1]}(s)\beta_{0}^{[1]}(t)+\alpha_{0}^{[2]}(s)\beta_{0}^{[2]}(t)$,
where 
%\begin{align*}
%&\alpha_{0}^{[1]}(t)=\beta_{0}^{[1]}(t)=4\sqrt{2}\sum_{i=1}^{4}(-1)^{i}i^{-2}\cos(i\pi t),\\
%&\alpha_{0}^{[2]}(t)=\beta_{0}^{[2]}(t)=4\sqrt{2}\sum_{i=1}^{4}(-1)^{i}i^{-2}\cos\big((i + 4)\pi t\big).
%\end{align*}
$\alpha_{0}^{[1]}(t)=\beta_{0}^{[1]}(t)=4\sqrt{2}\sum_{i=1}^{4}(-1)^{i}i^{-2}\cos(i\pi t)$, and $\alpha_{0}^{[2]}(t)=\beta_{0}^{[2]}(t)=\sqrt{0.4}\times 4\sqrt{2}\sum_{i=1}^{4}(-1)^{i}i^{-2}\cos\big((i + 4)\pi t\big)$.

Setting 6 uses a two-dimensional coefficient function borrowed from the GMRF literature \citep{happ2018impact}. We magnified their coefficient function by four times to make its Frobenius norm maintain at the same level as those of the other five settings. 

Under these six settings, the data are generated according to Models \eqref{eqn:2dflr} or \eqref{eqn:2dflr-other}, where $\mu_0 = 0$, and noise level $\sigma = 0.5$. %The coefficient functions are defined above and 
The predictor $X(s,t)$ follows a centered Gaussian process with the covariance structure described above. The sample 
size $n=2^5$, $2^6$, $2^7$, and $2^{8}$. The continuous functions are observed on a regular grid of length 100. %The number of eigenfunctions of the covariance operator is 200.
For each value of $r_c$ and each value of $n$, we repeat the experiments 100 times. The numerical results for different level of smoothness are similar. Hence, to save space, we present the results for all four choices of $r_c$ for Setting 1, and only for $r_c = 1$ for the other five settings. 
The kernel functions for the RKHS of FBLR %\carB{is}
{and the marginal RKHS of FLR+TPK are} the same,
\be
\label{eq:sim-kernel}
    K(s,t) = -B_4(|s-t|/2)/3 - B_4((s+t)/2)/3,
\ee
where $B_4(\cdot)$ is the 4th Bernoulli polynomial. This kernel indicates the Hilbert norm. 
%$\parallel f \parallel^2_\mathcal{H}=\int(f'')^2$. 

\subsection{Simulation results}
\label{ssec:sim-results}

Under Setting 1, we first examine the performance of FBLR. As discussed in Section \ref{ssec:algorithm-FBLR}, the implementation of our method needs special attention to the covariance structure estimation, tuning parameter selection, and proper initialization. We include both the true covariances $C_\alpha, C_\beta$ and the estimated ones $\hat{C_\alpha},\hat{C_\beta}$, with the truth as the oracle benchmark. We also consider two choices of tuning parameter selections, CV and iGCV. These choices lead to FBLR+CV+true, FBLR+iGCV+true, FBLR+CV+est, and FBLR+iGCV+est. We further compare three choices of initialization methods: Ridge after vectorization, PFPCR and MFPCR, which lead to multiple versions of our methods: Ridge$\to$FBLR, PFPCR$\to$FBLR and MFPCR$\to$FBLR, correspondingly.

\begin{figure}[!ht]
	\centerline{\includegraphics[width=\textwidth]{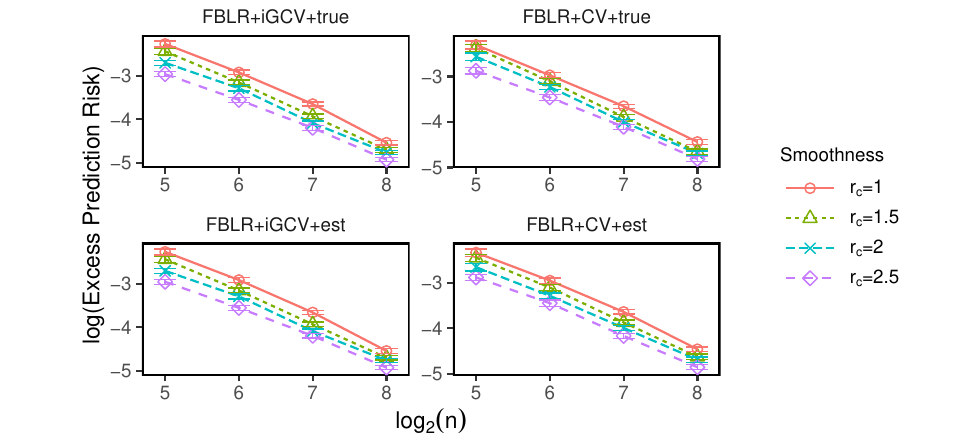}}
	\caption{Plots of the excess prediction risk vs the sample size $n$ with both axes in log scale under Setting 1. Four sample sizes and four values of $r_c$ are considered. The error bars are generated according to mean $\pm$ one SE. The four panels are all for FBLR approaches,  including FBLR+iGCV+true, FBLR+CV+true, FBLR+iGCV+est, and FBLR+CV+est. }
	\label{fig:r-error-blr}
\end{figure}

Figure \ref{fig:r-error-blr} shows the results of excess risk when considering tuning parameter selection via CV vs iGCV. % and true covariances $C_\alpha, C_\beta$ vs their estimates $\hat C_\alpha, \hat C_\beta$. 
Each panel demonstrates a linear relationship between log(risk) and $\log(n)$, and further reveals that the larger the $r_c$, the faster the convergence. These reconfirm the convergence rate developed theoretically in Theorem \ref{thm:upper}, where $r=r_c+1$. The four panels are almost identical, implying that the computationally-inexpensive iGCV performs similar as the computationally-expensive CV, and FBLR with the estimated covariances performs as well as with the true covariances. Hereafter, FBLR means FBLR+iGCV+est for these reasons. Since multiple $r_c$'s show similar messages, we will only use $r_c=1$ from now on. 

\begin{figure}[!ht]
	\centerline{\includegraphics[width=\textwidth]{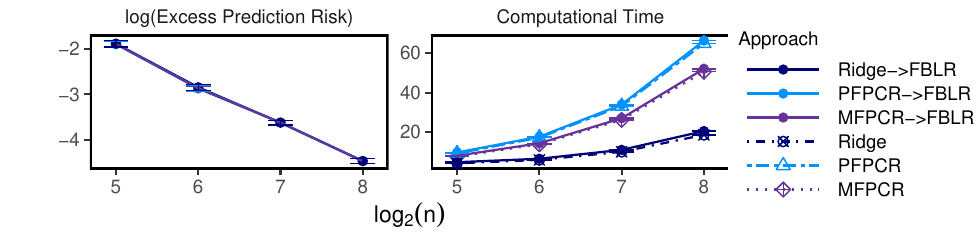}}
	\caption{Plots of the logarithm of excess prediction risk and computation time vs the sample size $n$ in log scale with $r_c=1$ under Setting 1. 
 %Three FBLR approaches with different initialization methods are considered here: Ridge after vectorization, PFPCR and MFPCR. 
 The error bars are mean $\pm$ one SE.}
	\label{fig:r-error-init}
\end{figure}

Figure \ref{fig:r-error-init} shows the performance of FBLR with three different initialization methods under Setting 1 with $r_c=1$. Initializing via Ridge, PFPCR or 
MFPCR produces indistinguishable results from the perspective of prediction risk. As for the computational time, it can be seen most of the time is spent on initialization, and FBLR itself takes little time to implement. Because of these observations and the fastest computation of Ridge, for the rest of this article, we will use Ridge as initialization, together with FBLR+iGCV+est. 

%For Settings 1-4, Figures \ref{fig:r-setting1to4} makes comparison of the prediction risk and computation time, respectively, between FBLR and the other existing methods, such as PFPCR, MFPCR, and GMRF. In the implementation of PFPCR and MFPCR, one needs to specify the maximum number $r^{\max}$ of principal components (PCs) in the code and the package will automatically find the optimal number of PCs among $1,2,\ldots,r^{\max}$. Because of Table \ref{tab:cases1to4}, the theoretically optimal number of PCs to be provided to estimate the coefficient functions are 4, 200, 8, and 204, respectively, for Settings 1-4. So in the implementation, we tried three choices of $r^{\max} \in \{4,8,\sqrt{n-1}\}$ for Setting 1, where $\sqrt{n-1}$ means we provide to the code the largest possible number of PCs given limited sample size. Figure \ref{fig:r-error-fpca} reveals that $r=\sqrt{n-1}$ is extremely time-consuming and even less accurate than the other two. So in Figures \ref{fig:r-error-setting1to4}-\ref{fig:r-time-setting1to4}, for the rest of this article, only $r\in \{4,8\}$ are compared with FBLR. 

\begin{figure}[!ht]
    \centerline{\includegraphics[width=0.99\textwidth]{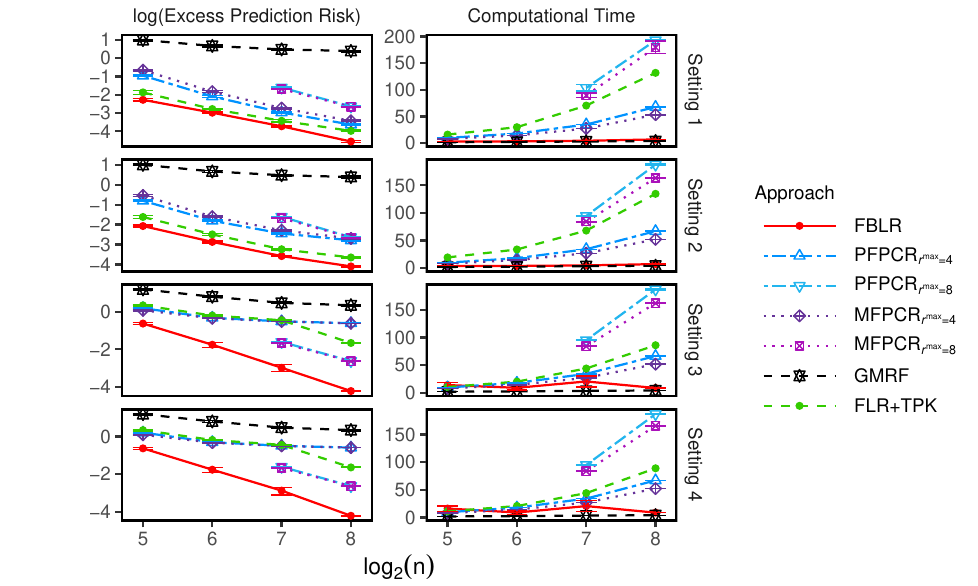}}
    \caption{Plots of the logarithm of the excess prediction risk and computation time vs the sample size $n$ in log scale for Settings 1-4 with $r_c = 1$. The error bars are generated according to mean $\pm$ one SE. ${\rm PFPCR}_{r_{\max}=8}$ and ${\rm MFPCR}_{r_{\max}=8}$ are only shown for $\log_2(n)=7,8$, because they require larger sample size.}
    \label{fig:r-setting1to4}
\end{figure}

For Settings 1-4, Figure \ref{fig:r-setting1to4} compares FBLR and three streams of existing methods, such as PFPCR, MFPCR, GMRF, and FLR+TPK, in terms of excess prediction risk and computation time. In the implementation of PFPCR and MFPCR, one needs to specify the maximum number $r^{\max}$ of PCs, and the package will automatically find the optimal number of PCs. 
According to Table \ref{tab:cases1to4}, the theoretically optimal numbers of PCs to be provided to estimate the coefficient functions in the noiseless case are 4, 200, 8, and 204, for Settings 1-4, respectively. We tried two choices of $r^{\max} \in \{4,8\}$. 
Additionally, in Appendix \ref{sec:exsim} of the Supplement, we examine the choice of $r^{\textrm{max}}=\lfloor \sqrt{n-1} \rfloor$, which means that we provide the code with the largest possible number of PCs that the software can handle. The brief message is that $r^{\max} = \lfloor \sqrt{n-1} \rfloor$ is extremely time consuming and even less accurate than the other two. So, for this section, only $r^{\max}\in \{4,8\}$ are compared with FBLR. 

In Figure \ref{fig:r-setting1to4}, it is unsurprising to see that GMRF performs the worst under all settings, albeit its fastest speed. Because it does not take advantage of the low-rank structure of the coefficient function. In terms of computational cost, FBLR is always much faster than 2D-FPCR with $r^{\textrm{max}}=8$ for all settings; much faster than 2D-FPCR with $r^{\textrm{max}}=4$ and FLR+ TPK under Settings 1-2; and faster than 2D-FPCR with $r^{\textrm{max}}=4$ and FLR+ TPK under Settings 3-4 for large sample sizes. Figure \ref{fig:r-setting1to4} also shows that the statistical performance of the prediction risk of FBLR dominates all the other methods under all four settings. 

% carl: Replacing some words to make the paragraph structure slightly more compact,
Let us compare FBLR and 2D-FPCR methods statistically. Under Setting 1, despite that ${\rm PFPCR}_{r_{\max}=4}$ and ${\rm MFPCR}_{r_{\max}=4}$ have the oracle knowledge of the true number of PCs and are perfectly aligned with the coefficient function, they are still worse than the penalized approach FBLR; ${\rm PFPCR}_{r_{\max}=8}$ and ${\rm MFPCR}_{r_{\max}=8}$ are worse than ${\rm PFPCR}_{r_{\max}=4}$ and ${\rm MFPCR}_{r_{\max}=4}$, because some unnecessary and noisy PCs are estimated. Under Setting 2, although more basis functions, 200 in total to be exact, are involved in the coefficient functions, $r^{\textrm{max}}=4$ still outperforms $r^{\textrm{max}}=8$ when $\log_2(n)=7$ because of smaller variance given the small sample size; $r^{\textrm{max}}=8$ and $r^{\textrm{max}}=4$ are comparable when the sample size reaches $\log_2(n)=8$. 
Under Settings 3-4, there is misalignment, and so 2D-FPCR with $r^{\textrm{max}}=4$ is completely off (the leading four PCs are orthogonal to the true coefficient functions). Therefore, $r^{\textrm{max}}=8$ performs better than $r^{\textrm{max}}=4$, but it is still not as accurate as FBLR.
In summary, when the true model is indeed bilinear \eqref{eqn:2dflr} under Settings 1-4, no matter misalignment exists or not, our penalized approach is more robust to the alignment structure than the PC-based approach, produces better prediction, and has less computational burden. 

Between FBLR and FLR+TPK, under Settings 1-2, FLR+TPK ranks the second and is significantly worse than FBLR; and under Settings 3-4, FLR+TPK is much worse than FBLR, even worse than 2D-FPCR with $r^{\textrm{max}}=8$. The phenomenon can be understood as follows. By the representer theorem, the estimate $\hat\beta(s,t)$ from FLR+TPK is a linear combination of the basis of the null space and $\int x_i(s_1,t_1)K_1((s_1,t_1),(s,t))ds_1dt_1$.  Since tensor product RKHS is used, the estimated function partially depends upon linear combinations of $\int x_i(s_1,t_1)K_1^1(s_1,s)K_1^2(t_1,t)ds_1dt_1$, ignoring some less important terms. Assume the kernels $K_1^1(s,t)$ and $K_1^2(s,t)$ have spectral decompositions $\sum_{k=1}^\infty s_k^1 \phi_k^1(s)\phi_k^1(t)$ and $\sum_{k=1}^\infty s_k^2 \phi_k^2(s)\phi_k^2(t)$, respectively.
%, the tensor product kernel has expression $\big(\sum_{k=1}^\infty s_k^K \phi_k^K(s_1)\phi_k^K(s)\big)\big(\sum_{k=1}^\infty s_k^K \phi_k^K(t_1)\phi_k^K(t)\big)$. 
Simplifying the representation theorem partially leads to linear combinations of 
\begin{eqnarray*}
\int x_i(s_1,t_1)\left(\sum_{k=1}^\infty s_k^1 \phi_k^1(s_1)\phi_k^1(s)\right)\left(\sum_{k=1}^\infty s_k^2 \phi_k^2(t_1)\phi_k^2(t)\right)ds_1dt_1 \\
=\sum_{j=1}^\infty\sum_{k=1}^\infty s_j^1 s_k^2 \left(\int x_i(s_1,t_1) \phi_j^1(s_1)  \phi_k^2(t_1)ds_1dt_1\right) \phi_j^1(s)\phi_k^2(t).
\end{eqnarray*}
Hence, the resulting estimator via FLR+TPK will be linear combinations of the products of two basis functions. However, the true coefficient function in Settings 1-4 is of the form of the product of two one-dimensional functions from \eqref{eq:slope_1to4}. The FLR+TPK does not take advantage of this knowledge. 
This is magnified even more for Settings 3-4. 

% \carlB{Under Settings 1-4, although FLR+TPK embeds a general metric space and obtains its estimator on a product RKHS which is naturally associated with the roughness penalty in a product form, enabling FLR to take advantage of the product form of slope functions. FBLR is still consistently doing better than FLR+TPK. It is worth noting that the Settings 3-4, the performance of FLR+TPK perform poor, due to the truth that FLR under RKHS framework with a product kernel is sensitive to the choice of the kernel and the subsequent RKHS. The RKHS associated with the kernel (\ref{eq:sim-kernel}) comprises the linear span of cosine basis functions that impose less penalty on cosine basis functions with small frequencies, as $K(s,t) = -B_4(|s-t|/2)/3 - B_4((s+t)/2)/3 = \sum_{k \geq 1}\frac{2}{(k\pi)^4} \cos(k \pi s) \cos(k \pi t)$. The product RKHS fails to produce an accurate estimator of the coefficient function under Settings 3-4.
% }

\vspace{-.5em}
\vspace{-.5em}

\begin{figure}[!ht]
    \centerline{\includegraphics[width=\textwidth]{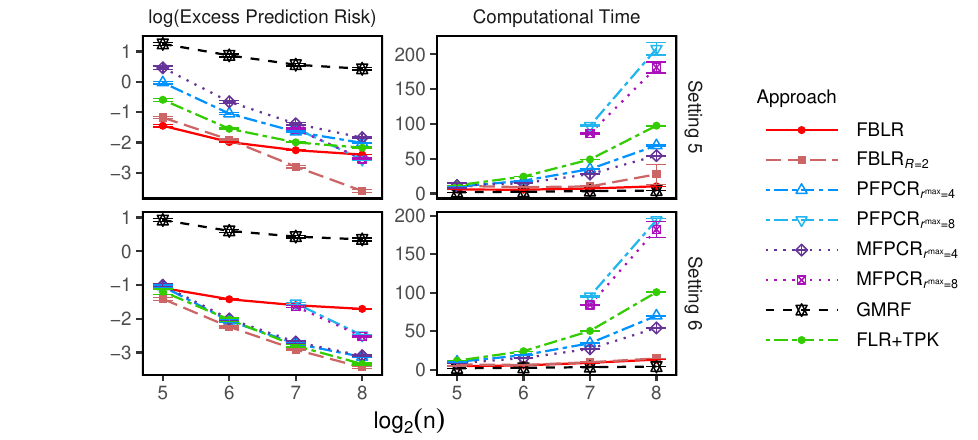}}
    \caption{Plots of the logarithm of the excess prediction risk and computation time vs the sample size $n$ in log scale for Settings 5-6 with $r_c = 1$. The error bars are generated according to mean $\pm$ one SE. ${\rm PFPCR}_{r_{\max}=8}$ and ${\rm MFPCR}_{r_{\max}=8}$ are only shown for $\log_2(n)=7,8$, because they require larger sample size.}
    \label{fig:r-setting5to6}
\end{figure}

Figure \ref{fig:r-setting5to6} shows the results for Settings 5-6, which follow Model \eqref{eqn:2dflr-2} and Model \eqref{eqn:2dflr-other}, respectively. Both settings don't satisfy the model assumption \eqref{eqn:2dflr}, and therefore puts our FBLR in a disadvantageous situation. We used the iterative deflation idea to apply FBLR twice: once on the original data $\{y_i, x_i\}$ and once on the residual $\{e_i, x_i\}$. We denote this approach as ${\rm FBLR}_{R=2}$, %for the estimation of Model \eqref{eqn:2dflr-2} with $R=2$, 
and obtain the estimate of the two-dimensional coefficient function of the form $\hat\beta(s,t)=\sum_{r=1}^{2}\hat\alpha_{0}^{[r]}(s)\hat\beta_{0}^{[r]}(t)$. Computationally, it is clear that FBLR, ${\rm FBLR}_{R=2}$, and GMRF are the fastest among all.

Under Setting 5, ${\rm FBLR}_{R=2}$ is supposed to be the best because it has the oracle knowledge of $R=2$. Indeed it performs better than PFPCR, MFPCR, GMRF and FLR+TPK for all sample sizes considered. It is also interesting to note that ${\rm FBLR}_{R=2}$ is better than ${\rm FBLR}_{R=1}$ for large sample sizes, but worse for small sample sizes, even though the true model consists of two terms. It implies that simple model pays off when limited by sample size: even though Model \eqref{eqn:2dflr} is more restrictive than Model \eqref{eqn:2dflr-2}, with limited data, estimating Model \eqref{eqn:2dflr} when the underlying truth is Model \eqref{eqn:2dflr-2} can still be beneficial.

Under Setting 6, the true two-dimensional coefficient function shown in Figure \ref{fig:r-beta0} is indeed not low-rank. Under this setting, ${\rm FBLR}_{R=2}$ performs the best among all estimators for all sample sizes, followed by FLR+TPK, ${\rm PFPCR}_{r^{\textrm{max}}=4}$, and ${\rm MFPCR}_{r^{\textrm{max}}=4}$. Note that the advantage of ${\rm FBLR}_{R=2}$ is significant because of the narrow SE shown in the figure. FBLR is worse than the other methods, except for small sample size. 2D-FPCRs with $r^{\textrm{max}}=8$ are worse than their counterparts with $r^{\textrm{max}}=4$, which suggests simpler model is preferred when PC regression is considered under Setting 6. In short, although the low-rank Model \eqref{eqn:2dflr-2} is more restrictive than the general Model \eqref{eqn:2dflr-other}, estimating Model \eqref{eqn:2dflr-2} with ${\rm FBLR}_{R=2}$ is still beneficial comparing to estimating the two-dimensional coefficient function in Model \eqref{eqn:2dflr-other} due to dimension reduction. 

\vspace{-.5em}
\vspace{-.5em}

In the Appendix, Section \ref{sec:exsim} provides extra simulations to study performance of various vectorization approaches for FLR, other less competitive methods, more analysis on 2D-FPCR, the impact of the choice of the kernel on the performance, and for sparse data.

\vspace{-1.5em}
\section{Real data analysis: Canadian weather}
\label{sec:realdata}

We perform real data analysis on two datasets, the Canadian weather data in this section and the LIDAR data in Appendix \ref{sec:rd lidar}.

The Canadian weather data\footnote{The data can be downloaded from the official website of the government of Canada at \url{https://climate.weather.gc.ca/historical_data/search_historic_data_e.html}.} has been widely used for FDA. Traditionally, it is typically used for 1D-FPCA, 1D-FPCR \citep{ramsay2005principal} or 1D-FLR \citep{cai2012minimax}, where each vector is of length 365, containing the daily temperature averaged over 24 hours \emph{and} averaged over a few years. We consider the matrices $x_i\in\mathbb{R}^{365 \times 24}$, where $x_i(s,t)$ is the temperature in the $t$-th hour of the $s$-th day of the year averaged over 2002-2021. Hence, it contains extra hourly information compared to 1D analysis. Following \citet{ramsay2005principal, cai2012minimax}, the response variable $y_i$ is the logarithm of the average annual precipitation over 2002-2021, and 35 weather stations are included.

We compare the performances of FBLR and some existing methods, including PFPCR, MFPCR, GMRF, FLR+TPK, Ridge after vectorization, and two variants of 1D-FLR. The first variant is to adopt the FLR method in \citet{cai2012minimax} after matrix vectorization, denoted by FLR+$\textrm{vec}$. (See Appendix \ref{sec:exsim-FLR-vec} for further discussion on the potential twist of the vectorization approach.) The second variant is to apply FLR on the vectors of length 365, which are obtained by averaging temperatures over 24 hours, $\sum_{t=1}^{24}x_i(s,t)/24$, denoted by FLR+$\textrm{ave}$. Both FBLR and FLR-related methods choose the kernel $K_1(s,t) = K_2(s,t) = 1-B_4(|s-t|)/24$, which is used in \citet{cai2012minimax} as well.
For PFPCR and MFPCR, we use $r^{\textrm{max}}=\lfloor\sqrt{n-1}\rfloor=5$. BLR (FBLR with no penalty) was not compared because the sample size is not large enough to estimate the parameters. 

\begin{figure}[!ht]
	\centerline{\includegraphics[width=\textwidth]{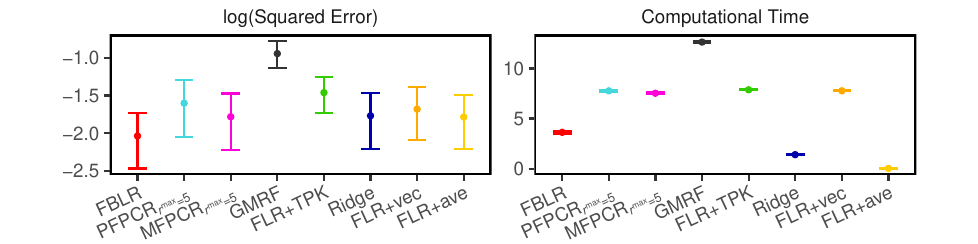}}
    \caption{Plots of the out-of-sample performance on the Canadian weather data, which include the testing error and computational time. The error bars are mean $\pm$ one SE.}
	\label{fig:rd_weather}
\end{figure}

We first compare all eight methods from the perspective of prediction accuracy and computational time in Figure \ref{fig:rd_weather}. The leave-one-out method is used to calculate the out-of-sample squared error. 
FBLR works the best compared to the other 2D and 1D methods, and the differences are all significant because the p-value of the paired t-test between FBLR and each of the other methods is less than 0.05. 
GMRF performs the worst, followed by FLR+TPK. MFPCR is similar to and PFPCR is worse than three 1D methods including Ridge, FLR+$\textrm{vec}$, and FLR+$\textrm{ave}$. 
Furthermore, FLR+$\textrm{ave}$, the traditional 1D FDA method by \citet{cai2012minimax}, performs worse than FBLR, suggesting that the temperature variations along the hour-of-the-day dimension contain extra information for predicting annual precipitation. 
For the computational time, it is unsurprising to see that FBLR takes longer than Ridge and FLR+$\textrm{ave}$, because FBLR uses Ridge as initialization, and FLR+$\textrm{ave}$ has a 1D predictor of length 365 instead of a 2D predictor of size 365 $\times$ 24. But the FBLR is much faster comparing to PFPCR, MFPCR, GMRF, FLR+TPK, and FLR+vec.

\begin{figure}[!ht]
	\centerline{\includegraphics[width= \textwidth]{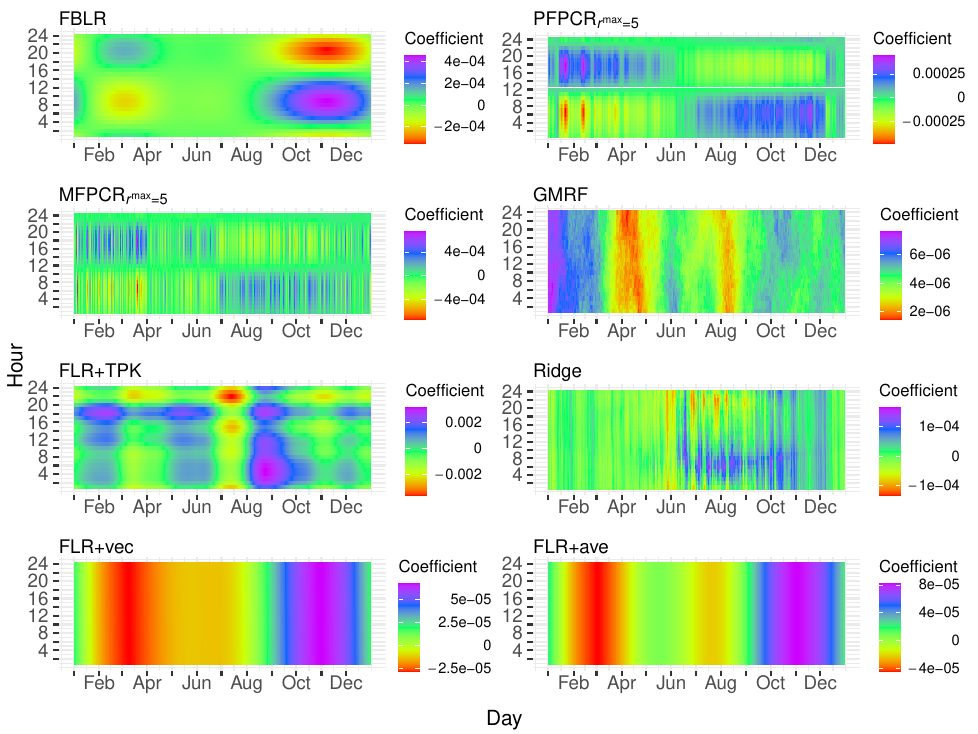}}
	\caption{Plots of the estimated 2D coefficient function $\hat\beta(s,t)$ in Model \eqref{eqn:2dflr-other} by eight methods for Canadian weather data. The $x$ and $y$ axes correspond to day and hour respectively. }
	\label{fig:rd_cw_betaHat}
\end{figure}

Figure \ref{fig:rd_cw_betaHat} displays the heat-maps of the estimated 2D coefficient function $\hat\beta(s,t)$ in Model \eqref{eqn:2dflr-other} for all approaches. For FLR+$\textrm{ave}$, the 2D function is generated from their estimated 1D coefficient function $\hat\beta(\cdot)$ in Model \eqref{eqn:1dflr} by repeating the yearly pattern for each hour, i.e., $\hat\beta(s,t) = \hat\beta(s)\times \frac{1}{24}$ for $s = 1,\ldots,365$ and $t = 1, \ldots, 24$.

%In Figure \ref{fig:rd_cw_betaHat}, 
As Figure \ref{fig:rd_cw_betaHat} shows, FBLR produces a smoother coefficient function estimation compared with the other 2D methods. 
Furthermore, despite PFPCR and MFPCR being designed to generate smooth estimations, the actual estimates are not as smooth as expected, especially for the day-of-the-year dimension. Among the 1D methods, Ridge estimation is non-smooth, and FLR+vec over-smoothes since the hour-of-day-dimension has almost no variation. 

% keep!!!!!!!!!!!!!!!
% \begin{figure}[!ht]
% 	\centerline{\includegraphics[width=.95\textwidth]{./FIGURES/rd_cw_slope.pdf}}
% 	\caption{Plots of the estimated 1D coefficient functions $\hat{\alpha}(\cdot)$ and $\hat{\beta}(\cdot)$ in Model \eqref{eqn:2dflr} for FBLR. The confidence intervals correspond to 95\% confidence level.}
% 	\label{fig:rd_cw_coef}
% \end{figure}

\begin{figure}[!ht]
	\centerline{\includegraphics[width=.8\textwidth]{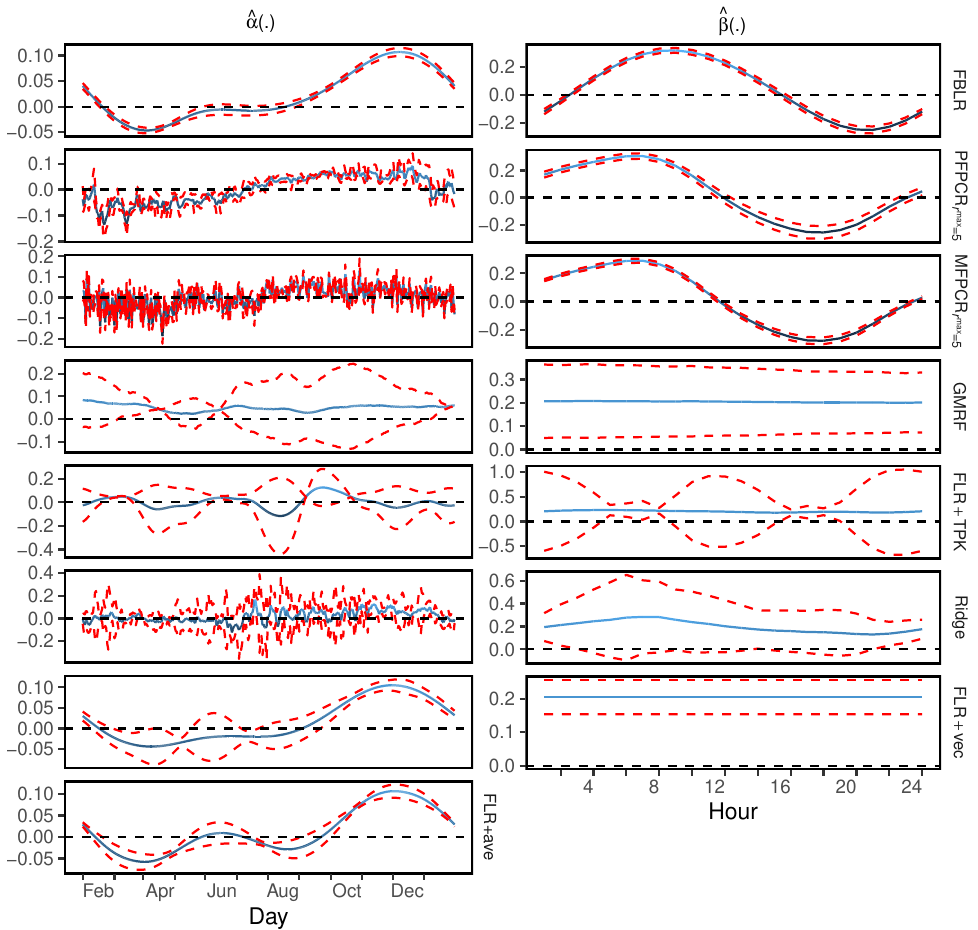}}
	\caption{Plots of the estimated 1D coefficient functions $\hat{\alpha}(\cdot)$ and $\hat{\beta}(\cdot)$ in Model \eqref{eqn:2dflr} by eight methods for Canadian weather data. The confidence intervals correspond to 95\% confidence level.}
	\label{fig:rd_cw_coef_all}
\end{figure}

Figure \ref{fig:rd_cw_coef_all} provides the visualization of 
the estimated coefficient functions $\hat{\alpha}(\cdot)$ and $\hat{\beta}(\cdot)$ in Model \eqref{eqn:2dflr}. For those methods that do not estimate the 1D coefficient functions directly, the leading left and right singular vectors of the estimated 2D coefficient functions are plotted.

The left panels of Figure \ref{fig:rd_cw_coef_all} show the day-of-the-year effect: PFPCR, MFPCR, and Ridge are not smooth; GMRF and FLR+TPK are not significant, since the confidence interval covers 0; the width of the confidence interval of FLR+$\textrm{vec}$ is very large, followed by FLR+$\textrm{ave}$, and FBLR is the narrowest; the shapes of FBLR and FLR+$\textrm{ave}$ \citep{cai2012minimax} are rather similar with a peak near November and a trough near March. Such a contrast between Spring and Fall appears at moisture-laden coastal locations \citep{donohoe2020seasonal}, which have warmer autumns and cooler springs than inland stations. As a consequence, they have more precipitation. 
However, there is slight difference between FBLR and FLR+$\textrm{ave}$ in that the confidence intervals throughout Summer contain zero for FBLR but not for FLR+$\textrm{ave}$. The results of FBLR imply that temperature variation in the summer has no effect on the precipitation. This makes sense, as the temperature in the summer at coastal and inland stations has no strong correlation with maritime effect. 

From the right panels of Figure \ref{fig:rd_cw_coef_all} for the time-of-the-day effect, it can be seen that GMRF, FLR+TPK, Ridge, and FLR+$\textrm{vec}$ estimations essentially suggest that there is no hourly effect, while FBLR, PFPCR, and MFPCR reveal and share a significant effect. The estimated $\hat{\beta}(\cdot)$'s by the latter three say that more precipitation occurs with warmer daytime and cooler nighttime.
In other words, more precipitation accompanies larger diurnal temperature variation, which promotes the local breeze circulations caused by land-water temperature differences \citep[e.g.,][]{yang2006mechanisms}, and is essential to convective precipitation in the areas near water bodies (e.g. sea, gulf, lake and river). 
% keep 
% Specifically, the temperature difference between the hot landmass and the cold water body causes onshore breezes to arise during the day, while this difference reversed at night and offshore breezes predominate. 
% Usually the diurnal variation of water surface temperature is so small \citep{sverdrup1942oceans}. As a result, the greater diurnal temperature variations over land, implying the greater land-water temperature contrasts, produce stronger localised diurnal breezes and corresponding stronger weather fronts, resulting in more dense cumulus clouds and higher probability of precipitation.
The residuals of FBLR are examined in Appendix \ref{sec:rd weather}, which suggests a fairly good fit of the data.

\section{Discussion}
\label{sec:dis}

This article studies the problem of regression of a scalar response on a two-dimensional functional predictor, which when observed on 2D grid becomes a matrix predictor. We propose a functional regression model where the two-dimensional coefficient function is assumed to adopt the form of a product of two 1D coefficient functions. We offer an iterative strategy for the circumstance when the two-dimensional coefficient function can be well approximated by the sum of a few products, which implies low rank for the matrix coefficient. We estimate the model via an innovative penalized approach and compare the penalized approach with the approach of regression on two-dimensional PCs and other methods. We show that the misalignment issue of the PC regression remains as in the 1D FLR case. Even if misalignment does not occur, the penalized approach still performs better than the PCR approach because of further shrinkage. Moreover, the penalized approach also exhibits a huge computational advantage. Real data application further demonstrates the ``stableness'' and smoothness of the penalized approach. 

There are a few meaningful directions for future extensions. The first is to rigorously understand Model \eqref{eqn:2dflr-2} with multiple terms, such as how to determine the optimal $R$. Empirically, one can choose this tuning parameter via cross-validation or similar approaches. Theoretically, this is a challenging topic worth studying further. 
The second one is to extend from scalar-on-matrix functional regression to scalar-on-tensor functional regression, which takes the form of multilinear instead of bilinear. In this case, a careful design of the penalty function is necessary and our work sheds light 
on this direction. The LIDAR task in Appendix \ref{sec:rd lidar} intrinsically requires such an extension to consider time, spatial range, and wavelength as a three-dimensional predictor. The fMRI data intrinsically requires such an extension to deal with four-dimensional functional input, corresponding to 3D brain and 1D time.
The third one is to extend to a generalized linear model for classification and other purposes. For example, one may want to classify whether a person has Alzheimer’s disease based on the fMRI data, which has spatial and temporal smooth input. 

Lastly, it is interesting to study the problem when data are not fully observed, from the sparse to the ultra dense regimes, as in \citet{li2010uniform, zhang2016sparse,guo2023sparse}. 
%\citet{li2018nonparametric}
We implement our procedure based on principal component analysis through conditional expectation (PACE) from \citet{yao2005functional}, and the simulation results for not-fully-observed data are given in Appendix Section \ref{sec:exsim-sparse}. Our conjecture for the theoretical property is that the convergence rate will remain the same for the ultra-dense scenario, but will involve a non-parametric term combined together with the current rate for the sparse and semi-dense scenarios. The rates shall be ranked from the fastest to the slowest for the ultra-dense, semi-dense, and sparse scenarios.

% In the current article, we mainly focus on the penalization approach for FBLR. In the literature, there is huge amount of work on the functional PCA based approach for one-way FLR. Although it has been proved that in the one-way case the penalization approach does not require the alignment of the eigenfunctions of the covariance and the reproducing kernel while functional PCA based does, whether the same holds remains unclear in the two-way case. Even the functional bilinear PCA without supervision, which intrinsically relates to functional tensor decomposition, is still in its infant stage with little theoretical analysis, let alone supervised regression with scores obtained from the functional bilinear PCA. In the two-way case, it will be rather meaningful to study the PCA based approach and compare it with our proposed penalization approach in the future, which will embody a more comprehensive understanding.

\section*{Acknowledgements}

We would like to thank the Action Editor and the anonymous referees for their detailed and insightful reviews, which helped to improve the paper substantially. Yang's research is supported in part by NSF grant IIS-1741390, Hong Kong grant GRF 17301620 and Hong Kong grant CRF C7162-20GF. Shen's research is supported in part by Hong Kong grant CRF C7162-20GF, China Strategy Key Grant 2022SQGH10861, HKU BRC grant, and HKU FBE Shenzhen Research Institutes grant. %Zhu's research is supported in part by ...

\section*{Supplementary materials}

The materials include proofs of main theorems and all lemmas, brief statement of the theorems and sketch of the proof for the case with two distinct domains, additional simulation results, additional results of the real data analysis of the Canadian weather, and the second real data example of the LIDAR data.

\begingroup
\linespread{.6}\selectfont
\bibliography{./TEX/ref}
\endgroup

%--------------------------------------------------------------------
\newpage

\setcounter{page}{1}
\setcounter{section}{0}
\begin{appendix}
\counterwithin{figure}{section}

\begin{center}
{\large Supplement to ``Optimal Functional Bilinear Regression with Two-way Functional Covariates via Reproducing Kernel Hilbert Space''}

\vspace{.5cm}
Dan Yang, Jianlong Shao, Haipeng Shen, and Hongtu Zhu 
\end{center}

{\noindent In this supplement, we provide proofs of the main theorems in Section \ref{sec:proof}, all lemmas and their detailed proofs in Sections \ref{sec:proof-main-lemma}-\ref{sec:proof-auxiliary-lemma}, and brief statement of the theorems and sketch of the proof for the case with two distinct domains in Section \ref{sec:theory-distinct}. %We also present more discussion on the penalty in Section \ref{sec:penalty} 
We also present additional simulation result in Section \ref{sec:exsim}, additional real data analysis on the Canadian weather in Section \ref{sec:rd weather}, and the real data example of LIDAR data in Section \ref{sec:rd lidar}.}

\vspace{-1em}
\section{Proofs of main theorems}
\label{sec:proof}

\subsection{Proof of Theorem \ref{thm:upper}}
In this section, we will show the proof of Theorem \ref{thm:upper}. We
start by providing a result from
\citet{Yuan+2010b} that will be extensively used in the proof.

\vspace{-.5em}
\begin{lemma}[Theorem 3 of \citet{Yuan+2010b}]
\label{lem:simultaneous-diagonalization}
For any function $f\in\H(K)$, it can be written as $ f = \sum_{k=1}^\infty f_k \omega_k$,
where $f_k = s_k^T \langle f, \omega_k\rangle_R$. Moreover, the
quadratic forms $\|\cdot\|_R^2$, $\|\cdot\|_K^2$, and $\|\cdot\|_0^2$
can be expressed as
% \begin{eqnarray*}
%   \|f\|_R^2 &=& \sum_{k=1}^\infty (1+\gamma_k^{-1}) f_k^2, \\
%   \|f\|_0^2 &=& \sum_{k=1}^\infty f_k^2, \\
%   \|f\|_K^2 &=& \sum_{k=1}^\infty \gamma_k^{-1} f_k^2.
% \end{eqnarray*}
$\|f\|_R^2 = \sum_{k=1}^\infty (1+\gamma_k^{-1}) f_k^2$, $\|f\|_0^2 = \sum_{k=1}^\infty f_k^2$, and $\|f\|_K^2 = \sum_{k=1}^\infty \gamma_k^{-1} f_k^2$.

\end{lemma}
Lemma \ref{lem:simultaneous-diagonalization} demonstrates that the norms $\|\cdot\|_R^2$, $\|\cdot\|_K^2$, and $\|\cdot\|_0^2$ can be expressed on the basis $\omega_k, k=1,2,...$, which was defined in Section \ref{ssec:theory-preliminary}. See \citet{Yuan+2010b} for the elementary proof of Lemma \ref{lem:simultaneous-diagonalization}.

Recall the definition of the excess prediction risk, 
% $  \calE (\halpha,\hbeta;\alpha_0,\beta_0)$
% =\expecs
%   \left(\int_{\T\times\T}X^*(s,t)
%   \big(\halpha(s)\hbeta(t)-\alpha_0(s)\beta_0(t)\big)~dsdt \right)^2$.
from the identity,
%\begin{equation*}
    $\halpha(s)\hbeta(t)-\alpha_0(s)\beta_0(t)
    = \\
\big( \halpha(s)-\alpha_0(s) \big) \big(\hbeta(t)-\beta_0(t)\big)
    +\big(\halpha(s)-\alpha_0(s)\big)\beta_0(t)
    +\big(\hbeta(t)-\beta_0(t)\big)\alpha_0(s)$,
%\end{equation*}
and the definition of $\|\cdot\|_0$ norm as in \eqref{eqn:zerosnorm} and the property \eqref{eqn:var-zeronorm}, it follows that the excess prediction risk can be further bounded by three terms,
\begin{equation}
\label{proof-upper-three-term}
\calE(\halpha,\hbeta;\alpha_0,\beta_0) \le
3\|\halpha-\alpha_0\|_0^2\|\hbeta-\beta_0\|_0^2
+3\|\halpha-\alpha_0\|_0^2\|\beta_0\|_0^2
+3\|\alpha_0\|_0^2\|\hbeta-\beta_0\|_0^2.
\end{equation}
Therefore, we only need to bound two terms $\|\halpha-\alpha_0\|_0^2$ and $\|\hbeta-\beta_0\|_0^2$. Due to symmetry in $\alpha$ and $\beta$, one bound on $\|\halpha-\alpha_0\|_0^2$ is sufficient for the problem. However, in what follows, we shall bound $\|\halpha-\alpha_0\|_a^2$ due to the necessity in the proof, where the norm $\|\cdot\|_a$ for $0\le a\le 1$ is defined by
$
    \|f\|_a^2 = \sum_{k=1}^\infty(1+\gamma_k^{-a})f_k^2,
$
when $f=\sum_{k=1}^\infty f_k\omega_k$. Clearly, $\|\cdot\|_a$ reduces to $\|\cdot\|_0$ by a factor of 2 when $a=0$ due to Lemma \ref{lem:simultaneous-diagonalization}.

Recall that the objective function for the optimization problem is
$  \ell_{n\lambda}(\alpha,\beta)=\ell_n(\alpha,\beta)+J(\alpha,\beta)$,
and the smoothness regularized estimator is obtained via
$
  (\halpha,\hbeta) = \arg\min \ell_{n\lambda}(\alpha,\beta).
$
Write
$
  \ell(\alpha,\beta) = \expec \ell_n(\alpha,\beta)$, and
  $\ell_\lambda(\alpha,\beta)= \expec \ell_{n \lambda}(\alpha,\beta)$,
to be the expectations of $\ell_{n}(\alpha,\beta)$ and
$\ell_{n\lambda}(\alpha,\beta)$ respectively. The convention is to use subscript $n$ to denote the sample version and without subscript for the population counterpart. Denote the minimizer of
$\ell_{\lambda}(\alpha,\beta)$ by $(\balpha,\bbeta)$, that is,
$
    (\balpha,\bbeta) = \arg\min \ell_\lambda(\alpha,\beta) = \arg\min
  \ell(\alpha,\beta) + J(\alpha,\beta)$.
To bound $\|\halpha-\alpha_0\|_a$, one can bound $\|\balpha-\alpha_0\|_a$ and $\|\halpha-\balpha\|_a$, which can be thought of as the deterministic error (or bias) and stochastic error (or variance) respectively.

To bound the stochastic error term, another pair $(\talpha,\tbeta)$ has to be introduced so that $\|\halpha-\balpha\|_a \le \|\halpha-\talpha\|_a + \|\talpha-\balpha\|_a$. Here $(\talpha,\tbeta)$ can be thought of as the expansion and is defined by
\vspace{-1em}
\begin{equation}
  \label{eq:tbeta-def}
  \left(\begin{array}{c}\talpha\\\tbeta\end{array}\right) =
    \left(\begin{array}{c}\balpha\\\bbeta\end{array}\right)
      - H^{-1}
      \left(\begin{array}{c}D_\alpha \ell_{n\lambda}(\balpha,\bbeta)\\D_\beta \ell_{n\lambda}(\balpha,\bbeta)\end{array}\right),
\end{equation}
where
\vspace{-.5em}
\begin{equation}
  \label{eq:H-definition}
  H = \left(
  \begin{array}{cc}
    D^2_{\alpha\alpha}\ell_\lambda(\balpha,\bbeta)&D^2_{\alpha\beta}\ell_\lambda(\balpha,\bbeta)\\
    D^2_{\beta\alpha}\ell_\lambda(\balpha,\bbeta)&D^2_{\beta\beta}\ell_\lambda(\balpha,\bbeta)\\
  \end{array}
  \right),
\end{equation}
where the following operators are defined. The first set of operators are the first- and second-order derivatives of the sample and population loss functions $\ell_n$ and $\ell$,
\begin{eqnarray*}
  D_\alpha \ell_n(\alpha,\beta)f &=& -\frac{2}{n}\sum_{i=1}^n\left(y_i-\int_{\T\times\T} x_i(s,t)\alpha(s)\beta(t) ~ dsdt\right)\left(\int_{\T\times\T} x_i(s,t)f(s)\beta(t) ~ dsdt\right),\\ [-.3em]
  D_\beta \ell_n(\alpha,\beta)f &=& -\frac{2}{n}\sum_{i=1}^n\left(y_i-\int_{\T\times\T} x_i(s,t)\alpha(s)\beta(t) ~ dsdt\right)\left(\int_{\T\times\T} x_i(s,t)\alpha(s)f(t) ~ dsdt\right),\\ [-.3em]
  D^2_{\alpha\alpha} \ell_n(\alpha,\beta)fg&=&\frac{2}{n}\sum_{i=1}^n\left(\int_{\T\times\T} x_i(s,t)f(s)\beta(t) ~ dsdt\right)\left(\int_{\T\times\T} x_i(s,t)g(s)\beta(t) ~ dsdt\right),\\ [-.3em]
  D^2_{\beta\beta} \ell_n(\alpha,\beta)fg&=&\frac{2}{n}\sum_{i=1}^n\left(\int_{\T\times\T} x_i(s,t)\alpha(s)f(t) ~ dsdt\right)\left(\int_{\T\times\T} x_i(s,t)\alpha(s)g(t) ~ dsdt\right),\\ [-.3em]
%\end{eqnarray*}
%\begin{eqnarray*}
  D^2_{\alpha\beta} \ell_n(\alpha,\beta)fg&=&-\frac{2}{n}\sum_{i=1}^n\left(y_i-\int_{\T\times\T} x_i(s,t)\alpha(s)\beta(t) ~ dsdt\right)\left(\int_{\T\times\T} x_i(s,t)f(s)g(t)~dsdt\right)\\ [-.3em]
  &&+\frac{2}{n}\sum_{i=1}^n\left(\int_{\T\times\T} x_i(s,t)f(s)\beta(t) ~ dsdt\right)\left(\int_{\T\times\T} x_i(s,t)\alpha(s)g(t) ~ dsdt\right),\\ [-.3em]
% \end{eqnarray*}
% \begin{eqnarray*}
  D_\alpha \ell(\alpha,\beta)f
  &=&-2\int_{\T^4}C(s_1,s_2)C(t_1,t_2)\big(\alpha_0(s_1)\beta_0(t_1)-\alpha(s_1)\beta(t_1)\big)f(s_2)\beta(t_2)~ds_1ds_2dt_1dt_2,\\ [-.1em]
  D_\beta \ell(\alpha,\beta)f
  &=&-2\int_{\T^4}C(s_1,s_2)C(t_1,t_2)\big(\alpha_0(s_1)\beta_0(t_1)-\alpha(s_1)\beta(t_1)\big)\alpha(s_2)f(t_2)~ds_1ds_2dt_1dt_2,\\ [-.3em]
  D^2_{\alpha\alpha} \ell(\alpha,\beta)fg&=&2\|\beta\|_0^2\int_{\T\times\T}C(s,t)f(s)g(t)~dsdt,\\ [-.3em]
  D^2_{\beta\beta} \ell(\alpha,\beta)fg&=&2\|\alpha\|_0^2\int_{\T\times\T}C(s,t)f(s)g(t)~dsdt,\\ [-.3em]
  D^2_{\alpha\beta} \ell(\alpha,\beta)fg&=&-2\left(\int_{\T\times\T}C(s,t)\alpha_0(s)f(t)dsdt\right)
  \left(\int_{\T\times\T}C(s,t)\beta_0(s)g(t)dsdt\right)\\ [-.3em]
  &&+4\left(\int_{\T\times\T}C(s,t)\alpha(s)f(t)dsdt\right)
  \left(\int_{\T\times\T}C(s,t)\beta(s)g(t)dsdt\right),
\end{eqnarray*}
and the second set of operators are first- and second-order derivatives of the sample and population objective functions $\ell_{n\lambda}$ and $\ell_\lambda$,
\begin{eqnarray*}
  D_\alpha \ell_{n\lambda}(\alpha,\beta) &=& D_\alpha \ell_n(\alpha,\beta) + D_\alpha J(\alpha,\beta),\\ [-.3em]
  D_\beta \ell_{n\lambda}(\alpha,\beta) &=& D_\beta \ell_n(\alpha,\beta) + D_\beta J(\alpha,\beta),\\ [-.3em]
  D^2_{\alpha\alpha} \ell_{n\lambda}(\alpha,\beta) &=& D^2_{\alpha\alpha} \ell_n(\alpha,\beta) + D^2_{\alpha\alpha} J(\alpha,\beta),\\ [-.3em]
  D^2_{\beta\beta} \ell_{n\lambda}(\alpha,\beta) &=& D^2_{\beta\beta} \ell_n(\alpha,\beta) + D^2_{\beta\beta} J(\alpha,\beta),\\ [-.3em]
  D^2_{\alpha\beta} \ell_{n\lambda}(\alpha,\beta) &=& D^2_{\alpha\beta} \ell_n(\alpha,\beta) + D^2_{\alpha\beta} J(\alpha,\beta),\\ [-.3em]
  D_\alpha \ell_{\lambda}(\alpha,\beta) &=& D_\alpha \ell(\alpha,\beta) + D_\alpha J(\alpha,\beta),\\ [-.3em]
  D_\beta \ell_{\lambda}(\alpha,\beta) &=& D_\beta \ell(\alpha,\beta) + D_\beta J(\alpha,\beta),\\ [-.3em]
  D^2_{\alpha\alpha} \ell_{\lambda}(\alpha,\beta) &=& D^2_{\alpha\alpha} \ell(\alpha,\beta) + D^2_{\alpha\alpha} J(\alpha,\beta),\\ [-.3em]
  D^2_{\beta\beta} \ell_{\lambda}(\alpha,\beta) &=& D^2_{\beta\beta} \ell(\alpha,\beta) + D^2_{\beta\beta} J(\alpha,\beta),\\ [-.3em]
  D^2_{\alpha\beta} \ell_{\lambda}(\alpha,\beta) &=& D^2_{\alpha\beta} \ell(\alpha,\beta) + D^2_{\alpha\beta} J(\alpha,\beta).
\end{eqnarray*}

% From the identity,
% \begin{equation*}
%   \halpha-\alpha_0 = (\halpha - \talpha) + (\talpha - \balpha) + (\balpha - \alpha_0),
% \end{equation*}
% it follows that
By the triangle inequality, we have
\begin{equation}
  \label{eqn:threeterms}
  \|\halpha-\alpha_0\|_a = \|(\halpha - \talpha) + (\talpha - \balpha)
  + (\balpha - \alpha_0)\|_a \le \|\halpha - \talpha\|_a + \|\talpha -
  \balpha\|_a + \|\balpha - \alpha_0\|_a.
\end{equation}
Lemmas \ref{lem:proof-bound1}, \ref{lem:proof-bound2} and \ref{lem:proof-bound3} in Section \ref{sec:proof-main-lemma} establish bounds for the three terms on the right hand side of \eqref{eqn:threeterms} respectively and together with \eqref{eqn:threeterms} imply that, if $\lambda=O(n^{-2r/(2r+1)})$,
\begin{eqnarray}
  \label{eqn:upperalphahat}
    \lim_{A\rightarrow\infty}\lim_{n\rightarrow\infty}\sup_{\substack{\alpha_0
        \in \H(K), \\ \beta_0 \in \H(K)}} \prob (\|\halpha-\alpha_0\|_0^2\ge
    An^{-\frac{2r}{2r+1}})&=&0,\\ [-.5em]
    \label{eqn:upperbetahat}
    \lim_{A\rightarrow\infty}\lim_{n\rightarrow\infty}\sup_{\substack{\alpha_0
        \in \H(K),\\ \beta_0 \in \H(K)}} \prob (\|\hbeta-\beta_0\|_0^2\ge
    An^{-\frac{2r}{2r+1}})&=&0.
\end{eqnarray}

Combining \eqref{eqn:upperalphahat}, \eqref{eqn:upperbetahat} and \eqref{proof-upper-three-term} completes the proof of Theorem \ref{thm:upper}. \hfill $\blacksquare$

{\bf Comment}: Note that the proof to 2D FBLR differs much from the proof to 1D FLR since FLR only requires expansion of $\tbeta$ whereas FBLR relies on 2D expansion \eqref{eq:tbeta-def} which complicates the proofs of the lemmas extensively. To save the readers some detour, expanding $\talpha,\tbeta$ separately without considering their interaction cannot lead to the full proof.

\subsection{Proof of Theorem \ref{thm:lower}}

Note that although it has been assumed throughout that the noise $\epsilon$ has mean zero and finite variance, for the proof of lower bound, it suffices to assume the normal case $\epsilon\sim N(0,\sigma^2)$. This is because any lower bound for the normal case yields a lower bound for the general case without normality.

We will invoke Lemma \ref{lem:tsybakov} in Section \ref{ssec:proof-auxiliary-lemma}.
%from \citet{Tsybakov2009}. 
To that end, we construct a parameter space $\Theta$ that contains elements $\theta = (\theta_{N+1},...,\theta_{2N})^T\in \{0,1\}^N$, where $N$ is the smallest integer such that $N\ge c_1 n^{1/(2r+1)}$ for some constant $c_1>0$ whose value will be specified later. The slope function $\alpha$ depends on $\theta$ through
\be
\label{eq:alpha-theta}
\alpha_\theta =
N^{-1/2}\sum_{k=N+1}^{2N}\theta_k\gamma_k^{1/2}\omega_k.
\ee
Then it is easy to verify that
\begin{eqnarray*}
\|\alpha_\theta\|_K^2 = N^{-1}\sum_{k=N+1}^{2N}\theta_k^2\gamma_k\|\omega_k\|_K^2
=N^{-1}\sum_{k=N+1}^{2N}\theta_k^2
\le N^{-1}\sum_{k=N+1}^{2N}1=1,
\end{eqnarray*}
which proves that $\alpha_\theta\in \H$.

Define the parameter space $\Theta$ as
$\Theta = \{\theta^{(0)},\theta^{(1)},...,\theta^{(M)}\}\subset \{0,1\}^N $. For $N\ge 8$, Gilbert Shannon Varshamov bound guarantees that the following conditions hold simultaneously:
\vspace{-.5em}
\begin{enumerate}
\itemsep -.5em 
\item $\theta^{(0)}=(0,...,0)^T$,
\item $H(\theta,\theta')>N/8$ for any pair $\theta\ne\theta'\in\Theta$, where $H$ is the Hamming distance,
\item The cardinality of the set is at least $M\ge 2^{N/8}$.
\end{enumerate}
\vspace{-.5em}

Denote by $P_\theta$ the joint distribution of $(X_i,Y_i),i=1,...,n$ given $\alpha_0=\alpha_\theta,\beta_0=\omega_1$, then the ratio of the density becomes
\be
 \log\frac{P_{\theta}}{P_{\theta'}} = \frac{2\sum_{i=1}^n\left(Y_i - \int X_i\alpha_\theta\omega_1\right)\int X_i(\alpha_\theta\omega_1-\alpha_{\theta'}\omega_1) + \sum_{i=1}^n \left(\int X_i(\alpha_\theta\omega_1-\alpha_{\theta'}\omega_1)\right)^2}{2\sigma^2}.
\ee
Based on this expression, the Kullback-Leibler distance between $P_\theta$ and $P_{\theta'}$ can be computed
$KL(P_\theta,P_{\theta'}) = \int\log\frac{P_{\theta}}{P_{\theta'}}P_\theta =  \frac{n}{2\sigma^2}\|\alpha_\theta-\alpha_{\theta'}\|_0^2\|\omega_1\|_0^2$.
Plugging in $\alpha_\theta$ \eqref{eq:alpha-theta} leads to
\bes
KL(P_\theta,P_{\theta'})
&=& \frac{n}{2\sigma^2}\left\| N^{-1/2}\sum_{k=N+1}^{2N}(\theta_k-\theta'_k)\gamma_k^{1/2}\omega_k\right\|_0^2
= \frac{n}{2N\sigma^2}\sum_{k=N+1}^{2N}(\theta_k-\theta'_k)^2\gamma_k\\ [-1em]
&\le& \frac{n\gamma_N}{2N\sigma^2}\sum_{k=N+1}^{2N}(\theta_k-\theta'_k)^2
\le \frac{n\gamma_N}{2N\sigma^2}H(\theta,\theta')
\le \frac{n\gamma_N}{2\sigma^2},
\ees
since the Hamming distance is bounded by the dimension. Due to the rate assumption of $\gamma$, the cardinality of the set $\Theta$, and the assumption on the size of $N$,
\bes
KL(P_\theta,P_{\theta'})
\le \frac{c_2nN^{-(2r)}}{2\sigma^2}
\le \frac{c_2c_1^{-2r+1}N^{2r+1}N^{-(2r)}}{2\sigma^2}
= \frac{c_2c_1^{2(-r)+1}N}{2\sigma^2}
\le \delta\log 2^{N/8}
\le \delta\log M,
\end{eqnarray*}
for any $0<\delta<1/8$ by taking $c_1$ large enough. This further proves that
\be
\label{eqn:tsybakov-kl}
 \frac{1}{M}\sum_{j=1}^M KL(P_{\theta^{(j)}},P_{\theta^{(0)}})\le \delta\log M,
\ee
which satisfies the second condition (ii) in Lemma \ref{lem:tsybakov}.

Turning to the first condition (i), define a distance between $(\alpha,\beta)$ and $(\alpha',\beta')$ as
$d\big((\alpha,\beta),(\alpha',\beta')\big)
=E\left(\int_{\T\times\T}X(s,t)\big(\alpha(s)\beta(t)-\alpha'(s)\beta'(t)\big)dsdt\right)^2$.
Due to the definition of $\|\cdot\|_0$ norm and the expression of $\alpha_\theta$, the distance can be lowered bounded by
\bes
d\big((\alpha_\theta,\beta_0),(\alpha_{\theta'},\beta_0)\big)
 &=&\|\alpha_\theta-\alpha_{\theta'}\|_0^2\|\beta_0\|_0^2
 =\left\|N^{-1/2}\sum_{k=N+1}^{2N}(\theta_k-\theta'_k)\gamma_k^{1/2}\omega_k\right\|_0^2\|\omega_1\|_0^2\\ [-0.8em]
 &=&N^{-1}\sum_{k=N+1}^{2N}(\theta_k-\theta'_k)^2\gamma_k
 \ge N^{-1}\gamma_{2N}\sum_{k=N+1}^{2N}(\theta_k-\theta'_k)^2
 = N^{-1}\gamma_{2N}H(\theta,\theta').
\ees
Because of the second requirement of the construction of the set $\Theta$, the rate assumption of $\gamma$, and the assumption on the size of $N$, we have
\bes
d\big((\alpha_\theta,\beta_0),(\alpha_{\theta'},\beta_0)\big)
 \ge \gamma_{2N}/8
 \ge c_3 2^{-2r-3}N^{-2r}
 \ge 2c_4\delta^{\frac{2r}{2r+1}}n^{-\frac{2r}{2r+1}}.
\ees
This inequality and \eqref{eqn:tsybakov-kl} together imply that
\bes
 \inf_{\halpha,\hbeta}\sup_{\theta\in\Theta}
 P_\theta\left(d\big((\halpha,\hbeta),(\alpha_\theta,\beta_0)\big)\ge c_4 \delta^{\frac{2r}{2r+1}}n^{-\frac{2r}{2r+1}}\right)
 \ge \frac{\sqrt{M}}{1+\sqrt{M}}\left(1-2\delta-\sqrt{\frac{2\delta}{\log M}}\right).
\ees
Letting $n\rightarrow \infty$, $\lim_{n\rightarrow\infty}\inf_{\halpha,\hbeta}\sup_{\theta\in\Theta}
 P_\theta\left(d\big((\halpha,\hbeta),(\alpha_\theta,\beta_0)\big)\ge c_4 \delta^{\frac{2r}{2r+1}}n^{-\frac{2r}{2r+1}}\right)
 \ge 1-2\delta$,
which further implies that $\lim_{a\rightarrow 0}\lim_{n\rightarrow\infty}\inf_{\halpha,\hbeta}\sup_{\theta\in\Theta}
 P_\theta\left(d\big((\halpha,\hbeta),(\alpha_\theta,\beta_0)\big)\ge an^{-2r/(2r+1)}\right)
 =1$. Realizing $\calE (\halpha,\hbeta;\alpha_0,\beta_0) = d\big((\halpha,\hbeta),(\alpha_0,\beta_0)\big)$ completes the proof. \hfill $\blacksquare$

\vspace{-1 em}
\section{Main lemmas and their proofs}
\label{sec:proof-main-lemma}

\subsection{Main lemmas}
\vspace{-.5em}
\begin{lemma}
\label{lem:proof-bound1}
If $\lambda=o(1),~0\le a\le 1$, then
\begin{equation*}
  \|\balpha - \alpha_0\|_a^2 = O(\lambda^{1-a}), \mbox{  and   }
\|\bbeta - \beta_0\|_a^2 = O(\lambda^{1-a}).
\end{equation*}
\end{lemma}

An immediate consequence of Lemma \ref{lem:proof-bound1} is $O(\|\balpha\|_a) = O(\|\alpha_0\|_a)=O(1)$ and $O(\|\bbeta\|_a) = O(\|\beta_0\|_a)=O(1)$ because of the following observation
\bes
    \|\balpha\|_a = \|\balpha-\alpha_0+\alpha_0\|_a\ge \|\alpha_0\|_a-\|\balpha-\alpha_0\|_a\succeq \|\alpha_0\|_a- o(1) = O(1),\\
    \|\balpha\|_a = \|\balpha-\alpha_0+\alpha_0\|_a\le \|\alpha_0\|_a+\|\balpha-\alpha_0\|_a\preceq \|\alpha_0\|_a+ o(1) = O(1),
\ees
and a parallel argument for $\bbeta$ holds. From now on, there will be multiple appearances of $\|\balpha\|_a, \|\bbeta\|_a$, which will be treated as constants.

\begin{lemma}
\label{lem:proof-bound2}
If $\lambda=o(1),~ 0\le a\le 1$ and $r>1/2$, then
\begin{equation*}
    \expec \|\talpha-\balpha\|_a^2 \preceq n^{-1}\lambda^{-(a+1/(2r))},
    \mbox{  and  } \expec \|\tbeta-\bbeta\|_a^2 \preceq n^{-1}\lambda^{-(a+1/(2r))}.
\end{equation*}
\end{lemma}
\begin{lemma}
\label{lem:proof-bound3}
If there exists some constant $c$ such that $1/(2r)<c \le 1$ and
$n^{-1}\lambda^{-(c+1/(2r))}=o(1)$, then
\begin{eqnarray*}
    \|\halpha-\talpha\|_a^2 = o_p(n^{-1}\lambda^{-(a+1/(2r))}),
    \mbox{  and   }\|\hbeta-\tbeta\|_a^2 = o_p(n^{-1}\lambda^{-(a+1/(2r))}).
\end{eqnarray*}
\end{lemma}

For the rest of Section \ref{sec:proof-main-lemma}, we will provide Proofs of Lemmas \ref{lem:proof-bound1}-\ref{lem:proof-bound3}.
%--------------------------------------------------------------------
%--------------------------------------------------------------------
%--------------------------------------------------------------------
%--------------------------------------------------------------------
%--------------------------------------------------------------------
\subsection{Proof of Lemma \ref{lem:proof-bound1}}
Expanding $\alpha$, $\alpha_0$, $\balpha$, $\beta$, $\beta_0$, and
$\bbeta$ on the basis $\{\omega_k: k=1,2,\ldots\}$ and denote
\begin{equation}
  \label{eq:proof-lemma-1-expansion}
  \begin{array}{lll}
    \alpha = \sum_{k=1}^\infty a_k \omega_k,~~~~&\alpha_0 = \sum_{k=1}^\infty a_{0k} \omega_k,~~~~&\balpha = \sum_{k=1}^\infty \ba_k \omega_k,\\
    \beta = \sum_{k=1}^\infty b_k \omega_k,~~~~ &\beta_0 = \sum_{k=1}^\infty b_{0k} \omega_k,~~~~ &\bbeta = \sum_{k=1}^\infty \bb_k \omega_k.
  \end{array}
\end{equation}
Substituting the expansions \eqref{eq:proof-lemma-1-expansion} into
$\ell(\alpha,\beta)$ and together with identities %in \eqref{eqn:kernelidentities}, 
\begin{equation}
\label{eqn:kernelidentities}
\langle\omega_j,\omega_k\rangle_{R} = \delta_{jk} (1/\gamma_k+1),~~
\langle C\omega_j,\omega_k\rangle_{\L_2} = \delta_{jk},~~ \mbox{and }
\langle \omega_j,\omega_k\rangle_{K} = \delta_{jk} /\gamma_k,
\end{equation}
it follows that
\begin{equation*}
  \ell(\alpha,\beta) = \sigma^2
  + \left(\sum_{k=1}^\infty a_k^2\right)\left(\sum_{k=1}^\infty b_k^2\right)
  + \left(\sum_{k=1}^\infty a_{0k}^2\right)\left(\sum_{k=1}^\infty b_{0k}^2\right)
  - 2\left(\sum_{k=1}^\infty a_{0k}a_k\right)\left(\sum_{k=1}^\infty b_{0k}b_k\right).
\end{equation*}
Similarly, the penalty term $J(\alpha,\beta)$ can be re-expressed as
\begin{equation*}
    J(\alpha,\beta) =
    \lambda\left(\sum_{k=1}^\infty a_k^2\right) \left(\sum_{k=1}^\infty\gamma_k^{-1}b_k^2\right)
    + \lambda\left(\sum_{k=1}^\infty\gamma_k^{-1}a_k^2\right)
    \left(\sum_{k=1}^\infty b_k^2\right)
    +\lambda^2\left(\sum_{k=1}^\infty\gamma_k^{-1}a_k^2\right)
    \left(\sum_{k=1}^\infty\gamma_k^{-1}b_k^2\right).
\end{equation*}
Minimizing $\ell(\alpha,\beta)+J(\alpha,\beta)$ with respect to $a_k$
and $b_k$ leads to
\begin{equation}
\label{eq:proof-lemma-1-expansion-ba-bb}
\ba_k = c\frac{a_{0k}}{1+\lambda\gamma_k^{-1}},\quad
\bb_k = c^{-1}\frac{b_{0k}}{1+\lambda\gamma_k^{-1}}, \quad k = 1,2, \ldots,
\end{equation}
where $c$ can be any nonzero real constant. For simplicity, we take
$c=1$ and hence $\balpha$ and $\bbeta$ can be written as follows, for all $k=1,2,\ldots$,
\begin{equation}
  \balpha = \sum_{k=1}^\infty \frac{a_{0k}}{1+\lambda\gamma_k^{-1}}
  \omega_k, \quad
  \bbeta = \sum_{k=1}^\infty \frac{b_{0k}}{1+\lambda\gamma_k^{-1}} \omega_k.
\end{equation}

Now we are ready to bound the $\|\balpha-\alpha_0\|_a^2$ term in view of the definition of $\|\cdot\|_a$ norm,
\bes
    \|\balpha - \alpha_0\|_a^2
    &=& \|\sum_{k=1}^\infty \frac{\lambda\gamma_k^{-1}a_{0k}}{1+\lambda\gamma_k^{-1}} \omega_k\|_a^2
    = \sum_{k=1}^\infty (1+\gamma_k^{-a})\left(\frac{\lambda\gamma_k^{-1}a_{0k}}{1+\lambda\gamma_k^{-1}}\right)^2\\
    &\le& \lambda^2\sup_k\frac{\gamma_k^{-1}(1+\gamma_k^{-a})}{(1+\lambda\gamma_k^{-1})^2}  \sum_{k=1}^\infty \gamma_k^{-1}a_{0k}^2
    = \lambda^2\|\alpha_0\|_K^2\sup_k\frac{\gamma_k^{-1}(1+\gamma_k^{-a})}{(1+\lambda\gamma_k^{-1})^2}.
\ees
Replacing maximum over non-negative integers by supremum over a continuous variable in $(0,\infty)$,
\bes
    \sup_k\frac{\gamma_k^{-1}(1+\gamma_k^{-a})}{(1+\lambda\gamma_k^{-1})^2}
    \le\sup_{x>0}\frac{x^{-1}(1+x^{-a})}{(1+\lambda x^{-1})^2}
%    &\le&\sup_{x>0}\frac{x^{-1}}{(1+\lambda x^{-1})^2}+ \sup_{x>0}\frac{x^{-1-a}}{(1+\lambda x^{-1})^2}\\
%    &=&\frac{1}{\inf_{x>0}(x^{1/2}+\lambda x^{-1/2})^2}+ \frac{1}{\inf_{x>0}(x^{(a+1)/2}+\lambda x^{-(1-a)/2})^2}\\
%    &=&O(\lambda^{-1})+O(\lambda^{-(a+1)})\\
    =O(\lambda^{-(a+1)}).
\ees
Combining the last two displays completes the proof of Lemma \ref{lem:proof-bound1}. \hfill $\blacksquare$
%--------------------------------------------------------------------
%--------------------------------------------------------------------
%--------------------------------------------------------------------
%--------------------------------------------------------------------
%--------------------------------------------------------------------
\subsection{Proof of Lemma \ref{lem:proof-bound2}}
For brevity, we first introduce a few more notations. Define a new norm
\begin{equation}
\label{eq:norm-lambda-def}
    \|\cdot\|_\lambda^2 = \|\cdot\|_0^2+\lambda\|\cdot\|_K^2,
\end{equation}
and write
\begin{equation}
  \bba = (\bar{a}_1,\bar{a}_2,\ldots)^T, \quad
\bbb = (\bar{b}_1,\bar{b}_2,\ldots)^T,
\end{equation}
for the vectors that contain the basis expansion coefficients of $\balpha,\bbeta$.

Since $\talpha,\tbeta$ defined in (\ref{eq:tbeta-def}) depends on $H^{-1}$, in order to bound $\|\talpha-\balpha\|_a^2 ,~\|\tbeta-\bbeta\|_a^2$, it is necessary to obtain the explicit form of $H^{-1}$. Note that $H$ takes a block form, we decompose $H$ defined in \eqref{eq:H-definition} and its inverse $H^{-1}$ accordingly as follows,
\begin{equation}
  \label{eq:H-inverse-block-def}
  H = \left(
  \begin{array}{cc}
    H_{11}&H_{12}\\
    H_{21}&H_{22}\\
  \end{array}
  \right),~~~~
  H^{-1} = \left(
  \begin{array}{cc}
    H^{11}&H^{12}\\
    H^{21}&H^{22}\\
  \end{array}
  \right).
\end{equation}
The two diagonal blocks of $H^{-1}$ are further decomposed into two
parts,
\begin{eqnarray}
\label{eq:H-inverse-block-def2}
    H^{11} = H^{11(1)}+ H^{11(2)},~~
    H^{22} = H^{22(1)}+ H^{22(2)},
\end{eqnarray}
where the first parts $^{(1)}$ are the leading terms contributing to the error and the second parts $^{(2)}$ and off diagonal blocks $G^{12},~G^{21}$ are negligible, which will be proved later.

Let $G_{kl}$, $k,l=1,2$, be matrices such that the $ij$-th entry of
$G_{kl}$ is given by,
$(G_{kl})_{ij} = H_{kl}w_iw_j, \quad i,j=1,2,\ldots$.
Write $G^{kl}$ and $G^{kk(l)}$, $k,l=1,2$, in a similar way. Define
$G$ and $G^{-1}$ as matrix counterparts of $H$ and $H^{-1}$
respectively,
\begin{equation}
  \label{eq:G-inverse-block-def}
  G = \left(
  \begin{array}{cc}
    G_{11}&G_{12}\\
    G_{21}&G_{22}\\
  \end{array}
  \right),~~~~
  G^{-1} = \left(
  \begin{array}{cc}
    G^{11}&G^{12}\\
    G^{21}&G^{22}\\
  \end{array}
  \right).
\end{equation}
and $G^{11}$ and $G^{22}$ are further decomposed as,
\begin{eqnarray}
  \label{eq:G-inverse-block-def2}
  G^{11} = G^{11(1)}+ G^{11(2)},~~
  G^{22} = G^{22(1)}+ G^{22(2)}.
\end{eqnarray}
All detailed expressions for each term in $G^{-1}$ are provided in
Lemma \ref{lem:G-inverse} in Section \ref{ssec:proof-auxiliary-lemma}.

Now we are ready to establish the upper bounds for
$\|\talpha-\balpha\|_a$ and $\|\tbeta-\bbeta\|_a$. From the
definitions of $\talpha$ and $\tbeta$ in \eqref{eq:tbeta-def}, the
difference $\talpha-\balpha$ can be written as
$\balpha-\talpha = H^{11(1)} D_\alpha \ell_{n\lambda}(\balpha,\bbeta)
  + H^{11(2)} D_\alpha \ell_{n\lambda}(\balpha,\bbeta)
  + H^{12} D_\beta \ell_{n\lambda}(\balpha,\bbeta)$,
and hence it follows that
\begin{equation}
\label{eq:proof-lemma-2-three-term}
    \|\talpha-\balpha\|_a \le
    \|H^{11(1)} D_\alpha \ell_{n\lambda}(\balpha,\bbeta)\|_a
    + \|H^{11(2)} D_\alpha \ell_{n\lambda}(\balpha,\bbeta)\|_a
    + \|H^{12} D_\beta \ell_{n\lambda}(\balpha,\bbeta)\|_a.
\end{equation}
We now derive the upper bound for each term of the right hand side in
\eqref{eq:proof-lemma-2-three-term}.

For the first term of \eqref{eq:proof-lemma-2-three-term}, we have
\begin{eqnarray}
    \expec\|H^{11(1)} D_\alpha \ell_{n\lambda}(\balpha,\bbeta)\|_a^2
    \label{eq:proof-lemma-2-term-1-part1}
    &=&2^{-2}\|\bbeta\|_\lambda^{-4}
    \expec\|\diag\big((1+\lambda\gamma_k^{-1})^{-1}\big)
    D_\alpha \ell_{n\lambda}(\balpha,\bbeta)\|_a^2\\
    \label{eq:proof-lemma-2-term-1-part2}
    &=&2^{-2}\|\bbeta\|_\lambda^{-4} \expec\sum_{k=1}^\infty(1+\gamma_k^{-a})(1+\lambda\gamma_k^{-1})^{-2}(D_\alpha \ell_{n\lambda}(\balpha,\bbeta)\omega_k)^2 \quad \\ %add \quad to prevent overlapping
    \label{eq:proof-lemma-2-term-1-part3}
    &\preceq&n^{-1}\sum_{k=1}^\infty(1+\gamma_k^{-a})(1+\lambda\gamma_k^{-1})^{-2}\\
    \label{eq:proof-lemma-2-term-1-part4}
    &\preceq&n^{-1}\sum_{k=1}^\infty(1+k^{2ar})(1+\lambda k^{2r})^{-2}\\
    \label{eq:proof-lemma-2-term-1-part5}
    &\asymp& n^{-1}\lambda^{-(a+1/(2r))},
\end{eqnarray}
where (\ref{eq:proof-lemma-2-term-1-part1}) relies on Lemma
\ref{lem:G-inverse}, (\ref{eq:proof-lemma-2-term-1-part2}) can be
obtained by the definition of $\|\cdot\|_0$ norm,
(\ref{eq:proof-lemma-2-term-1-part3}) comes from Lemma
\ref{lem:first-order-derivative},
(\ref{eq:proof-lemma-2-term-1-part4}) is based upon the rate
assumptions on $\gamma_k$ (\ref{eq:gamma-order}), and
(\ref{eq:proof-lemma-2-term-1-part5}) holds if $4r>2ar+1$,
which is valid so long as $r>1/2$. The last line is obtained by
replacing summation by integral approximation and the beta
function. %(See, for instance, johnstone p. 75 (3.48)).

The second term of (\ref{eq:proof-lemma-2-three-term}) can be treated as follows
\begin{eqnarray}
    &&\expec\|H^{11(2)} D_\alpha \ell_{n\lambda}(\balpha,\bbeta)\|_a^2\nonumber\\
    \label{eq:proof-lemma-2-term-2-part1}
    &=&\expec\|4^{-1}\|\balpha\|_\lambda^{-2}\|\bbeta\|_\lambda^{-2}\bba\bba^TD_\alpha \ell_{n\lambda}(\balpha,\bbeta)\|_a^2\\ [-0.3em]
    \label{eq:proof-lemma-2-term-2-part6}
    &=&4^{-2}\|\balpha\|_\lambda^{-4}\|\bbeta\|_\lambda^{-4}\|\balpha\|_a^2\expec\big(\bba^T D_\alpha \ell_{n\lambda}(\balpha,\bbeta)\big)^2\nonumber
    = 4^{-2}\|\balpha\|_\lambda^{-4}\|\bbeta\|_\lambda^{-4}\|\balpha\|_a^2\expec\left(\sum_{k=1}^\infty \bar a_kD_\alpha \ell_{n\lambda}(\balpha,\bbeta)\omega_k\right)^2\nonumber\\ [-0.3em]
    \label{eq:proof-lemma-2-term-2-part2}
    &\le&4^{-2}\|\balpha\|_\lambda^{-4}\|\bbeta\|_\lambda^{-4}\|\balpha\|_a^2
    \sum_{k=1}^\infty (1+\gamma_k^{-c})\bar a_k^2
    \sum_{k=1}^\infty (1+\gamma_k^{-c})^{-1}
    \expec(D_\alpha \ell_{n\lambda}(\balpha,\bbeta)\omega_k)^2\\ [-0.3em]
    \label{eq:proof-lemma-2-term-2-part3}
    &\preceq&\|\balpha\|_c^2n^{-1}\sum_{k=1}^\infty (1+\gamma_k^{-c})^{-1}\\ [-0.3em]
    \label{eq:proof-lemma-2-term-2-part4}
    &\asymp&n^{-1}\\
    \label{eq:proof-lemma-2-term-2-part5}
    &=&o(\expec\|H^{11(1)} D_\alpha \ell_{n\lambda}(\balpha,\bbeta)\|_a^2),
\end{eqnarray}
where (\ref{eq:proof-lemma-2-term-2-part1}) plugs in the expression of $G^{11(2)}$ from Lemma \ref{lem:G-inverse}, (\ref{eq:proof-lemma-2-term-2-part2}) is an application of Cauchy-Schwarz inequality, (\ref{eq:proof-lemma-2-term-2-part3}) makes use of Lemma \ref{lem:first-order-derivative}, (\ref{eq:proof-lemma-2-term-2-part4}) always holds provide that $c>1/2r$, and (\ref{eq:proof-lemma-2-term-2-part5}) demonstrates that this term is dominated by the first term of (\ref{eq:proof-lemma-2-three-term}) when $\lambda=o(1)$.

Similarly, for the third term of (\ref{eq:proof-lemma-2-three-term}),
\begin{eqnarray}
    \expec\|H^{12} D_\beta \ell_{n\lambda}(\balpha,\bbeta)\|_a^2
    &=&\expec\|4^{-1}\|\balpha\|_\lambda^{-2}\|\bbeta\|_\lambda^{-2}\bba\bbb^TD_\beta \ell_{n\lambda}(\balpha,\bbeta)\|_a^2\nonumber\\
    &=& 4^{-2}\|\balpha\|_\lambda^{-4}\|\bbeta\|_\lambda^{-4}\|\balpha\|_a^2
    \expec\big(\bbb^T D_\beta \ell_{n\lambda}(\balpha,\bbeta)\big)^2\nonumber\\
    &\asymp&n^{-1}
    = o(\expec\|H^{11(1)} D_\alpha \ell_{n\lambda}(\balpha,\bbeta)\|_a^2) \label{eq:proof-lemma-2-term-3-part1},
\end{eqnarray}
noticing the symmetry of the third last line of the display and (\ref{eq:proof-lemma-2-term-2-part6}).

Combining (\ref{eq:proof-lemma-2-three-term},
\ref{eq:proof-lemma-2-term-1-part5},
\ref{eq:proof-lemma-2-term-2-part5},
\ref{eq:proof-lemma-2-term-3-part1}), we have now proved Lemma
\ref{lem:proof-bound2} for $\alpha$, the bound for $\beta$ can be
retrieved in parallel. \hfill $\blacksquare$
%--------------------------------------------------------------------
%--------------------------------------------------------------------
%--------------------------------------------------------------------
%--------------------------------------------------------------------
%--------------------------------------------------------------------
\subsection{Proof of Lemma \ref{lem:proof-bound3}}

In pursuance of $\|\halpha-\talpha\|_0^2, ~~\|\hbeta-\tbeta\|_0^2$, we revisit $\talpha,~\tbeta$ and $\halpha,\hbeta$.
By definition of $\talpha,\tbeta$ defined in (\ref{eq:tbeta-def}),
$H\left(\begin{array}{c}\balpha-\talpha\\\bbeta-\tbeta\end{array}\right) =
    \left(\begin{array}{c}D_\alpha \ell_{n\lambda}(\balpha,\bbeta)\\D_\beta \ell_{n\lambda}(\balpha,\bbeta)\end{array}\right)$.
Plugging the definition of $H$ in (\ref{eq:H-definition}), we have
\begin{eqnarray*}
    \left(
    \begin{array}{cc}
    D^2_{\alpha\alpha}\ell_\lambda(\balpha,\bbeta)&D^2_{\alpha\beta}\ell_\lambda(\balpha,\bbeta)\\
    D^2_{\beta\alpha}\ell_\lambda(\balpha,\bbeta)&D^2_{\beta\beta}\ell_\lambda(\balpha,\bbeta)\\
    \end{array}
    \right)
    \left(\begin{array}{c}\balpha-\talpha\\\bbeta-\tbeta\end{array}\right) =
    \left(\begin{array}{c}D_\alpha \ell_{n\lambda}(\balpha,\bbeta)\\D_\beta \ell_{n\lambda}(\balpha,\bbeta)\end{array}\right),
\end{eqnarray*}
which is equivalent to
\begin{eqnarray}
\label{eq:proof-lemma-3-tilde-relation}
    \begin{array}{c}
    D^2_{\alpha\alpha}\ell_\lambda(\balpha,\bbeta)(\balpha-\talpha) + D^2_{\alpha\beta}\ell_\lambda(\balpha,\bbeta)(\bbeta-\tbeta) = D_\alpha \ell_{n\lambda}(\balpha,\bbeta),\\
    D^2_{\beta\alpha}\ell_\lambda(\balpha,\bbeta)(\balpha-\talpha) + D^2_{\beta\beta}\ell_\lambda(\balpha,\bbeta)(\bbeta-\tbeta) = D_\beta \ell_{n\lambda}(\balpha,\bbeta).
    \end{array}
\end{eqnarray}
Taylor expansion of $D_\alpha \ell_{n\lambda}(\halpha,\hbeta), ~D_\beta \ell_{n\lambda}(\halpha,\hbeta)$ around $D_\alpha \ell_{n\lambda}(\balpha,\bbeta), ~D_\beta \ell_{n\lambda}(\balpha,\bbeta)$ implies that
\begin{eqnarray}
\label{eq:proof-lemma-3-hat-relation}
    \begin{array}{c}
    0 = D_\alpha \ell_{n\lambda}(\halpha,\hbeta)
    = D_\alpha \ell_{n\lambda}(\balpha,\bbeta)
    + D_{\beta\alpha}^2 \ell_{n\lambda}(\balpha,\bbeta)(\hbeta-\bbeta)
    + D_{\alpha\alpha}^2 \ell_{n\lambda}(\balpha,\bbeta)(\halpha-\balpha)+ R_\alpha,\\
    0 = D_\beta \ell_{n\lambda}(\halpha,\hbeta)
    = D_\beta \ell_{n\lambda}(\balpha,\bbeta)
    + D_{\beta\alpha}^2 \ell_{n\lambda}(\balpha,\bbeta)(\halpha-\balpha)
    + D_{\beta\beta}^2 \ell_{n\lambda}(\balpha,\bbeta)(\hbeta-\bbeta)+ R_\beta,
\end{array}
\end{eqnarray}
where the higher order residual terms are
\begin{eqnarray*}
    R_\alpha = \frac{1}{2}D_{\alpha\beta\beta}^3 \ell_{n\lambda}(\balpha,\bbeta)(\hbeta-\bbeta)^2
    + D_{\alpha\alpha\beta}^3 \ell_{n\lambda}(\balpha,\bbeta)(\halpha-\balpha)(\hbeta-\bbeta)
    + \frac{1}{2}D_{\alpha\alpha\beta\beta}^4 \ell_{n\lambda}(\balpha,\bbeta)(\halpha-\balpha)(\hbeta-\bbeta)^2,\\ [-0.3em]
    R_\beta = \frac{1}{2}D_{\alpha\alpha\beta}^3 \ell_{n\lambda}(\balpha,\bbeta)(\halpha-\balpha)^2
    + D_{\alpha\beta\beta}^3 \ell_{n\lambda}(\balpha,\bbeta)(\halpha-\balpha)(\hbeta-\bbeta)
    + \frac{1}{2}D_{\alpha\alpha\beta\beta}^4 \ell_{n\lambda}(\balpha,\bbeta)(\halpha-\balpha)^2(\hbeta-\bbeta).
\end{eqnarray*}

Integrating (\ref{eq:proof-lemma-3-tilde-relation},\ref{eq:proof-lemma-3-hat-relation}), we arrive at
\begin{eqnarray*}
    &&D_{\alpha\alpha}^2 \ell_{\lambda}(\balpha,\bbeta)(\halpha-\talpha) + D_{\alpha\beta}^2 \ell_{\lambda}(\balpha,\bbeta)(\hbeta-\tbeta)\\ [-0.3em]
    &=&D_{\alpha\alpha}^2 \ell_{\lambda}(\balpha,\bbeta)(\halpha-\balpha) + D_{\alpha\beta}^2 \ell_{\lambda}(\balpha,\bbeta)(\hbeta-\bbeta)
    + D_{\alpha\alpha}^2 \ell_{\lambda}(\balpha,\bbeta)(\balpha-\talpha) + D_{\alpha\beta}^2 \ell_{\lambda}(\balpha,\bbeta)(\bbeta-\tbeta)\\ [-0.3em]
    &=&D_{\alpha\alpha}^2 \ell_{\lambda}(\balpha,\bbeta)(\halpha-\balpha) + D_{\alpha\beta}^2 \ell_{\lambda}(\balpha,\bbeta)(\hbeta-\bbeta)
    - D_{\alpha\alpha}^2 \ell_{n\lambda}(\balpha,\bbeta)(\halpha-\balpha) - D_{\alpha\beta}^2 \ell_{n\lambda}(\balpha,\bbeta)(\hbeta-\bbeta)-R_\alpha\\ [-0.3em]
    &=&D_{\alpha\alpha}^2 \ell(\balpha,\bbeta)(\halpha-\balpha) + D_{\alpha\beta}^2 \ell(\balpha,\bbeta)(\hbeta-\bbeta)
    - D_{\alpha\alpha}^2 \ell_{n}(\balpha,\bbeta)(\halpha-\balpha) - D_{\alpha\beta}^2 \ell_{n}(\balpha,\bbeta)(\hbeta-\bbeta)-R_\alpha,\\ [-0.3em]
    &&D_{\beta\alpha}^2 \ell_{\lambda}(\balpha,\bbeta)(\halpha-\talpha)
    + D_{\beta\beta}^2 \ell_{\lambda}(\balpha,\bbeta)(\hbeta-\tbeta)\\ [-0.3em]
    &=&D_{\beta\alpha}^2 \ell(\balpha,\bbeta)(\halpha-\balpha)
    + D_{\beta\beta}^2 \ell(\balpha,\bbeta)(\hbeta-\bbeta)
    - D_{\beta\alpha}^2 \ell_{n}(\balpha,\bbeta)(\halpha-\balpha)
    - D_{\beta\beta}^2 \ell_{n}(\balpha,\bbeta)(\hbeta-\bbeta)-R_\beta,
\end{eqnarray*}
which equates the following given the definition of $H$ in (\ref{eq:H-definition})
\begin{eqnarray*}
    H
    \left(
    \begin{array}{c}
    \halpha-\talpha\\
    \hbeta-\tbeta\\
    \end{array}
    \right)
    &=&
    \left(
    \begin{array}{c}
    D_{\alpha\alpha}^2 \ell(\balpha,\bbeta)(\halpha-\balpha)
    - D_{\alpha\alpha}^2 \ell_{n}(\balpha,\bbeta)(\halpha-\balpha) \\
    D_{\beta\alpha}^2 \ell(\balpha,\bbeta)(\halpha-\balpha)
    - D_{\beta\alpha}^2 \ell_{n}(\balpha,\bbeta)(\halpha-\balpha)\\
    \end{array}
    \right)\\ [-0.3em]
    &&+
    \left(
    \begin{array}{c}
    D_{\alpha\beta}^2 \ell(\balpha,\bbeta)(\hbeta-\bbeta)
    - D_{\alpha\beta}^2 \ell_{n}(\balpha,\bbeta)(\hbeta-\bbeta)\\
    D_{\beta\beta}^2 \ell(\balpha,\bbeta)(\hbeta-\bbeta)
    - D_{\beta\beta}^2 \ell_{n}(\balpha,\bbeta)(\hbeta-\bbeta)\\
    \end{array}
    \right)
    -
    \left(
    \begin{array}{c}
    R_\alpha\\
    R_\beta\\
    \end{array}
    \right).
\end{eqnarray*}
Multiplying both sides of the last display by $H^{-1}$ leads to
\begin{eqnarray}
\label{eq:proof-lemma-3-6-term}
    \begin{array}{lll}
    \halpha-\talpha
    &=& H^{11(1)}\big(D_{\alpha\alpha}^2 \ell(\balpha,\bbeta)(\halpha-\balpha)
    - D_{\alpha\alpha}^2 \ell_{n}(\balpha,\bbeta)(\halpha-\balpha)\big)\\ [-0.2em]
    &+& H^{11(2)}\big(D_{\alpha\alpha}^2 \ell(\balpha,\bbeta)(\halpha-\balpha)
    - D_{\alpha\alpha}^2 \ell_{n}(\balpha,\bbeta)(\halpha-\balpha)\big)\\ [-0.2em]
    &+& H^{12}\big(D_{\beta\alpha}^2 \ell(\balpha,\bbeta)(\halpha-\balpha)
    - D_{\beta\alpha}^2 \ell_{n}(\balpha,\bbeta)(\halpha-\balpha)\big)\\ [-0.2em]
    &+& H^{11(1)}\big(D_{\alpha\beta}^2 \ell(\balpha,\bbeta)(\hbeta-\bbeta)
    - D_{\alpha\beta}^2 \ell_{n}(\balpha,\bbeta)(\hbeta-\bbeta)\big)\\ [-0.2em]
    &+& H^{11(2)}\big(D_{\alpha\beta}^2 \ell(\balpha,\bbeta)(\hbeta-\bbeta)
    - D_{\alpha\beta}^2 \ell_{n}(\balpha,\bbeta)(\hbeta-\bbeta)\big)\\ [-0.2em]
    &+& H^{12}\big(D_{\beta\beta}^2 \ell(\balpha,\bbeta)(\hbeta-\bbeta)
    - D_{\beta\beta}^2 \ell_{n}(\balpha,\bbeta)(\hbeta-\bbeta)\big)\\ [-0.2em]
    &-& H^{11}\big(R_\alpha\big)
    - H^{12}\big(R_\beta\big).
    \end{array}
\end{eqnarray}

To study the bound of $\|\halpha-\talpha\|_a$, one only needs to analyze the $\|\cdot\|_a$ norm of the eight terms in (\ref{eq:proof-lemma-3-6-term}) separately, among which the first six terms can be bounded by Lemmas \ref{lem:proof-lemma-3-leading-term} and \ref{lem:proof-lemma-3-non-leading-term} and the last two terms are of smaller order. Therefore,
    $\|\halpha-\talpha\|_a^2 = O(n^{-1}\lambda^{-(a+1/(2r))}\|\halpha-\balpha\|_c^2+n^{-1}\lambda^{-(a+1/(2r))}\|\hbeta-\bbeta\|_c^2)$.
In particular, we obtain the following when letting $a=c$, 
$\|\halpha-\talpha\|_c^2 = O(n^{-1}\lambda^{-(c+1/(2r))}\|\halpha-\balpha\|_c^2+n^{-1}\lambda^{-(c+1/(2r))}\|\hbeta-\bbeta\|_c^2)$.
Under the condition $n^{-1}\lambda^{-(c+1/(2r))}=o(1)$,
applying the triangle inequality yields
$    \|\talpha-\balpha\|_c^2\ge\|\halpha-\balpha\|_c^2-\|\halpha-\talpha\|_c^2 = \big(1-o(1)\big)\|\halpha-\balpha\|_c^2 - o(1)\|\hbeta-\bbeta\|_c^2$.
Hence,
    $\|\halpha-\balpha\|_c^2=O(\|\talpha-\balpha\|_c^2+\|\tbeta-\bbeta\|_c^2)$,
which implies $
    \|\halpha-\talpha\|_c^2 = O(n^{-1}\lambda^{-(c+1/(2r))}\|\talpha-\balpha\|_c^2+n^{-1}\lambda^{-(c+1/(2r))}\|\tbeta-\bbeta\|_c^2)$.
Together with Lemma \ref{lem:proof-bound2} and a parallel argument for $\beta$, completes the proof of Lemma \ref{lem:proof-bound3}. \hfill $\blacksquare$

\section{Auxiliary lemmas and their proofs}
\label{sec:proof-auxiliary-lemma}

\subsection{Auxiliary lemmas}
\label{ssec:proof-auxiliary-lemma}
\begin{lemma}(Expression of $G^{-1}$)
\label{lem:G-inverse}
Suppose $G^{-1}$ is decomposed as in \eqref{eq:G-inverse-block-def}
and \eqref{eq:G-inverse-block-def2}. Then it adopts the following form
\begin{eqnarray*}
    G^{11(1)} &=& 2^{-1}\|\bbeta\|_\lambda^{-2}\diag\big((1+\lambda\gamma_k^{-1})^{-1}\big),\\
    G^{11(2)} &=& -4^{-1}\|\balpha\|_\lambda^{-2}\|\bbeta\|_\lambda^{-2}\bba\bba^T,\\
    G^{22(1)} &=& 2^{-1}\|\balpha\|_\lambda^{-2}\diag\big((1+\lambda\gamma_k^{-1})^{-1}\big),\\
    G^{22(2)} &=& -4^{-1}\|\balpha\|_\lambda^{-2}\|\bbeta\|_\lambda^{-2}\bbb\bbb^T,\\
    G^{12} &=&-4^{-1} \|\balpha\|_\lambda^{-2}\|\bbeta\|_\lambda^{-2}\bba\bbb^T,\\
    G^{21} &=&-4^{-1} \|\balpha\|_\lambda^{-2}\|\bbeta\|_\lambda^{-2}\bbb\bba^T.
\end{eqnarray*}
\end{lemma}

%%%%%%%%%%%%%%%%%%%%%%%%%%%%%%%%%%%%%%%%%%%%%%%%%%%%%%%%%%%%%%%%%%%%%%
\begin{lemma}(Properties of first order operators)
\label{lem:first-order-derivative}
The first order operators have the following properties, for $k=1,2,\ldots$,
\begin{equation*}
    \expec \left( D_\alpha \ell_{n\lambda}(\balpha,\bbeta)\omega_k \right)^2 = O(n^{-1}),\quad
    \expec \left( D_\beta \ell_{n\lambda}(\balpha,\bbeta)\omega_k    \right)^2 = O(n^{-1}).
\end{equation*}
\end{lemma}

%%%%%%%%%%%%%%%%%%%%%%%%%%%%%%%%%%%%%%%%%%%%%%%%%%%%%%%%%%%%%%%%%%%%%%
\begin{lemma}
\label{lem:fourth-moment-bound}
For any $f_i \in \L_2(\T)$, $i=1,\ldots,4$, the following inequality
holds,
\begin{eqnarray*}
    \expec \left(\left(\int_{\T\times\T} X(s,t)f_1(s)f_2(t) ~ dsdt\right)\left(\int_{\T\times\T} X(s,t)f_3(s)f_4(t)~dsdt\right)\right)^2\le M\|f_1\|_0^2\|f_2\|_0^2\|f_3\|_0^2\|f_4\|_0^2,
\end{eqnarray*}
where the constant $M$ is defined in \eqref{eq:4th-moment-condition}.
\end{lemma}

%%%%%%%%%%%%%%%%%%%%%%%%%%%%%%%%%%%%%%%%%%%%%%%%%%%%%%%%%%%%%%%%%%%%%%
\begin{lemma}(Property of second order derivative)
\label{lem:second-order-derivative}
The second order derivative operator can be bounded by
\begin{eqnarray*}
    E(D_{\alpha\alpha}^2 \ell(\balpha,\bbeta)\omega_j\omega_k
    - D_{\alpha\alpha}^2 \ell_{n}(\balpha,\bbeta)\omega_j\omega_k)^2\asymp n^{-1},\\ [-0.2em]
    E(D_{\beta\beta}^2 \ell(\balpha,\bbeta)\omega_j\omega_k
    - D_{\beta\beta}^2 \ell_{n}(\balpha,\bbeta)\omega_j\omega_k)^2\asymp n^{-1},\\ [-0.2em]
    E(D_{\beta\alpha}^2 \ell(\balpha,\bbeta)\omega_j\omega_k
    - D_{\beta\alpha}^2 \ell_{n}(\balpha,\bbeta)\omega_j\omega_k)^2\asymp n^{-1},\\ [-0.2em]
    E(D_{\alpha\beta}^2 \ell(\balpha,\bbeta)\omega_j\omega_k
    - D_{\alpha\beta}^2 \ell_{n}(\balpha,\bbeta)\omega_j\omega_k)^2\asymp n^{-1}.
\end{eqnarray*}
\end{lemma}

%%%%%%%%%%%%%%%%%%%%%%%%%%%%%%%%%%%%%%%%%%%%%%%%%%%%%%%%%%%%%%%%%%%%%%
\begin{lemma}
\label{lem:proof-lemma-3-leading-term}
The two leading terms in (\ref{eq:proof-lemma-3-6-term}) can be bounded by
\begin{eqnarray*}
    \|G^{11(1)}\big(D_{\alpha\alpha}^2 \ell(\balpha,\bbeta)(\halpha-\balpha)
    - D_{\alpha\alpha}^2 \ell_{n}(\balpha,\bbeta)(\halpha-\balpha)\big)\|_a^2
    &\asymp&n^{-1}\lambda^{-(a+1/(2r))}\|\halpha-\balpha\|_c^2,\\ [-0.2em]
    \|G^{11(1)}\big(D_{\alpha\beta}^2 \ell(\balpha,\bbeta)(\hbeta-\bbeta)
    - D_{\alpha\beta}^2 \ell_{n}(\balpha,\bbeta)(\hbeta-\bbeta)\big)\|_a^2 &\asymp&n^{-1}\lambda^{-(a+1/(2r))}\|\hbeta-\bbeta\|_c^2.
\end{eqnarray*}
\end{lemma}

%%%%%%%%%%%%%%%%%%%%%%%%%%%%%%%%%%%%%%%%%%%%%%%%%%%%%%%%%%%%%%%%%%%%%%
\begin{lemma}
\label{lem:proof-lemma-3-non-leading-term}
The four non-leading terms in (\ref{eq:proof-lemma-3-6-term}) can be bounded by
\begin{eqnarray*}
    \|G^{12}\big(D_{\beta\alpha}^2 \ell(\balpha,\bbeta)(\halpha-\balpha)
    - D_{\beta\alpha}^2 \ell_{n}(\balpha,\bbeta)(\halpha-\balpha)\big)\|_a^2
    &\asymp&n^{-1}\|\halpha-\balpha\|_c^2,\\ [-0.2em]
    \|G^{12}\big(D_{\beta\beta}^2 \ell(\balpha,\bbeta)(\hbeta-\bbeta)
    - D_{\beta\beta}^2 \ell_{n}(\balpha,\bbeta)(\hbeta-\bbeta)\big)\|_a^2
    &\asymp&n^{-1}\|\hbeta-\bbeta\|_c^2,\\ [-0.2em]
    \|G^{11(2)}\big(D_{\alpha\alpha}^2 \ell(\balpha,\bbeta)(\halpha-\balpha)
    - D_{\alpha\alpha}^2 \ell_{n}(\balpha,\bbeta)(\halpha-\balpha)\big)\|_a^2
    &\asymp&n^{-1}\|\halpha-\balpha\|_c^2,\\ [-0.2em]
    \|G^{11(2)}\big(D_{\alpha\beta}^2 \ell(\balpha,\bbeta)(\hbeta-\bbeta)
    - D_{\alpha\beta}^2 \ell_{n}(\balpha,\bbeta)(\hbeta-\bbeta)\big)\|_a^2 &\asymp&n^{-1}\|\hbeta-\bbeta\|_c^2.
\end{eqnarray*}
\end{lemma}

%%%%%%%%%%%%%%%%%%%%%%%%%%%%%%%%%%%%%%%%%%%%%%%%%%%%%%%%%%%%%%%%%%%%%%
\begin{lemma}(Theorem 2.5 of \citet{Tsybakov2009})
\label{lem:tsybakov}
Assume that $M\ge 2$ and suppose that the parameter space $\Theta$ contains $\theta_0,\theta_1,...,\theta_M$ such that
(i) $d(\theta_j,\theta_k)\ge 2s >0, \forall 0\le j<k \le M$,\\
(ii) $\forall j=1,...,M$
\bes
    \frac{1}{M}\sum_{j=1}^M KL(P_{\theta_j},P_{\theta_0})\le \delta\log M,
\ees
with $0<\delta< 1/8$. Then
\bes
    \inf_{\hat\theta}\sup_{\theta\in\Theta}P_\theta(d(\hat\theta,\theta)\ge s)\ge \frac{\sqrt{M}}{1+\sqrt{M}}\left(1-2\delta-\sqrt{\frac{2\delta}{\log M}}\right)>0.
\ees
\end{lemma}
The readers are referred to \citet{Tsybakov2009} for the proof of the lemma.

%%%%%%%%%%%%%%%%%%%%%%%%%%%%%%%%%%%%%%%%%%%%%%%%%%%%%%%%%%%%%%%%%%%%%%
\subsection{Proofs of auxiliary lemmas}
\label{ssec:proof-auxiliary-lemma-proof}
{\bf Proof of Lemma \ref{lem:G-inverse}}

Recall the penalty function 
    $J(\alpha,\beta) =
    \lambda\|\alpha\|_0^2\|\beta\|_K^2+\lambda\|\beta\|_0^2\|\alpha\|_K^2+\lambda^2\|\alpha\|_K^2\|\beta\|_K^2$.
The operators related to the penalty function can be defined as
follows,
\begin{eqnarray*}
  D^2_{\alpha\alpha}J(\alpha,\beta)fg &=& 2\lambda\|\beta\|_0^2\langle f, g\rangle_K + 2\lambda\|\beta\|_K^2\langle Cf, g\rangle_{\L^2} + 2\lambda^2\|\beta\|_K^2\langle f, g\rangle_K,\\
  D^2_{\beta\beta}J(\alpha,\beta)fg &=& 2\lambda\|\alpha\|_0^2\langle f, g\rangle_K + 2\lambda\|\alpha\|_K^2\langle Cf, g\rangle_{\L^2} + 2\lambda^2\|\alpha\|_K^2\langle f, g\rangle_K,\\
  D^2_{\alpha\beta}J(\alpha,\beta)fg &=& 4\lambda\langle C\alpha, f\rangle_{\L^2} \langle \beta, g\rangle_K + 4\lambda\langle C\beta, g\rangle_{\L^2} \langle \alpha, f\rangle_K + 4\lambda^2\langle \alpha, f\rangle_K \langle \beta, g\rangle_K.
\end{eqnarray*}
Therefore, 
  $(G_{11})_{jk}=D^2_{\alpha\alpha} \ell_{\lambda}(\balpha,\bbeta)\omega_j\omega_k
  =2\|\bbeta\|_\lambda^2(1+\lambda\gamma_k^{-1}) \delta_{jk}$,
and similarly,
    $(G_{22})_{jk}=D^2_{\beta\beta} \ell_{\lambda}(\balpha,\bbeta)\omega_j\omega_k
    = 2\|\balpha\|_\lambda^2(1+\lambda\gamma_k^{-1}) \delta_{jk}$,
where the definition of the 
$\|\cdot\|_\lambda$ norm is given in
\eqref{eq:norm-lambda-def}.

For the off diagonal blocks $G_{12}$ and $G_{21}$, recall the
expansions of $\alpha_0$, $\balpha$, $\beta_0$, and $\bbeta$ in
\eqref{eq:proof-lemma-1-expansion} and the coefficients of $\balpha$
and $\bbeta$ in \eqref{eq:proof-lemma-1-expansion-ba-bb}, we get
$
    (G_{12})_{jk}=D^2_{\alpha\beta} \ell_{\lambda}(\balpha,\bbeta)\omega_j\omega_k
    = -2a_{0j}b_{0k} + 4\bar{a}_j\bar{b}_k + 4\lambda\gamma_k^{-1}\bar{a}_j\bar{b}_k + 4\lambda\gamma_j^{-1}\bar{a}_j\bar{b}_k + 4\lambda^2\gamma_j^{-1}\gamma_k^{-1}\bar{a}_j\bar{b}_k
    = 2a_{0j}b_{0k}$,
and
$
  (G_{21})_{jk}=D^2_{\beta\alpha} \ell_{\lambda}(\balpha,\bbeta)\omega_j\omega_k
= 2a_{0k}b_{0j}$.
Put the above results in matrix form, it is clear that
\begin{equation*}
  \begin{array}{rl}
    G_{11}=2\|\bbeta\|_\lambda^2\diag\big((1+\lambda\gamma_k^{-1})\big), \quad&
    G_{12}=2\azero\bzero^T, \\ [-0.2em]
    G_{22}=2\|\balpha\|_\lambda^2\diag\big((1+\lambda\gamma_k^{-1})\big),
    \quad & G_{21}=2\bzero\azero^T.
  \end{array}
\end{equation*}

The blocks in $G^{-1}$ can be computed from the block matrix inversion
formula as follows,
\begin{equation*}
\begin{array}{lr}
    G^{11} = (G_{11}-G_{12}G_{22}^{-1}G_{21})^{-1}, \quad &
    G^{12} = -G_{11}^{-1}G_{12}G^{22},    \\ [-0.2em]
    G^{22} = (G_{22}-G_{21}G_{11}^{-1}G_{12})^{-1}, \quad &
    G^{21} = -G_{22}^{-1}G_{21}G^{11},
\end{array}
\end{equation*}
which will be resolved one by one. We begin with the
$G_{12}G_{22}^{-1}G_{21}$ term,
\begin{eqnarray*}
    G_{12}G_{22}^{-1}G_{21}
    &=& (2 \azero \bzero^T)(2^{-1} \|\balpha\|_\lambda^{-2} \diag
    \big( (1 + \lambda \gamma_k^{-1})^{-1} \big) (2 \bzero \azero^T) \\ [-0.2em]
    &=&  2 \|\balpha\|_\lambda^{-2} \big(\bzero^T \diag
    \big( (1 + \lambda \gamma_k^{-1})^{-1} \big) \bzero \big) \azero
    \azero^T \\
    &=& 2 \|\balpha\|_\lambda^{-2}\|\bbeta\|_\lambda^2\azero\azero^T,
\end{eqnarray*}
and hence,
\begin{equation*}
  G_{11}-G_{12}G_{22}^{-1}G_{21}
  = 2\|\bbeta\|_\lambda^2\diag\big((1+\lambda\gamma_k^{-1})\big)
  - 2 \|\balpha\|_\lambda^{-2}\|\bbeta\|_\lambda^2\azero\azero^T.
\end{equation*}
From the Woodbury matrix inversion identity, it follows that
\begin{equation*}
  G^{11} = (G_{11}-G_{12}G_{22}^{-1}G_{21})^{-1}
  = 2^{-1}\|\bbeta\|_\lambda^{-2}\diag\big((1+\lambda\gamma_k^{-1})^{-1}\big)
  -4^{-1}\|\balpha\|_\lambda^{-2}\|\bbeta\|_\lambda^{-2}\bba\bba^T.
\end{equation*}
The first term on the right hand side is defined as $G^{11(1)}$ and
the second one as $G^{11(2)}$.

The $G^{22}$ term can be calculated in a similar fashion and we have
\begin{equation*}
  G^{22}= 2^{-1}\|\balpha\|_\lambda^{-2}\diag\big((1+\lambda\gamma_k^{-1})^{-1}\big)
  -4^{-1}\|\balpha\|_\lambda^{-2}\|\bbeta\|_\lambda^{-2}\bbb\bbb^T.
\end{equation*}
Similarly, the $G^{22(1)}$ and $G^{22(2)}$ terms are defined as the
first and the second term on the right hand side of the above
equation.
As for the $G^{12}$ term, it follows that
\begin{eqnarray*}
    G^{12} &=& -G_{11}^{-1}G_{12}G^{22}\\
    &=& -2^{-1}\|\balpha\|_\lambda^{-2}\|\bbeta\|_\lambda^{-2}
    \diag\big((1+\lambda\gamma_k^{-1})^{-1}\big)\azero\bzero^T
    \left(\diag\big((1+\lambda\gamma_k^{-1})^{-1}\big)-
    2^{-1}\|\bbeta\|_\lambda^{-2}\bbb\bbb^T\right)\\
    &=&-4^{-1} \|\balpha\|_\lambda^{-2}\|\bbeta\|_\lambda^{-2}\bba\bbb^T,
\end{eqnarray*}
and similarly, $G^{21} =-4^{-1} \|\balpha\|_\lambda^{-2}\|\bbeta\|_\lambda^{-2}\bbb\bba^T$.
The proof of Lemma \ref{lem:G-inverse} is complete. \hfill $\blacksquare$

%%--------------------------------------------------------------------
{\bf Proof of Lemma \ref{lem:first-order-derivative}}

Since $(\balpha,\bbeta)$ is the minimizer of $\ell_{\lambda}(\alpha,
\beta)$, the stationary condition ensures that $D_\beta \ell_{\lambda}
(\balpha, \bbeta)=0$, and hence
$D_\beta \ell_{n\lambda}(\balpha,\bbeta)
    = D_\beta \ell_{n\lambda}(\balpha,\bbeta)  - D_\beta \ell_{\lambda}(\balpha,\bbeta)
    = D_\beta \ell_n(\balpha,\bbeta)  - D_\beta \ell(\balpha,\bbeta)$.
Therefore, for any positive interger $k$, we have
\begin{eqnarray*}
   && \expec(D_\beta \ell_{n\lambda}(\balpha,\bbeta)\omega_k)^2
    = \expec(D_\beta \ell_n(\balpha,\bbeta)\omega_k  - D_\beta \ell(\balpha,\bbeta)\omega_k)^2\\ 
    &=&\frac{4}{n}\var\left(\left(Y-\int_{\T\times\T}X(s,t)\balpha(s)\bbeta(t)~dsdt\right)
    \left(\int_{\T\times\T}X(s,t)\balpha(s)\omega_k(t)~dsdt\right)\right)\\
    &\le&\frac{4}{n}\expec\left(\left(Y-\int_{\T\times\T}X(s,t)\balpha(s)\bbeta(t)~dsdt\right)
    \left(\int_{\T\times\T}X(s,t)\balpha(s)\omega_k(t)~dsdt\right)\right)^2\\
    &=&\frac{4\sigma^2}{n}\expec\left(\int_{\T\times\T}X(s,t)\balpha(s)\omega_k(t)~dsdt\right)^2\\
    &+&\frac{4}{n}\expec\left(\left(\int_{\T\times\T}X(s,t)\big(\balpha(s)\bbeta(t)-\alpha_0(s)\beta_0(t)\big)~dsdt\right)
    \left(\int_{\T\times\T}X(s,t)\balpha(s)\omega_k(t)~dsdt\right)\right)^2,
\end{eqnarray*}

For the first term, since $\|\omega_k\|_0^2=1$, we have
\begin{equation}
  \label{eqn:4sigmasq-n-first}
  \expec\left(\int_{\T\times\T}X(s,t)\balpha(s)\omega_k(t)~dsdt\right)^2
  = \|\balpha\|_0^2 \|\omega_k\|_0^2 = \|\balpha\|_0^2.
\end{equation}
For the second term, by Cauchy-Schwarz inequality, it follows that
\begin{eqnarray*}
  &&\expec\left(\left(\int_{\T\times\T}X(s,t)\big(\balpha(s)\bbeta(t)-\alpha_0(s)\beta_0(t)\big)~dsdt\right)
  \left(\int_{\T\times\T}X(s,t)\balpha(s)\omega_k(t)~dsdt\right)\right)^2\\ [-0.3em]
  &\le&\left(\expec\left(\int_{\T\times\T}X(s,t)\big(\balpha(s)\bbeta(t)-\alpha_0(s)\beta_0(t)\big)~dsdt\right)^4
  \expec\left(\int_{\T\times\T}X(s,t)\balpha(s)\omega_k(t)~dsdt\right)^4\right)^{1/2}\\ [-0.3em]
  &\le&M\expec\left(\int_{\T\times\T}X(s,t)\big(\balpha(s)\bbeta(t)-\alpha_0(s)\beta_0(t)\big)~dsdt\right)^2
  \expec\left(\int_{\T\times\T}X(s,t)\balpha(s)\omega_k(t)~dsdt\right)^2
\end{eqnarray*}
\vspace{-.3em}
\begin{eqnarray}
  \label{eqn:4-n-second-one}
  &=&M\|\balpha\|_0^2\expec\left(\int_{\T\times\T}X(s,t)\big(\balpha(s)\bbeta(t)-\alpha_0(s)\beta_0(t)\big)~dsdt\right)^2,
\end{eqnarray}
where the second inequality uses the fourth moment condition
\eqref{eq:4th-moment-condition}.
% The identity
% \begin{equation*}
%     \balpha(s)\bbeta(t)-\alpha_0(s)\beta_0(t) = \\
%     \big(\balpha(s)-\alpha_0(s)\big)\big(\bbeta(t)-\beta_0(t)\big)
%     +\big(\balpha(s)-\alpha_0(s)\big)\beta_0(t)
%     +\big(\bbeta(t)-\beta_0(t)\big)\alpha_0(s)
% \end{equation*}
% yields
It is easy to see
$
    \expec\left(\int_{\T\times\T}X(s,t)\big(\balpha(s)\bbeta(t)-\alpha_0(s)\beta_0(t)\big)~dsdt\right)^2 
     \le 3\|\balpha-\alpha_0\|_0^2\|\bbeta-\beta_0\|_0^2
    +3\|\balpha-\alpha_0\|_0^2\|\beta_0\|_0^2
    +3\|\alpha_0\|_0^2\|\bbeta-\beta_0\|_0^2$.
% \begin{eqnarray*}
%     && \expec\left(\int_{\T\times\T}X(s,t)\big(\balpha(s)\bbeta(t)-\alpha_0(s)\beta_0(t)\big)~dsdt\right)^2 \\
%      &\le& 3\|\balpha-\alpha_0\|_0^2\|\bbeta-\beta_0\|_0^2
%     +3\|\balpha-\alpha_0\|_0^2\|\beta_0\|_0^2
%     +3\|\alpha_0\|_0^2\|\bbeta-\beta_0\|_0^2.
% \end{eqnarray*}
By Lemma \ref{lem:proof-bound1}, it is clear that
%\vspace{-3em}
\begin{equation}
  \label{eqn:4-n-second-two}
    \expec\left(\int_{\T\times\T}X(s,t)\big(\balpha(s)\bbeta(t)-\alpha_0(s)\beta_0(t)\big)~dsdt\right)^2 = O(\lambda).
\end{equation}
Now \eqref{eqn:4sigmasq-n-first}, \eqref{eqn:4-n-second-one} and
\eqref{eqn:4-n-second-two} together yield
$\expec(D_\beta \ell_{n\lambda}(\balpha,\bbeta)\omega_k)^2 \le
\frac{4\sigma^2}{n}\|\balpha\|_0^2 +
\frac{4}{n}M\|\balpha\|_0^2O(\lambda) = O(n^{-1})$. 
An identical argument proves that $\expec(D_\alpha
\ell_{n\lambda}(\balpha,\bbeta)\omega_k)^2 = O(n^{-1})$.
\hfill $\blacksquare$

%---------------------------------------------------------------------
{\bf Proof of Lemma \ref{lem:fourth-moment-bound}}

By Cauchy-Schwarz inequality and the fourth moment condition
\eqref{eq:4th-moment-condition}, it follows that,
\begin{eqnarray*}
    &&\expec\left(\left(\int_{\T\times\T} X(s,t)f_1(s)f_2(t) ~ dsdt\right)\left(\int_{\T\times\T} X(s,t)f_3(s)f_4(t)~dsdt\right)\right)^2\\ [-0.3em]
    &\le&\left(\expec\left(\int_{\T\times\T} X(s,t)f_1(s)f_2(t) ~ dsdt\right)^4\expec\left(\int_{\T\times\T} X(s,t)f_3(s)f_4(t)~dsdt\right)^4\right)^{1/2}\\ [-0.3em]
    &\le&M\expec\left(\int_{\T\times\T} X(s,t)f_1(s)f_2(t) ~ dsdt\right)^2\expec\left(\int_{\T\times\T} X(s,t)f_3(s)f_4(t)~dsdt\right)^2\\ [-0.3em]
    &=&M\|f_1\|_0^2\|f_2\|_0^2\|f_3\|_0^2\|f_4\|_0^2.
\end{eqnarray*}
\hfill $\blacksquare$

%---------------------------------------------------------------------
{\bf Proof of Lemma \ref{lem:second-order-derivative}}

Since $D_{\alpha\alpha}^2 \ell(\balpha,\bbeta) = ED_{\alpha\alpha}^2 \ell_{n}(\balpha,\bbeta)$,
\begin{eqnarray*}
    &&\expec(D_{\alpha\alpha}^2 \ell(\balpha,\bbeta)\omega_j\omega_k
    - D_{\alpha\alpha}^2 \ell_{n}(\balpha,\bbeta)\omega_j\omega_k)^2\\ [-0.3em]
    &=&\expec\left(\frac{1}{n}\sum_{i=1}^n\left(\int_{\T\times\T} x_i(s,t)\omega_j(s)\bbeta(t) ~ dsdt\right)\left(\int_{\T\times\T} x_i(s,t)\omega_k(s)\bbeta(t) ~ dsdt\right)\right.\\ [-0.3em]
    &&-\left.\|\bbeta\|_0^2\int_{\T\times\T}C(s,t)\omega_j(s)\omega_k(t)~dsdt\right)^2\\ [-0.3em]
    &=&\frac{1}{n}\var\left(\left(\int_{\T\times\T} X(s,t)\omega_j(s)\bbeta(t) ~ dsdt\right)\left(\int_{\T\times\T} X(s,t)\omega_k(s)\bbeta(t) ~ dsdt\right)\right)\\ [-0.3em]
% \end{eqnarray*}
% \begin{eqnarray*}   
    &\le&\frac{1}{n}\expec\left(\left(\int_{\T\times\T} X(s,t)\omega_j(s)\bbeta(t) ~ dsdt\right)\left(\int_{\T\times\T} X(s,t)\omega_k(s)\bbeta(t) ~ dsdt\right)\right)^2
    \le \frac{M}{n}\|\bbeta\|_0^4
    \asymp O(n^{-1}),
\end{eqnarray*}
where the second inequality is a simple application of Lemma \ref{lem:fourth-moment-bound}. $E(D_{\beta\beta}^2 \ell(\balpha,\bbeta)\omega_j\omega_k
- D_{\beta\beta}^2 \ell_{n}(\balpha,\bbeta)\omega_j\omega_k)^2\asymp n^{-1}$ can be proved likewise.
Notice that $D_{\beta\alpha}^2 \ell(\balpha,\bbeta) = ED_{\beta\alpha}^2 \ell_{n}(\balpha,\bbeta)$,
\begin{eqnarray*}
    &&\expec(D_{\beta\alpha}^2 \ell(\balpha,\bbeta)\omega_j\omega_k
    - D_{\beta\alpha}^2 \ell_{n}(\balpha,\bbeta)\omega_j\omega_k)^2\\  [-0.3em]
%    &=&\expec\left(-\frac{2}{n}\sum_{i=1}^n\left(y_i-\int_{\T\times\T} x_i(s,t)\balpha(s)\bbeta(t) ~ dsdt\right)\left(\int_{\T\times\T} x_i(s,t)\omega_j(s)\omega_k(t)~dsdt\right).\right.\\
%    &&+\frac{2}{n}\sum_{i=1}^n\left(\int_{\T\times\T} x_i(s,t)\omega_j(s)\bbeta(t) ~ dsdt\right)\left(\int_{\T\times\T} x_i(s,t)\balpha(s)\omega_k(t) ~ dsdt\right),\\
%    &&-2\left(\int_{\T\times\T}C(s,t)\alpha_0(s)\omega_j(t)dsdt\right)
%    \left(\int_{\T\times\T}C(s,t)\beta_0(s)\omega_k(t)dsdt\right)\\
%    &&\left.+4\left(\int_{\T\times\T}C(s,t)\balpha(s)\omega_j(t)dsdt\right)
%    \left(\int_{\T\times\T}C(s,t)\bbeta(s)\omega_k(t)dsdt\right)\right)^2\\
    &=&\frac{4}{n}\var\left(-\left(Y-\int_{\T\times\T} X(s,t)\balpha(s)\bbeta(t) ~ dsdt\right)\left(\int_{\T\times\T} X(s,t)\omega_j(s)\omega_k(t)~dsdt\right)\right.\\ [-0.3em]
    &&\left.+\left(\int_{\T\times\T} X(s,t)\omega_j(s)\bbeta(t) ~ dsdt\right)\left(\int_{\T\times\T} X(s,t)\balpha(s)\omega_k(t) ~ dsdt\right)\right)\\ [-0.3em]
    &\le&\frac{4}{n}\expec\left(-\left(Y-\int_{\T\times\T} X(s,t)\balpha(s)\bbeta(t) ~ dsdt\right)\left(\int_{\T\times\T} X(s,t)\omega_j(s)\omega_k(t)~dsdt\right)\right.\\  [-0.3em]
    &&\left.+\left(\int_{\T\times\T} X(s,t)\omega_j(s)\bbeta(t) ~ dsdt\right)\left(\int_{\T\times\T} X(s,t)\balpha(s)\omega_k(t) ~ dsdt\right)\right)^2.
\end{eqnarray*}
Plugging in the expression of $Y$ and applying Cauchy-Schwarz inequality one more time,
\begin{eqnarray*}
    &&E(D_{\beta\alpha}^2 \ell(\balpha,\bbeta)\omega_j\omega_k
    - D_{\beta\alpha}^2 \ell_{n}(\balpha,\bbeta)\omega_j\omega_k)^2\\  [-0.3em] 
    &\le&\frac{4}{n}E\left(-\epsilon\left(\int_{\T\times\T}X(s,t)\omega_j(s)\omega_k(t)~dsdt\right)\right.\\  [-0.3em]
    &&-\left(\int_{\T\times\T} X(s,t)\alpha_0(s)\beta_0(t) ~ dsdt\right)\left(\int_{\T\times\T} X(s,t)\omega_j(s)\omega_k(t)~dsdt\right)\\  [-0.3em]
    &&+ \left(\int_{\T\times\T} X(s,t)\balpha(s)\bbeta(t) ~ dsdt\right)\left(\int_{\T\times\T} X(s,t)\omega_j(s)\omega_k(t)~dsdt\right)\\ [-0.3em]
    &&\left.+\left(\int_{\T\times\T} X(s,t)\omega_j(s)\bbeta(t) ~ dsdt\right)\left(\int_{\T\times\T} X(s,t)\balpha(s)\omega_k(t) ~ dsdt\right)\right)^2\\ [-0.3em]
% \end{eqnarray*}
% \begin{eqnarray*}   
    &\le&\frac{16}{n}\left\{
    E\left(\epsilon\left(\int_{\T\times\T}X(s,t)\omega_j(s)\omega_k(t)~dsdt\right)\right)^2\right.\\ [-0.2em]
    &&+E\left(\left(\int_{\T\times\T} X(s,t)\alpha_0(s)\beta_0(t) ~ dsdt\right)\left(\int_{\T\times\T} X(s,t)\omega_j(s)\omega_k(t)~dsdt\right)\right)^2\\ [-0.2em]
    &&+E\left(\left(\int_{\T\times\T} X(s,t)\balpha(s)\bbeta(t) ~ dsdt\right)\left(\int_{\T\times\T} X(s,t)\omega_j(s)\omega_k(t)~dsdt\right)\right)^2\\ [-0.2em]
    &&+\left.E\left(\left(\int_{\T\times\T} X(s,t)\omega_j(s)\bbeta(t) ~ dsdt\right)\left(\int_{\T\times\T} X(s,t)\balpha(s)\omega_k(t) ~ dsdt\right)\right)^2\right\}\\ [-0.2em]
    &\le& \frac{16}{n}(\sigma^2+M\|\alpha_0\|_0^2\|\beta_0\|_0^2
    +M\|\balpha\|_0^2\|\bbeta\|_0^2
    +M\|\balpha\|_0^2\|\bbeta\|_0^2)
    \asymp n^{-1},
\end{eqnarray*}
where the last inequality invokes Lemma \ref{lem:fourth-moment-bound} three times. $E(D_{\alpha\beta}^2 \ell(\balpha,\bbeta)\omega_j\omega_k
- D_{\alpha\beta}^2 \ell_{n}(\balpha,\bbeta)\omega_j\omega_k)^2\asymp n^{-1}$ can be proved analogously. \hfill $\blacksquare$

%--------------------------------------------------------------------
{\bf Proof of Lemma \ref{lem:proof-lemma-3-leading-term}}

Recall the expansion (\ref{eq:proof-lemma-1-expansion}), we additionally write the expansion of $\halpha,\hbeta$ as
\be
\label{eq:proof-lemma-expansion-hat}
    \halpha = \sum_{k=1}^\infty \ha_k \omega_k,~~
    \hbeta = \sum_{k=1}^\infty \hb_k \omega_k.
\ee

To bound the first term in (\ref{eq:proof-lemma-3-6-term}), recall the expression of the $G^{11(1)}$ in Lemma \ref{lem:G-inverse} and note the definition of the $\|\cdot\|_a$ norm
\begin{eqnarray*}
    &&\|G^{11(1)}\big(D_{\alpha\alpha}^2 \ell(\balpha,\bbeta)(\halpha-\balpha)
    - D_{\alpha\alpha}^2 \ell_{n}(\balpha,\bbeta)(\halpha-\balpha)\big)\|_a^2\\ [-0.2em]
    &=& 4^{-1}\|\bbeta\|_\lambda^{-4}\sum_{k=1}^\infty (1+\gamma_k^{-a})(1+\lambda\gamma_k^{-1})^{-2}(D_{\alpha\alpha}^2 \ell(\balpha,\bbeta)(\halpha-\balpha)\omega_k
    - D_{\alpha\alpha}^2 \ell_{n}(\balpha,\bbeta)(\halpha-\balpha)\omega_k)^2.
\end{eqnarray*}
Plugging in the expansion of functions $\halpha,\balpha$ in (\ref{eq:proof-lemma-1-expansion}) and (\ref{eq:proof-lemma-expansion-hat}), we get
\begin{eqnarray*}
    &&\|G^{11(1)}\big(D_{\alpha\alpha}^2 \ell(\balpha,\bbeta)(\halpha-\balpha)
    - D_{\alpha\alpha}^2 \ell_{n}(\balpha,\bbeta)(\halpha-\balpha)\big)\|_a^2\\ [-0.2em]
    &=& 4^{-1}\|\bbeta\|_\lambda^{-4}\sum_{k=1}^\infty (1+\gamma_k^{-a})(1+\lambda\gamma_k^{-1})^{-2}
    \left\{\sum_{j=1}^\infty(\ha_j-\ba_j)(D_{\alpha\alpha}^2 \ell(\balpha,\bbeta)\omega_j\omega_k
    - D_{\alpha\alpha}^2 \ell_{n}(\balpha,\bbeta)\omega_j\omega_k)\right\}^2.
\end{eqnarray*}
Cauchy-Schwarz inequality produces
\begin{eqnarray*}
    &&\|G^{11(1)}\big(D_{\alpha\alpha}^2 \ell(\balpha,\bbeta)(\halpha-\balpha)
    - D_{\alpha\alpha}^2 \ell_{n}(\balpha,\bbeta)(\halpha-\balpha)\big)\|_a^2\\  [-0.3em]
    &\le&4^{-1}\|\bbeta\|_\lambda^{-4}\sum_{k=1}^\infty (1+\gamma_k^{-c})(1+\lambda\gamma_k^{-1})^{-2}
    \left\{\sum_{j=1}^\infty(1+\gamma_j^{-a})(\ha_j-\ba_j)^2\right\}\\  [-0.3em]
    &&\left\{\sum_{j=1}^\infty(1+\gamma_j^{-c})^{-1}\left(D_{\alpha\alpha}^2 \ell(\balpha,\bbeta)\omega_j\omega_k
    - D_{\alpha\alpha}^2 \ell_{n}(\balpha,\bbeta)\omega_j\omega_k\right)^2\right\}.
\end{eqnarray*}
Lemma \ref{lem:second-order-derivative} generates
\begin{eqnarray*}
    &&\|G^{11(1)}\big(D_{\alpha\alpha}^2 \ell(\balpha,\bbeta)(\halpha-\balpha)
    - D_{\alpha\alpha}^2 \ell_{n}(\balpha,\bbeta)(\halpha-\balpha)\big)\|_a^2\\  [-0.3em]
    &\preceq& \|\bbeta\|_\lambda^{-4}\sum_{k=1}^\infty (1+\gamma_k^{-a})(1+\lambda\gamma_k^{-1})^{-2}
    \left(\sum_{j=1}^\infty(1+\gamma_j^{-c})(\ha_j-\ba_j)^2\right)\left(\sum_{j=1}^\infty(1+\gamma_j^{-c})^{-1}n^{-1}\right).
\end{eqnarray*}
By the definition of $\|\cdot\|_c$ norm, 
$\sum_{j=1}^\infty(1+\gamma_j^{-c})(\ha_j-\ba_j)^2=\|\halpha-\balpha\|_c^2$,
and
$\sum_{j=1}^\infty(1+\gamma_j^{-c})^{-1}<\infty$,
whenever $c>1/2r$, and
$\sum_{k=1}^\infty (1+\gamma_k^{-a})(1+\lambda\gamma_k^{-1})^{-2} = O(\lambda^{-(a+1/(2r))})$,
we achieve
\bes
    \|G^{11(1)}\big(D_{\alpha\alpha}^2 \ell(\balpha,\bbeta)(\halpha-\balpha)
    - D_{\alpha\alpha}^2 \ell_{n}(\balpha,\bbeta)(\halpha-\balpha)\big)\|_a^2
    \asymp n^{-1}\lambda^{-(a+1/(2r))}\|\halpha-\balpha\|_c^2.
\ees

Following the same spirit,
\begin{eqnarray*}
    &&\|G^{11(1)}\big(D_{\alpha\beta}^2 \ell(\balpha,\bbeta)(\hbeta-\bbeta)
    - D_{\alpha\beta}^2 \ell_{n}(\balpha,\bbeta)(\hbeta-\bbeta)\big)\|_a^2 \\  [-0.4em]
    &=& 4^{-1}\|\bbeta\|_\lambda^{-4}\sum_{k=1}^\infty (1+\gamma_k^{-a})(1+\lambda\gamma_k^{-1})^{-2}(D_{\alpha\beta}^2 \ell(\balpha,\bbeta)(\hbeta-\bbeta)\omega_k
    - D_{\alpha\beta}^2 \ell_{n}(\balpha,\bbeta)(\hbeta-\bbeta)\omega_k)^2\\  [-0.4em]
    &=& 4^{-1}\|\bbeta\|_\lambda^{-4}\sum_{k=1}^\infty (1+\gamma_k^{-a})(1+\lambda\gamma_k^{-1})^{-2}\\  [-0.3em]
    &&\left\{\sum_{j=1}^\infty(\hb_j-\bb_j)(D_{\alpha\beta}^2 \ell(\balpha,\bbeta)\omega_j\omega_k
    - D_{\alpha\beta}^2 \ell_{n}(\balpha,\bbeta)\omega_j\omega_k)\right\}^2
\end{eqnarray*}
\begin{eqnarray*}   
    &\le&4^{-1}\|\bbeta\|_\lambda^{-4}\sum_{k=1}^\infty (1+\gamma_k^{-c})(1+\lambda\gamma_k^{-1})^{-2}
    \left\{\sum_{j=1}^\infty(1+\gamma_j^{-a})(\hb_j-\bb_j)^2\right\}\\  [-0.3em]
    &&\left\{\sum_{j=1}^\infty(1+\gamma_j^{-c})^{-1}\left(D_{\alpha\beta}^2 \ell(\balpha,\bbeta)\omega_j\omega_k
    - D_{\alpha\beta}^2 \ell_{n}(\balpha,\bbeta)\omega_j\omega_k\right)^2\right\}\\ [-0.3em]
    &\preceq&\sum_{k=1}^\infty (1+\gamma_k^{-a})(1+\lambda\gamma_k^{-1})^{-2}\|\hbeta-\bbeta\|_c^2
    \left\{\sum_{j=1}^\infty(1+\gamma_j^{-c})^{-1}n^{-1}\right\}\\ [-0.3em]
    &\asymp&n^{-1}\lambda^{-(a+1/(2r))}\|\hbeta-\bbeta\|_c^2,
\end{eqnarray*}
which finalizes the proof of Lemma \ref{lem:proof-lemma-3-leading-term}.  \hfill $\blacksquare$

%--------------------------------------------------------------------
{\bf Proof of Lemma \ref{lem:proof-lemma-3-non-leading-term}}

Plug in the expression of the $G^{12}$ in Lemma \ref{lem:G-inverse},
\begin{eqnarray*}
    &&\|G^{12}\big(D_{\beta\alpha}^2 \ell(\balpha,\bbeta)(\halpha-\balpha)
    - D_{\beta\alpha}^2 \ell_{n}(\balpha,\bbeta)(\halpha-\balpha)\big)\|_a^2\\ [-0.2em]
    &=&\|-4^{-1} \|\balpha\|_\lambda^{-2}\|\bbeta\|_\lambda^{-2}\bba\bbb^T\big(D_{\beta\alpha}^2 \ell(\balpha,\bbeta)(\halpha-\balpha)
    - D_{\beta\alpha}^2 \ell_{n}(\balpha,\bbeta)(\halpha-\balpha)\big)\|_a^2,
\end{eqnarray*}
which can be simplified as
$    4^{-2} \|\balpha\|_\lambda^{-4}\|\bbeta\|_\lambda^{-4}\|\balpha\|_a^2(\bbb^T\big(D_{\beta\alpha}^2 \ell(\balpha,\bbeta)(\halpha-\balpha)
    - D_{\beta\alpha}^2 \ell_{n}(\balpha,\bbeta)(\halpha-\balpha)\big))^2$.
Replace $\halpha-\balpha$ by its expansion (\ref{eq:proof-lemma-1-expansion}) and (\ref{eq:proof-lemma-expansion-hat}),
\begin{eqnarray*}
    &&\|G^{12}\big(D_{\beta\alpha}^2 \ell(\balpha,\bbeta)(\halpha-\balpha)
    - D_{\beta\alpha}^2 \ell_{n}(\balpha,\bbeta)(\halpha-\balpha)\big)\|_a^2\\ [-0.8em]
    &=&4^{-2} \|\balpha\|_\lambda^{-4}\|\bbeta\|_\lambda^{-4}\|\balpha\|_a^2
    \left\{\sum_{k=1}^\infty\sum_{j=1}^\infty\bb_k(\ha_j-\ba_j)(D_{\beta\alpha}^2 \ell(\balpha,\bbeta)\omega_j\omega_k
    - D_{\beta\alpha}^2 \ell_{n}(\balpha,\bbeta)\omega_j\omega_k)\right\}^2.
\end{eqnarray*}
Due to Cauchy-Schwarz inequality,
\begin{eqnarray*}
    &&\|G^{12}\big(D_{\beta\alpha}^2 \ell(\balpha,\bbeta)(\halpha-\balpha)
    - D_{\beta\alpha}^2 \ell_{n}(\balpha,\bbeta)(\halpha-\balpha)\big)\|_a^2\\ [-0.8em]
    &\le&4^{-2} \|\balpha\|_\lambda^{-4}\|\bbeta\|_\lambda^{-4}\|\balpha\|_a^2
    \left\{\sum_{k=1}^\infty\sum_{j=1}^\infty(1+\gamma_k^{-c})(1+\gamma_j^{-c})\bb_k^2(\ha_j-\ba_j)^2\right\}\\ [-0.8em]
    &&\left\{\sum_{k=1}^\infty\sum_{j=1}^\infty(1+\gamma_k^{-c})^{-1}(1+\gamma_j^{-c})^{-1}(D_{\beta\alpha}^2 \ell(\balpha,\bbeta)\omega_j\omega_k
    - D_{\beta\alpha}^2 \ell_{n}(\balpha,\bbeta)\omega_j\omega_k)^2\right\}.
\end{eqnarray*}
Bounding the second order derivative operator by Lemma \ref{lem:second-order-derivative} gives
\begin{eqnarray*}
    &&\|G^{12}\big(D_{\beta\alpha}^2 \ell(\balpha,\bbeta)(\halpha-\balpha)
    - D_{\beta\alpha}^2 \ell_{n}(\balpha,\bbeta)(\halpha-\balpha)\big)\|_a^2\\ [-0.3em]
    &\preceq&n^{-1}\|\bbeta\|_c^2\|\halpha-\balpha\|_c^2
    \left\{\sum_{k=1}^\infty\sum_{j=1}^\infty(1+\gamma_k^{-c})^{-1}(1+\gamma_j^{-c})^{-1}\right\}
    \asymp n^{-1}\|\halpha-\balpha\|_c^2,
\end{eqnarray*}
because $1+\gamma_k^{-c}$ is summable provided $c>1/2r$.

The proof of the remaining three terms has similar flavor: reexpress the inverse of the second order derivative operator, simplify the expression, expand the functions by basis, apply Cauchy-Schwarz inequality, bound the second order derivative operator, and tidy the formula as follows
\begin{eqnarray*}
    &&\|G^{12}\big(D_{\beta\beta}^2 \ell(\balpha,\bbeta)(\hbeta-\bbeta)
    - D_{\beta\beta}^2 \ell_{n}(\balpha,\bbeta)(\hbeta-\bbeta)\big)\|_a^2\\ [-0.3em]
    &=& \|-4^{-1} \|\balpha\|_\lambda^{-2}\|\bbeta\|_\lambda^{-2}\bba\bbb^T\big(D_{\beta\beta}^2 \ell(\balpha,\bbeta)(\hbeta-\bbeta)
    - D_{\beta\beta}^2 \ell_{n}(\balpha,\bbeta)(\hbeta-\bbeta)\big)\|_a^2\\ [-0.3em]
    &=&4^{-2} \|\balpha\|_\lambda^{-4}\|\bbeta\|_\lambda^{-4}\|\balpha\|_a^2(\bbb^T\big(D_{\beta\beta}^2 \ell(\balpha,\bbeta)(\hbeta-\bbeta)
    - D_{\beta\beta}^2 \ell_{n}(\balpha,\bbeta)(\hbeta-\bbeta)\big))^2\\ [-0.6em]
    &=&4^{-2} \|\balpha\|_\lambda^{-4}\|\bbeta\|_\lambda^{-4}\|\balpha\|_a^2
    \left\{\sum_{k=1}^\infty\sum_{j=1}^\infty\bb_k(\hb_j-\bb_j)(D_{\beta\beta}^2 \ell(\balpha,\bbeta)\omega_j\omega_k
    - D_{\beta\beta}^2 \ell_{n}(\balpha,\bbeta)\omega_j\omega_k)\right\}^2\\ [-0.6em]
    &\preceq&\|\bbeta\|_c^2\|\hbeta-\bbeta\|_c^2\left\{\sum_{k=1}^\infty\sum_{j=1}^\infty(1+\gamma_k^{-c})^{-1}(1+\gamma_j^{-c})^{-1}n^{-1}\right\} 
    \asymp n^{-1}\|\hbeta-\bbeta\|_c^2, \\ [-0.3em]
    &&\|G^{11(2)}\big(D_{\alpha\alpha}^2 \ell(\balpha,\bbeta)(\halpha-\balpha)
    - D_{\alpha\alpha}^2 \ell_{n}(\balpha,\bbeta)(\halpha-\balpha)\big)\|_a^2\\ [-0.3em]
    &=&\|-4^{-1} \|\balpha\|_\lambda^{-2}\|\bbeta\|_\lambda^{-2}\bba\bba^T \big(D_{\alpha\alpha}^2 \ell(\balpha,\bbeta)(\halpha-\balpha)
    - D_{\alpha\alpha}^2 \ell_{n}(\balpha,\bbeta)(\halpha-\balpha)\big)\|_a^2\\ [-0.3em]
    &=&4^{-2}\|\balpha\|_\lambda^{-4}\|\bbeta\|_\lambda^{-4}\|\balpha\|_a^2
    \left(\bba^T \big(D_{\alpha\alpha}^2 \ell(\balpha,\bbeta)(\halpha-\balpha)
    - D_{\alpha\alpha}^2 \ell_{n}(\balpha,\bbeta)(\halpha-\balpha)\big)\right)^2\\ [-0.4em]
    &=& 4^{-2}\|\balpha\|_\lambda^{-4}\|\bbeta\|_\lambda^{-4}\|\balpha\|_a^2
    \left\{\sum_{k=1}^\infty\sum_{j=1}^\infty\ba_k(\ha_j-\ba_j)(D_{\alpha\alpha}^2 \ell(\balpha,\bbeta)\omega_j\omega_k
    - D_{\alpha\alpha}^2 \ell_{n}(\balpha,\bbeta)\omega_j\omega_k)\right\}^2\\ [-0.4em]
    &\preceq&\|\balpha\|_c^2\|\halpha-\balpha\|_c^2\left\{\sum_{k=1}^\infty\sum_{j=1}^\infty(1+\gamma_k^{-c})^{-1}(1+\gamma_j^{-c})^{-1}n^{-1}\right\}
    \asymp n^{-1}\|\halpha-\balpha\|_c^2,
\end{eqnarray*}
\begin{eqnarray*}
    &&\|G^{11(2)}\big(D_{\alpha\beta}^2 \ell(\balpha,\bbeta)(\hbeta-\bbeta)
    - D_{\alpha\beta}^2 \ell_{n}(\balpha,\bbeta)(\hbeta-\bbeta)\big)\|_a^2\\ [-0.3em]
    &=& \|-4^{-1} \|\balpha\|_\lambda^{-2}\|\bbeta\|_\lambda^{-2}\bba\bba^T \big(D_{\alpha\beta}^2 \ell(\balpha,\bbeta)(\hbeta-\bbeta)
    - D_{\alpha\beta}^2 \ell_{n}(\balpha,\bbeta)(\hbeta-\bbeta)\big)\|_a^2\\ [-0.2em]
    &=& 4^{-2}\|\balpha\|_\lambda^{-4}\|\bbeta\|_\lambda^{-4}\|\bba\|_a^2(\bba^T \big(D_{\alpha\beta}^2 \ell(\balpha,\bbeta)(\hbeta-\bbeta)
    - D_{\alpha\beta}^2 \ell_{n}(\balpha,\bbeta)(\hbeta-\bbeta)\big))^2\\ [-0.4em]
    &=& 4^{-2}\|\balpha\|_\lambda^{-4}\|\bbeta\|_\lambda^{-4}\|\balpha\|_a^2
    \left\{\sum_{k=1}^\infty\sum_{j=1}^\infty\ba_k(\hb_j-\bb_j)(D_{\alpha\alpha}^2 \ell(\balpha,\bbeta)\omega_j\omega_k
    - D_{\alpha\alpha}^2 \ell_{n}(\balpha,\bbeta)\omega_j\omega_k)\right\}^2\\ [-0.4em]
    &\preceq&\|\balpha\|_c^2\|\hbeta-\bbeta\|_c^2\left\{\sum_{k=1}^\infty\sum_{j=1}^\infty(1+\gamma_k^{-c})^{-1}(1+\gamma_j^{-c})^{-1}n^{-1}\right\}
    \asymp n^{-1}\|\hbeta-\bbeta\|_c^2.
\end{eqnarray*}
\hfill $\blacksquare$

\section{Theoretical results for the distinct version of two domains}
\label{sec:theory-distinct}

In Section \ref{sec:theory}, for simplicity and notational convenience, we assumed everything related to the two domains are the same, including $\T_1 = \T_2 = \T, ~K_1=K_2=K, ~C_\alpha=C_\beta=C$ and $\lambda_\alpha=\lambda_\beta=\lambda$. In this section, we briefly sketch the preliminary, theorems, and proofs for the distinct version. 

Recall in Section \ref{ssec:theory-preliminary}, we defined the following successively: kernel $R$ associated with the norm $\|\cdot\|_R:=(\|\cdot\|_0^2 + \|\cdot\|_K^2)^{1/2}$, where $\|\cdot\|_0$ and $\|\cdot\|_K$ depend on $C$ and $K$ respectively; linear operators $\L_R, \L_{R^{1/2}}$ and $\L_T = \L_{R^{1/2}CR^{1/2}}$; eigenvalues $s_k^T$'s of $\L_T$; basis functions $\omega_k$, and values $\gamma_k = (1/s_k^T-1)^{-1}$. Suppose all of these quantities are defined again successively and differently for two domains with subscripts $\alpha$ and $\beta$ corresponding to the two domains. Then we have $\gamma_{k,\alpha}$, $\gamma_{k,\beta}$, $\omega_{k,\alpha}$, and $\omega_{k,\beta}$.
The functions, $\omega_{k,\alpha}$ and $\omega_{k,\beta}$, are essential in the proof, since we will expand all of the functions of interest onto these basis functions. The decay rates of $\gamma_{k,\alpha}$ and $\gamma_{k,\beta}$ play a prominent role in the convergence rate.

We will impose the following condition instead of Condition 1 in Section \ref{ssec:theory-preliminary}:\\
{\bf Condition 3:} the values $\gamma_{k,\alpha}$ and $\gamma_{k,\beta}$ satisfy the following decay rates,
\begin{equation}
  \label{eq:gamma-order-distinct}
  \gamma_{k,\alpha}\asymp k^{-2r_\alpha},\,
  \gamma_{k,\beta}\asymp k^{-2r_\beta},
\end{equation}
for some constants $0<r_\alpha,r_\beta<\infty$.\\

Theorems \ref{thm:upper-distinct} and \ref{thm:lower-distinct} state the results of the matching upper and lower bounds for the distinct version and hence the optimality of our proposed estimator.
\begin{theorem}
  \label{thm:upper-distinct}
  Under Conditions 2-3, the smoothness regularization estimators $(\halpha,\hbeta)$ defined in \eqref{eqn:twowayobj} with Candidate 3, %$\lambda=O(n^{-2(r_c+r_k)/(2(r_c+r_k)+1)})$
  $\lambda_\alpha=O(n^{-2r_\alpha/(2r_\alpha+1)})$ and $\lambda_\beta=O(n^{-2r_\beta/(2r_\beta+1)})$ satisfies
  \begin{equation*}
    \lim_{A\rightarrow\infty}\lim_{n\rightarrow\infty}\sup_{\alpha_0 \in
      \H(K),\; \beta_0 \in \H(K)}
    \prob \left(\mathcal{E}(\halpha,\hbeta;\alpha_0,\beta_0)\ge
    A%n^{-\frac{2(r_c+r_k)}{2(r_c+r_k)+1}} 
    \max\left\{n^{-\frac{2r_\alpha}{2r_\alpha+1}},
    n^{-\frac{2r_\beta}{2r_\beta+1}}\right\}
    \right)=0.
  \end{equation*}
\end{theorem}

\begin{theorem}
  \label{thm:lower-distinct}
  Under the same assumptions as in Theorem \ref{thm:upper-distinct}, for any estimate $(\talpha,\tbeta)$ based on the observations $\{(x_i(\cdot,\cdot),y_i): i=1,2,\ldots,n\}$, we have the following lower bound,
  \begin{equation*}
    \lim_{a\rightarrow 0}\lim_{n\rightarrow\infty}\inf_{\talpha,\tbeta}\sup_{\alpha_0 \in
      \H(K),\; \beta_0 \in \H(K)} \prob\left(\mathcal{E}(\talpha,\tbeta;\alpha_0,\beta_0)\ge
    a%n^{-\frac{2(r_c+r_k)}{2(r_c+r_k)+1}}
    %n^{-\frac{2r}{2r+1}}
    \max\left\{n^{-\frac{2r_\alpha}{2r_\alpha+1}},
    n^{-\frac{2r_\beta}{2r_\beta+1}}\right\}
    \right)=1.
  \end{equation*}
\end{theorem}

For the lower bound, proof of Theorem \ref{thm:lower-distinct} is the same as proof of Theorem \ref{thm:lower} when applying the same argument twice to $\alpha$ and $\beta$ separately. 

For the upper bound, proof of Theorem \ref{thm:upper-distinct} follows the same logic as the proof of Theorem \ref{thm:upper}. Note that Lemma \ref{lem:simultaneous-diagonalization} still holds with extra subscripts $\alpha,\beta$. Proof of Theorem \ref{thm:upper} relies upon \eqref{proof-upper-three-term}, proof of Theorem \ref{thm:upper-distinct} relies upon the distinct version of \eqref{proof-upper-three-term}, which is 
\begin{equation}
\label{proof-upper-three-term-distinct}
\calE(\halpha,\hbeta;\alpha_0,\beta_0) \le
3\|\halpha-\alpha_0\|_{0\alpha}^2\|\hbeta-\beta_0\|_{0\beta}^2
+3\|\halpha-\alpha_0\|_{0\alpha}^2\|\beta_0\|_{0\beta}^2
+3\|\alpha_0\|_{0\alpha}^2\|\hbeta-\beta_0\|_{0\beta}^2,
\end{equation}
where the norms on the right hand side are defined in \eqref{eqn:zerosnorm}.

Define the norms $\|\cdot \|_{a\alpha}$ and $\|\cdot \|_{a\beta}$ similarly as for $\|\cdot \|_{a}$. Previously for the version with same domain, to bound \eqref{proof-upper-three-term}, we bound the three terms on the right hand side of \eqref{eqn:threeterms}. Now for the distinct version, to bound \eqref{proof-upper-three-term-distinct}, we bound the three terms on the right hand sides of \eqref{eqn:threeterms-distinct-alpha} and \eqref{eqn:threeterms-distinct-beta}
\begin{eqnarray}
  \label{eqn:threeterms-distinct-alpha}
  \|\halpha-\alpha_0\|_{a\alpha} \le \|\halpha - \talpha\|_{a\alpha} + \|\talpha -
  \balpha\|_{a\alpha} + \|\balpha - \alpha_0\|_{a\alpha},\\
  \label{eqn:threeterms-distinct-beta}
  \|\hbeta-\beta_0\|_{a\beta} \le \|\hbeta - \tbeta\|_{a\beta} + \|\tbeta -
  \bbeta\|_{a\beta} + \|\bbeta - \beta_0\|_{a\beta}.  
\end{eqnarray}

Lemmas \ref{lem:proof-bound1}, \ref{lem:proof-bound2} and \ref{lem:proof-bound3} provide the bounds of the three terms on the right hand side of \eqref{eqn:threeterms} when the two domains are similar. Lemmas \ref{lem:proof-bound1-distinct}, \ref{lem:proof-bound2-distinct} and \ref{lem:proof-bound3-distinct} replace Lemmas \ref{lem:proof-bound1}, \ref{lem:proof-bound2} and \ref{lem:proof-bound3} for the distinct version.

\begin{lemma}
\label{lem:proof-bound1-distinct}
If $\lambda_\alpha=\lambda_\beta=o(1),~0\le a\le 1$, then
\begin{equation*}
  \|\balpha - \alpha_0\|_{a\alpha}^2 = O(\lambda_\alpha^{1-a}), \mbox{  and   }
\|\bbeta - \beta_0\|_{a\beta}^2 = O(\lambda_\beta^{1-a}).
\end{equation*}
\end{lemma}

\begin{lemma}
\label{lem:proof-bound2-distinct}
If $\lambda_\alpha=\lambda_\beta=o(1),~ 0\le a\le 1$ and $r_\alpha, r_\beta>1/2$, then
\begin{equation*}
    \expec \|\talpha-\balpha\|_{a\alpha}^2 \preceq n^{-1}\lambda_\alpha^{-(a+1/(2r_\alpha))},
    \mbox{  and  } \expec \|\tbeta-\bbeta\|_{a\beta}^2 \preceq n^{-1}\lambda_\beta^{-(a+1/(2r_\beta))}.
\end{equation*}
\end{lemma}
\begin{lemma}
\label{lem:proof-bound3-distinct}
If there exists some constant $c$ such that $\max\{1/(2r_\alpha),1/(2r_\beta)\}<c \le 1$ and
$n^{-1}\lambda_\alpha^{-(c+1/(2r_\alpha))}=o(1)$, $n^{-1}\lambda_\beta^{-(c+1/(2r_\beta))}=o(1)$, then
\begin{eqnarray*}
    \|\halpha-\talpha\|_{a\alpha}^2 = o_p(n^{-1}\lambda_\alpha^{-(a+1/(2r_\alpha))}),
    \mbox{  and   }\|\hbeta-\tbeta\|_{a\beta}^2 = o_p(n^{-1}\lambda_\beta^{-(a+1/(2r_\beta))}).
\end{eqnarray*}
\end{lemma}

Since the roles of $\alpha$'s and $\beta$'s can be switched, for the distinct version, we only need to prove Lemmas \ref{lem:proof-bound1-distinct}, \ref{lem:proof-bound2-distinct} and \ref{lem:proof-bound3-distinct} related to $\alpha$'s. 

% For the case of identical domain, we bound $\|\halpha-\alpha_0\|_a^2$, where the norm $\|\cdot\|_a$ for $0\le a\le 1$ is defined by
% $
%     \|f\|_a^2 = \sum_{k=1}^\infty(1+\gamma_k^{-a})f_k^2,
% $
% when $f=\sum_{k=1}^\infty f_k\omega_k$. For the case of distinct version, we need to bound two different norms $\|\cdot\|_{a,\alpha}$ and $\|\cdot\|_{a,\beta}$ for $\halpha-\alpha$ and $\hbeta-\beta$ respectively. Here, the two norms are defined by 
% $
%     \|f\|_{a,\alpha}^2 = \sum_{k=1}^\infty(1+\gamma_{k,\alpha}^{-a})f_k^2,
% $
% and 
% $
%     \|f\|_{a,\beta}^2 = \sum_{k=1}^\infty(1+\gamma_{k,\beta}^{-a})f_k^2.
% $

The proof of Lemma \ref{lem:proof-bound1-distinct} remains almost the same as the proof of Lemma \ref{lem:proof-bound1} except for adding subscripts $_\alpha$ and $_\beta$. 

The proof of Lemma \ref{lem:proof-bound2-distinct} for the distinct version relies upon the upper bounds of the three terms on the right hand side of \eqref{eq:proof-lemma-2-three-term}, which is dominated by the first term. This first term is still bounded by the same rate as in \eqref{eq:proof-lemma-2-term-1-part5} when extra subscript is added, i.e., $r$ becomes $r_\alpha$ and $\lambda$ becomes $\lambda_\alpha$. The reason is that \eqref{eq:proof-lemma-2-term-1-part5} depends upon Lemma \ref{lem:G-inverse} and Lemma \ref{lem:first-order-derivative}. 
Lemma \ref{lem:first-order-derivative} still holds for the distinct version and in Lemma \ref{lem:G-inverse}, the two relevant equations become $G^{11(1)} = 2^{-1}\|\bbeta\|_{\lambda_\beta}^{-2}\diag\big((1+\lambda_\alpha\gamma_{k,\alpha}^{-1})^{-1}\big)$ and $G^{22(1)} = 2^{-1}\|\balpha\|_{\lambda_\alpha}^{-2}\diag\big((1+\lambda_\beta\gamma_{k,\beta}^{-1})^{-1}\big)$. 

The proof of Lemma \ref{lem:proof-bound3-distinct} for the distinct version still hinges upon \eqref{eq:proof-lemma-3-6-term} and its $\beta$ version. The terms on the right hand side of \eqref{eq:proof-lemma-3-6-term} need to be bounded by Lemmas \ref{lem:proof-lemma-3-leading-term} and \ref{lem:proof-lemma-3-non-leading-term}. Lemma \ref{lem:proof-lemma-3-non-leading-term} still applies for the distinct version when $\|\cdot\|_{c}$ is replaced by $\|\cdot\|_{c\alpha}$ or $\|\cdot\|_{c\beta}$. Lemma \ref{lem:proof-lemma-3-leading-term} holds when adding subscripts appropriately to the RHS. Therefore,
$\|\halpha-\talpha\|_{a\alpha}^2 = O(n^{-1}\lambda_\alpha^{-(a+1/(2r_\alpha))}\|\halpha-\balpha\|_{c\alpha}^2+n^{-1}\lambda_\beta^{-(a+1/(2r_\beta))}\|\hbeta-\bbeta\|_{c\beta}^2)$.
In particular, we obtain the following when letting $a=c$, 
$\|\halpha-\talpha\|_{c\alpha}^2 = O(n^{-1}\lambda_\alpha^{-(a+1/(2r_\alpha))}\|\halpha-\balpha\|_{c\alpha}^2+n^{-1}\lambda_\beta^{-(a+1/(2r_\beta))}\|\hbeta-\bbeta\|_{c\beta}^2)$.
Under the conditions $n^{-1}\lambda_\alpha^{-(c+1/(2r_\alpha))}=o(1)$ and $n^{-1}\lambda_\beta^{-(c+1/(2r_\beta))}=o(1)$,
applying the triangle inequality yields
$\|\talpha-\balpha\|_{c\alpha}^2\ge\|\halpha-\balpha\|_{c\alpha}^2-\|\halpha-\talpha\|_{c\alpha}^2 = \big(1-o(1)\big)\|\halpha-\balpha\|_{c\alpha}^2 - o(1)\|\hbeta-\bbeta\|_{c\beta}^2$.
Hence, $\|\halpha-\balpha\|_{c\alpha}^2=O(\|\talpha-\balpha\|_{c\alpha}^2+\|\tbeta-\bbeta\|_{c\beta}^2)$,
which implies $\|\halpha-\talpha\|_{c\alpha}^2 = O(n^{-1}\lambda_\alpha^{-(c+1/(2r_\alpha))}\|\talpha-\balpha\|_{c\alpha}^2+n^{-1}\lambda_\alpha^{-(c+1/(2r_\alpha))}\|\tbeta-\bbeta\|_{c\beta}^2)$.
Together with Lemma \ref{lem:proof-bound2-distinct} and a parallel argument for $\beta$, completes the proof of Lemma \ref{lem:proof-bound3-distinct}.

\section{Additional simulation results}
\label{sec:exsim}
In this section, we provide additional simulation results as a supplement to Section \ref{ssec:sim-results}. 

\subsection{Additional simulation results on 1D FLR+vectorization}
\label{sec:exsim-FLR-vec}

We assess the performance of FLR after various types of vectorization. Given the matrix-valued predictor, another natural but undesirable choice is to perform vectorization first and then apply existing methods which apply to vector-valued data. 
It is known in the literature of tensor data analysis that such vectorization is sub-optimal. With the additional feature of the functional data, to make comprehensive comparison with existing methods, we still include the 1D FLR of \citet{cai2012minimax} after vectorization. In the literature, the default vectorization approach, denoted by $\textrm{vec}$ here, is to stack all the columns of a matrix one by one. In the context of functional data, because of the requirement of smoothness, there are a few other ways of vectorization. For example, it might be worthwhile to consider flipping the even-numbered column vectors upside down and then stack all the columns together, which is denoted by $\textrm{vec}^*$. Furthermore, maybe rows are smoother, and so stacking rows (and potentially flipping even-numbered rows) is more appropriate. 
These considerations lead to four ways of vectorization and result in FLR+$\textrm{vec}(X(s,t))$, FLR+$\textrm{vec}(X^T(s,t))$, FLR+$\textrm{vec}^*(X(s,t))$ and FLR+$\textrm{vec}^*(X^T(s,t))$.

\begin{figure}[!ht]
	\centerline{\includegraphics[width=\textwidth]{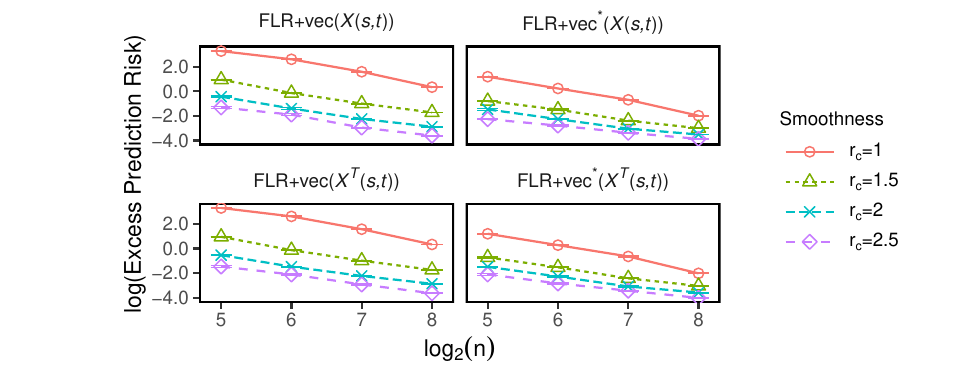}}
	\caption{Plots of the excess prediction risk vs the sample size with both axes in log scale under Setting 1. Four sample sizes and four values of $r_c$ are considered. The error bars correspond to mean $\pm$ one SE. The four panels are for four types of vectorization methods.}
	\label{fig:r-error-flr}
\end{figure}

The comparison of the four different vectorization approaches is shown in Figure \ref{fig:r-error-flr}. Because in the simulation setup, the covariances $C_\alpha,C_\beta$ and coefficient functions $\alpha_0(\cdot),\beta_0(\cdot)$ are symmetric for the two domains, transposing $X$ or not, i.e., stacking rows or columns, does not matter. However, concatenating head-to-tail or head-to-head does matter as revealed in the figure, where the performance of $\textrm{vec}^*$ dominates that of $\textrm{vec}$. 

For these reasons, for the rest of this section, FLR refers to FLR+$\textrm{vec}^*(X(s,t))$; in Section \ref{sec:realdata} on the Canadian weather data, FLR refers to FLR+$\textrm{vec}(X^T(s,t))$, because it is natural to connect the last hour in the current day to the first hour in the next day; in Section Appendix \ref{sec:rd lidar} on the LIDAR data, where the two domains are different, we consider two choices: FLR+$\textrm{vec}^*(X(s,t))$ and FLR+$\textrm{vec}^*(X^T(s,t))$.

\subsection{Additional simulation results on other competitors}
\label{sec:exsim-competitors}

\begin{figure}[!ht]
    \centerline{ \includegraphics[width=\textwidth]{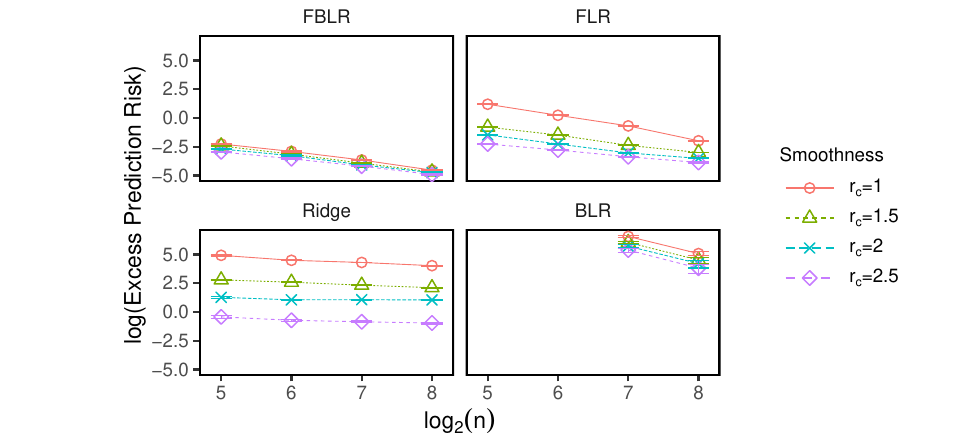}}
    \caption{Plots of the excess prediction risk vs 
    the sample size in log scale for Settings 1. Four approaches are included. The error bars are generated according to mean $\pm$ one SE. BLR is only shown for $\log_2(n)=7,8$, because it requires a larger sample size.}
    \label{fig:r-error-4mdl}
\end{figure}

Besides GMRF, PFPCR, MFPCR, FLR+TPK, we also compare the performance of FBLR with three more existing methods: FLR, Ridge regression after plain vectorization, and bilinear regression (BLR), which is a special case of FBLR when $\lambda_\alpha=\lambda_\beta=0$, under Setting 1. Figure \ref{fig:r-error-4mdl} shows the clear advantage of FBLR over them due to their lack of smoothness or lack of matrix structure.

\subsection{Additional simulation results on 2D-FPCR}
\label{sec:exsim-2D-FPCR}

\begin{figure}[!ht]
	\centerline{\includegraphics[width= \textwidth]{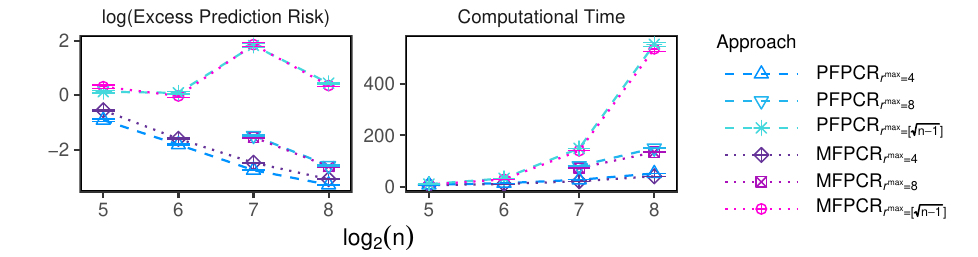}}
	\caption{Plots of the excess prediction risk and computational time vs the sample size in log scale with $r_c=1$ under Setting 1. %Six approaches of PFPCR or MFPCR with different number of components $r^{\max}$ are considered here. 
 The error bars correspond to mean $\pm$ one SE. }
	\label{fig:r-error-fpca}
\end{figure}

We examine the choice of $r^{\max}$ in 2D-FPCR.
Figure \ref{fig:r-error-fpca} demonstrates the prediction risk and computation time of PFPCR and MFPCR with three options of $r^{\max}\in \{4,8,\lfloor\sqrt{n-1}\rfloor\}$. Here, $\lfloor\sqrt{n-1}\rfloor$ is the largest possible value for $r^{\max}$ in \citet{chen2017modelling}. It is clear that $r^{\max}=\lfloor\sqrt{n-1}\rfloor$ is far more computationally expensive and even less accurate than the other two. Because given that the true coefficient function under Setting 1 only consists of the leading four basis functions, estimating more than four PCs will hurt the performance. Therefore, in Section \ref{sec:sim}, only $r^{\max} \in \{4,8\}$ are compared with FBLR.

\subsection{Additional simulation results on FBLR with a different choice of kernel}
\label{sec:exsim-kernel-choice}

It is known that the estimation procedures that use the RKHS frameworks depend on the choice of the kernels, such as 1D FLR, FLR+TPK, and our FBLR. In this section, we use a simulation study to investigate how much influence the choice of the kernel has on the performance of FBLR. For the simulation study in Section \ref{sec:sim} and the real data of LIDAR in Section \ref{sec:rd lidar}, we use kernel \eqref{eq:sim-kernel}; for real data of Canadian weather in Section \ref{sec:realdata},  we use $K(s,t) = 1-B_4(|s-t|)/24$, both choices of kernels were used in \citet{cai2012minimax}. 
In this section, we consider a universal Gaussian kernel $K(s,t) = \exp \left(-\frac{(s - t)^2}{2\sigma^2}\right)$, and refer to this approach as FBLR+GK. The parameter $\sigma$ is selected through cross-validation. We provide simulation results for all Settings 1-6 (with $r_c = 1$) in Section \ref{ssec:sim-results}. For Settings 5-6, we implement ${\rm FBLR}_{R=2}$ with the Gaussian kernel, denoted as ${\rm FBLR}_{R=2}$+GK.

\begin{figure}[!ht]
	\centerline{\includegraphics[width= .9\textwidth]{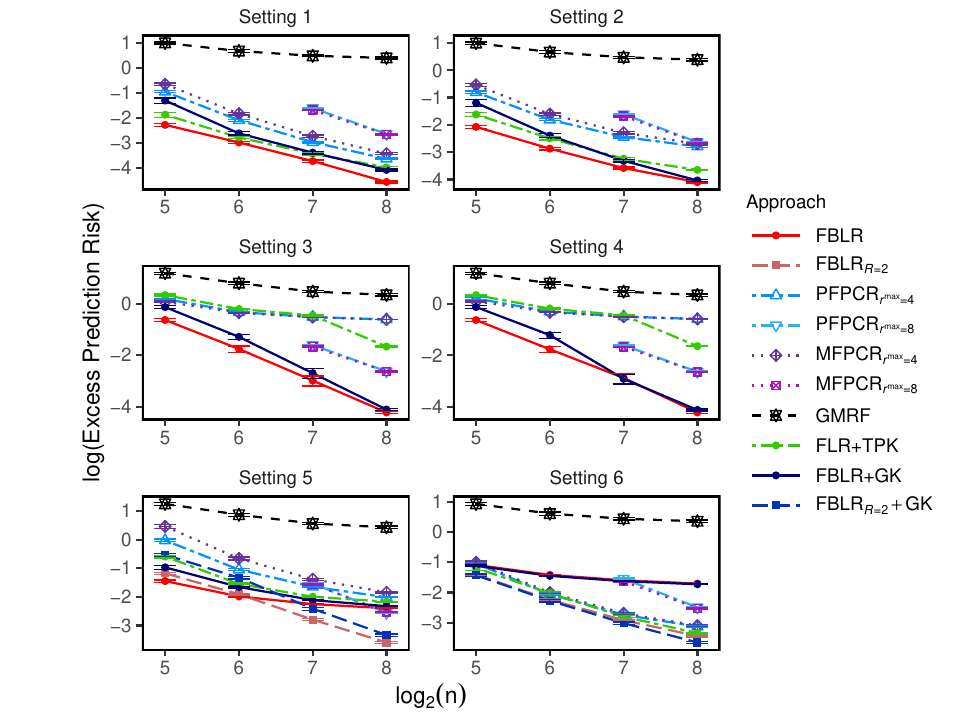}}
	\caption{Plots of the excess prediction risk vs the sample size in log scale with $r_c=1$ under Settings 1-6. The error bars correspond to mean $\pm$ one SE. }
	\label{fig:r-error+GK}
\end{figure}

% Conclusions:
% 1. Gaussian kernel is worse than the original kernel
% 2. FBLR+GK work better than other 2D FLR methods.
Figure \ref{fig:r-error+GK} demonstrates the prediction risk of FBLR+GK and ${\rm FBLR}_{R=2}$+GK, along with all the approaches shown in Section \ref{ssec:sim-results}. 

For Settings 1-4, where the true model follows a bilinear form (\ref{eqn:2dflr}), both $\alpha(\cdot)$ and $\beta(\cdot)$ are composed of multiple cosine basis functions. The aforementioned kernel \eqref{eq:sim-kernel} matches the linear span of these basis functions. See \citet{cai2012minimax} for more details on the RKHS associated with the kernel \eqref{eq:sim-kernel}. It is expected that FBLR+GK performs worse than FBLR (with kernel \eqref{eq:sim-kernel}). However, FBLR+GK consistently outperforms PFPCR, MFPCR, GMRF, and FLR+TPK, except for Settings 1-2 with very small sample size where FBLR+GK is slightly worse than FLR+TPK. Note that here, FLR+TPK still uses the tensor product kernel that depends on kernel \eqref{eq:sim-kernel}.

For Settings 5-6, where the true models are based on Model (\ref{eqn:2dflr-other}), the true coefficient function $\beta_0(\cdot,\cdot)$ is a 2D function and hence we also consider ${\rm FBLR}_{R=2}$+GK. 
Under Setting 5, the true coefficient function still depends on cosine basis function, therefore, FBLR+GK and ${\rm FBLR}_{R=2}$+GK are slightly worse than FBLR and ${\rm FBLR}_{R=2}$ respectively. But the comparison between FBLR+GK, ${\rm FBLR}_{R=2}$+GK and all the other procedures remains the same as the comparison between FBLR, ${\rm FBLR}_{R=2}$ and all the other procedures. In short, excluding FBLR and ${\rm FBLR}_{R=2}$, FBLR+GK is the best among all for small sample sizes while ${\rm FBLR}_{R=2}$+GK is the best among all for large sample sizes.
Under Setting 6, where the two-dimensional coefficient function is not low-rank and we do not know the true basis function. It is seen that FBLR and FBLR+GK yield similar results, as do ${\rm FBLR}_{R=2}$ and ${\rm FBLR}_{R=2}$+GK. ${\rm FBLR}_{R=2}$ and ${\rm FBLR}_{R=2}$+GK dominate all the others. 

In summary, no matter for FBLR with or without deflation, the implementation of the methodology certainly requires the specific choice of the kernel function, but the impact of the kernel function on the performance of the FBLR and its iterative deflation version is minimal and sometimes negligible. More importantly, the small impact of the kernel choice does not overshadow the strength of FBLR over other methods.

\subsection{Additional simulation results on FBLR for not-fully-observed data}
\label{sec:exsim-sparse}

In this section, we study the numerical performance of our procedure when the data are not fully observed. We examine all six settings in Section \ref{sec:sim}. For each setting, we generate $n$, varying $n$, samples of the two-dimensional functional predictor on a 100 $\times$ 100 regular grid within $[0,1]$ $\times$ $[0,1]$. 
The predictor is fully observed in the first domain $\T_1$, but not fully observed in the second domain $\T_2$, being recorded at $L$ random locations. The observations are denoted as $\{X_i(s,T_{ij}), 1 \leq i \leq n, 1 \leq s \leq 100, 1 \leq j \leq L\}$. We consider three sampling frequencies ($L = 10, 30, 50$) to achieve varying levels of sparsity.

To extend FBLR to not-fully-observed data, we first applied the principal component analysis through conditional expectation (PACE) method proposed by \citet{yao2005functional} to impute data. Then we apply FBLR or ${\rm FBLR}_{R=2}$ to the resulting dense data on the $100 \times 100$ grid.

Note that PACE is applicable to one-dimensional functional data. To adapt it for our two-dimensional data, we first convert $n$ samples of sizes $100\times L$ to $100 n$ samples of one-dimensional input of size $L$. Applying PACE will lead to $100 n$ samples of one-dimensional input of size $100$. We finally reshape the data back to $n$ samples of sizes $100\times 100$.

\begin{figure}[!ht]
	\centerline{\includegraphics[width= .9\textwidth]{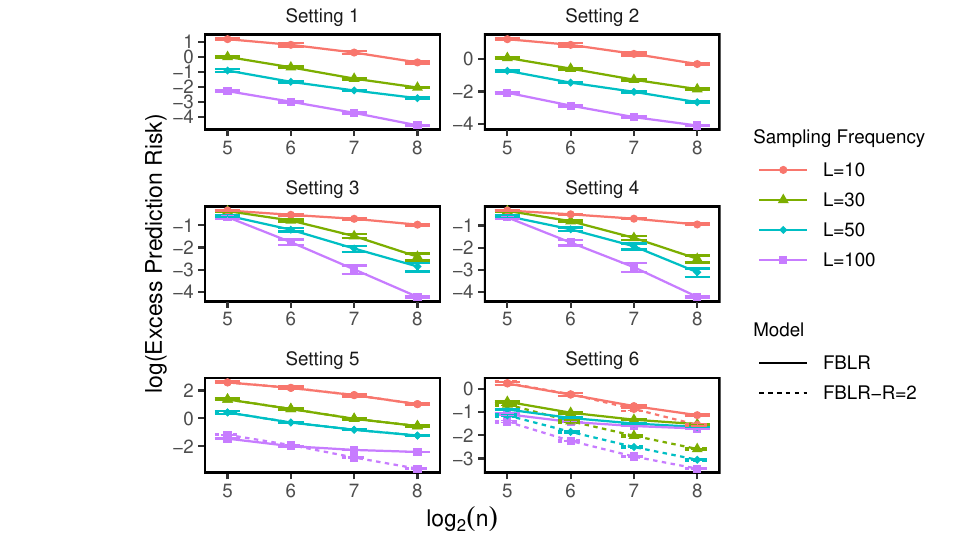}}
	\caption{Plots of the excess prediction risk and vs the sample size in log scale with $r_c=1$ under Settings 1-6. The error bars correspond to mean $\pm$ one SE. }
	\label{fig:r-error-sparse}
\end{figure}

The results are presented in Figure \ref{fig:r-error-sparse}, together with dense data ($L = 100$). For Settings 1-4, FBLR performance is shown. For Settings 5-6, the performance of FBLR and ${\rm FBLR}_{R=2}$ are both shown. The performance of either FBLR or ${\rm FBLR}_{R=2}$ improves as sample size $n$ increases or data get denser. Interestingly, in Setting 5, although the true coefficient function is rank 2, ${\rm FBLR}_{R=2}$ outperforms FBLR for dense data ($L=100$) but not for non-fully-observed data ($L=10,30,50$). This further demonstrates that when data get sparser, simpler model might be preferred. In Setting 6, the true coefficient function does not have low rank, ${\rm FBLR}_{R=2}$ does outperform FBLR for all sparsity levels.

%Notably, the functional predictor is two-way, whereas PACE requires the predictor to be one-way. Recall that the predictors have a separable covariance structure, which implies that all the functional curves along the first dimension $\T_1$, $X_i(s,.)$, share the same covariance structure. Therefore, before using PACE, we reshaped the data by pooling the functional curves along the first dimension $\T_1$, resulting in one-way functional predictors denoted as $\{X_k^*(T_{kj}), 1 \leq k \leq 100n \}$.  In this context, $X_{100*(i-1)+s}^*(T_{100*(i-1)+s,j}) = X_{i}(s,T_{ij})$.

%We adopt the sparse design from previous studies \citep{guo2023sparse}. 

% \dyR{ for carl: 
% \begin{eqnarray*}
%     \alpha_0(t) = 4\sqrt{2}\sum_{i=1}^{200}(-1)^i i^{-2} \cos \big((i + 0) \pi t\big)\\
%     \beta_0(t) = 4\sqrt{2}\sum_{i=1}^{200}(-1)^i i^{-4} \cos \big((i + 0) \pi t\big)    
% \end{eqnarray*}
% }

\section{Additional real data analysis on Canadian weather}
\label{sec:rd weather}

\begin{figure}[!ht]
 	\centerline{\includegraphics[width=.8\textwidth]{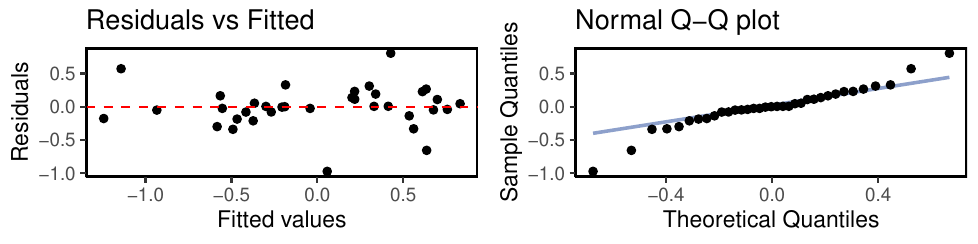}}
 	\caption{The diagnosis plots of the residuals of FBLR for the Canadian weather data.}
	\label{fig:rd_cw_resid}
\end{figure}

This section provides more information on the application to the Canadian weather data as a supplement to Section \ref{sec:realdata}. Figure \ref{fig:rd_cw_resid} provides the residual diagnosis of FBLR, which suggests a fairly good fit of the data. 
% keep
% The left panel shows that the residuals are distributed randomly around the zero and form an approximate horizontal band, indicating the linear relationship and homogeneous error variance. Furthermore, there is no any residual visibly away from the random pattern of the residuals verifying that there are no outliers. The right panel displays the normal Q-Q plots, suggesting the residuals are normally distributed. The Q-Q plot also shows that FBLR may overestimate the precipitation at Kamloops station, corresponding to the rightmost point, while underestimating the precipitation at Watson station and Prince Rupert station, the two leftmost points. This can be ascribed to their special topological positions. The rain clouds typically just pass over Kamloops because it lies deep in the Thompson River Valley. Watson and Prince Rupert, on the other hand, are both located on the windward slope, where moist air is forced to ascend, resulting in orographic precipitation. These findings are similar to those reported in \citet{ramsay2005principal, cai2012minimax}.

\section{Real data analysis: LIDAR}
\label{sec:rd lidar}

\begin{figure}[!ht]
    \centerline{\includegraphics[width=.35\textwidth]{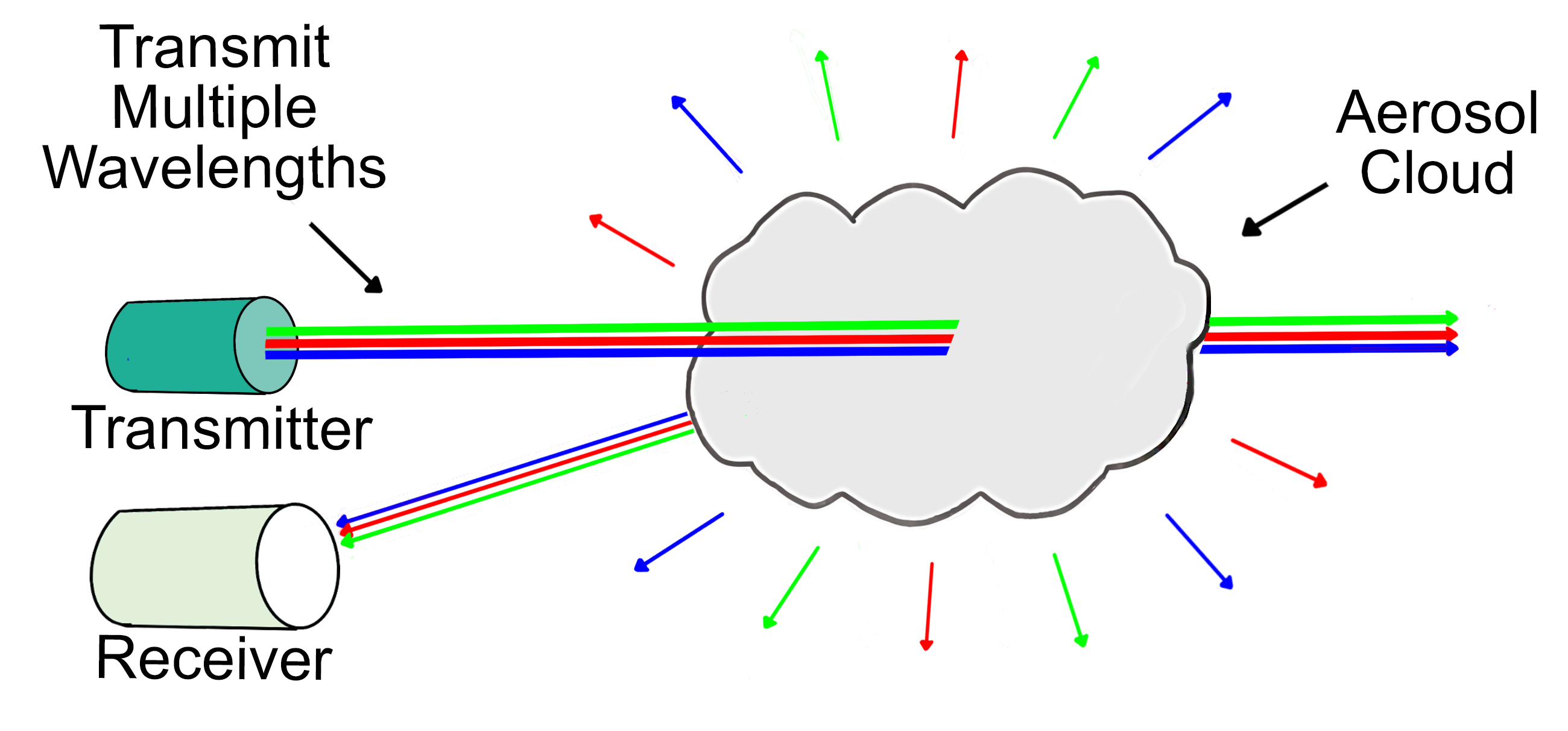}}
    \caption{A comic describing the LIDAR data generation. }
    \label{fig:comic}
\end{figure}

We demonstrate the performance of FBLR and other methods on the LIDAR data. The goal is to discriminate biological threat aerosol clouds in the atmosphere from non-biological interferent aerosol clouds such as dust or smoke.
We use the same dataset as in \citet{xun2013parameter}, where there are 28 aerosol clouds, half being biological and the other half non-biological. 
For each aerosol cloud at each time point, a set of 19 wavelength pulses is transmitted. 
The LIDAR receiver collects a fraction of the total optical power back-scattered over 60 equally-spaced range points (excluding background) at time $1,2, \ldots, 20$ for wavelength $1, 2, \ldots, 19$. See Figure \ref{fig:comic} for an illustration of the data generation process. 

\begin{figure}[!ht]
    \centerline{\includegraphics[width=\textwidth]{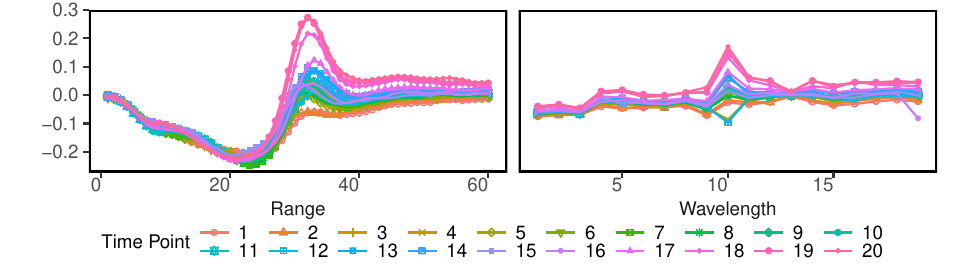}}
    \caption{Snapshots of one biological sample in the LIDAR data. 20 curves correspond to 20 time points. The left panel is the received signal over all 60 ranges at wavelength value 11. The right panel is the received signal over all 19 wavelengths at range value 45.}
    \label{fig:rd_eda}
\end{figure}

Figure \ref{fig:rd_eda} provides some visualization for one biological sample. It shows that the signal is smooth along three domains: time, range, and wavelength. 
In the literature of chemical biology, researchers have used approaches such as support vector machines after feature engineering or partial differential equation to solve this problem. Because of the smoothness of the surface, we will apply functional methods instead. 

% keep
% this type of classification is traditionally done in two steps, for instance, in \citet{warren2009detection}: 1. estimate a 1D vector, which captures the dependence of physically-meaningful quantity on the wavelength, from the 3D input through some complicated methodology; 2. use standard machine learning algorithm such as Bayes or support vector machine to make classification based on this 1D vector from Step 1. 

\begin{figure}
    \centerline{\includegraphics[width=\textwidth]{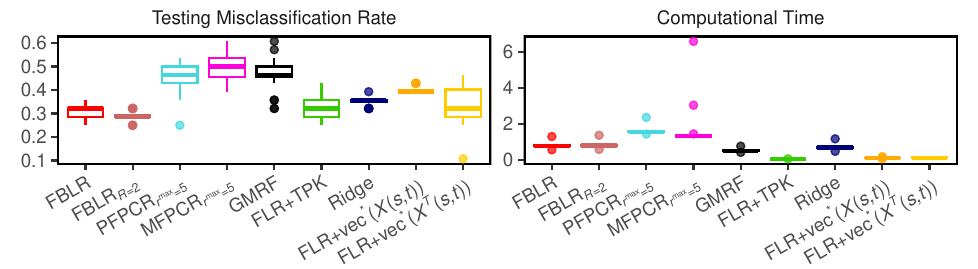}}
    \caption{Boxplots of the testing misclassification rates and computation time of 20 time points for multiple approaches in the LIDAR data.}
    \label{fig:rd_lidar}
\end{figure}

Ideally, one should use 3D input of size $60\times 19\times 20$ as the predictor to make prediction. Given that FBLR and most existing methods only apply to 2D data, we will perform the analysis 20 times corresponding to 20 time points separately. 
For each time point, we have 28 observations with input $x_i$ of size $60 \times 19$ and response $y_i$ taking value 0 or 1, standing for non-biological or biological aerosol, respectively. We adopt the regression approach to make classification: assign to class 1 if and only if the predicted response is larger than .5. 
We use leave-one-out method to compute the out-of-sample testing misclassification rate. The boxplots of 20 testing misclassification rates for 20 time points and computational time for nine approaches are given in Figure \ref{fig:rd_lidar}. 

Both FBLR and ${\rm FBLR}_{R=2}$ are included. We perform the test on the separability of the covariance by using \citet{aston2017tests} and the separability is verified for all time points. 
As explained in Appendix \ref{sec:exsim-FLR-vec}, FLR with two types of vectorization is compared with. For FBLR-related and FLR-related methods, we use the kernel in \eqref{eq:sim-kernel} again. For PFPCR and MFPCR, we set $r^{\max}=\lfloor\sqrt{n-1}\rfloor=5$. BLR is not included because the sample size is not large enough.
Figure \ref{fig:rd_lidar} shows that ${\rm FBLR}_{R=2}$ is the best, followed by FBLR, and then FLR+TPK. The other 2D methods such as PFPCR, MFPCR and GMRF are close to random guesses and even worse than the 1D methods such as FLR+$\textrm{vec}^*(X(s,t))$, FLR+$\textrm{vec}^*(X^T(s,t))$ and Ridge.

\begin{figure}[!ht]
    \centerline{\includegraphics[width=\textwidth]{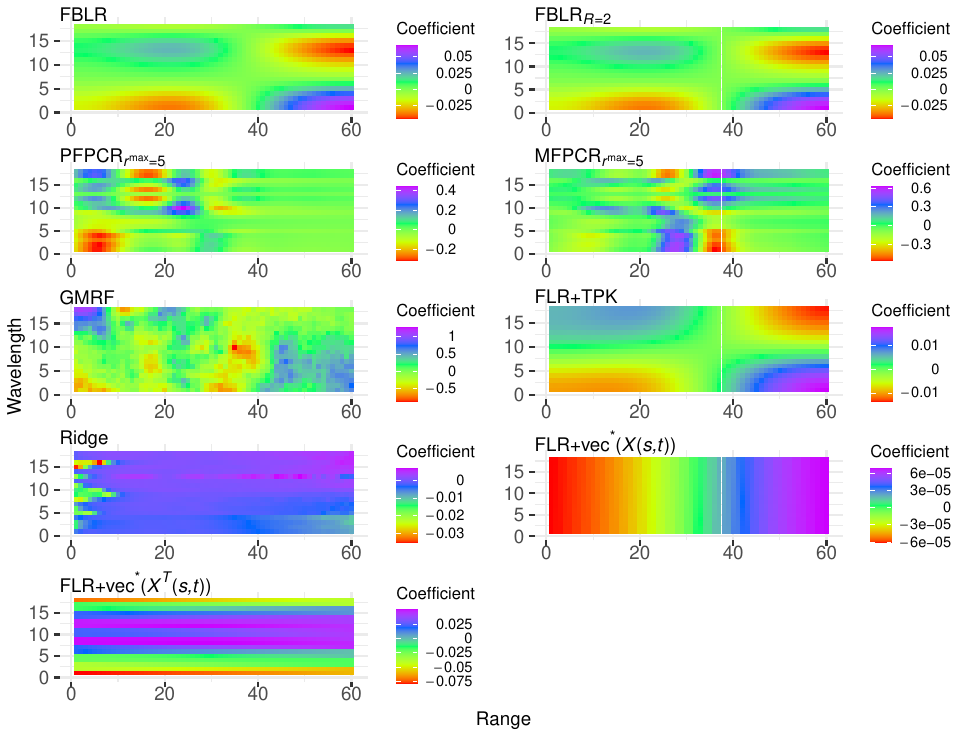}}
    \caption{Plots of the estimated 2D coefficient function $\hat\beta(\cdot,\cdot)$ in Model \eqref{eqn:2dflr-other} for LIDAR data at time point 2. The two axes correspond to range and wavelength, respectively.}
    \label{fig:rd_betaHat}
\end{figure}

\begin{figure}[!ht]
	\centerline{\includegraphics[width=\textwidth]{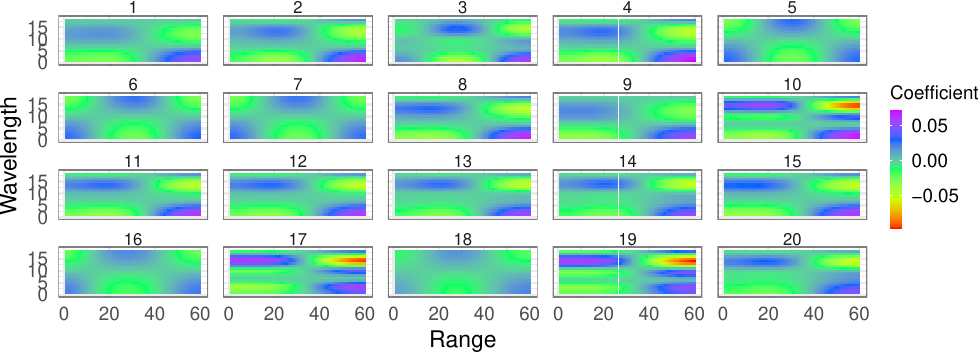}}
	\caption{Plots of the estimated 2D coefficient functions $\hat\beta(\cdot,\cdot)$ in Model \eqref{eqn:2dflr-other} by ${\rm FBLR}_{R=2}$ for LIDAR data at 20 time points. The two axes correspond to range and wavelength.}
	\label{fig:r-error-fblr2}
\end{figure}

\begin{figure}[!ht]
    \centerline{\includegraphics[width=\textwidth]{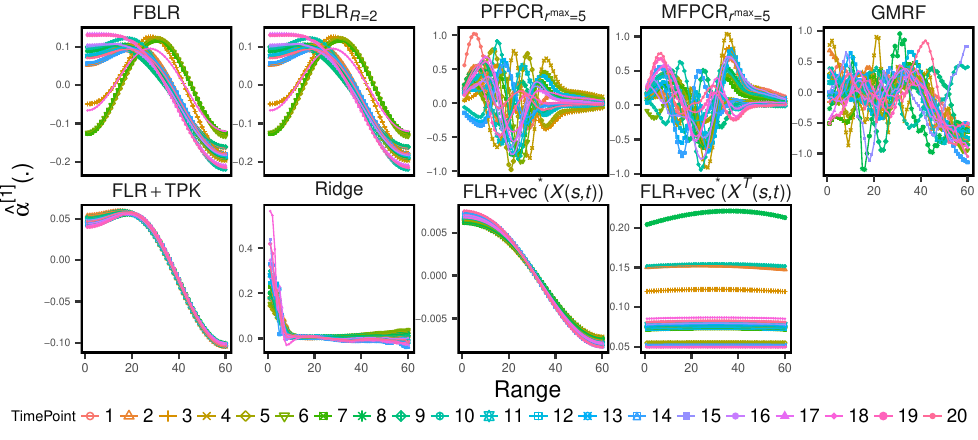}}
    \caption{Plots of the estimated 1D coefficient function $\hat\alpha^{[1]}(\cdot)$ that corresponds to the range domain in Model \eqref{eqn:2dflr}. 20 curves in each panel correspond to 20 time points.}
    \label{fig:rd_leftHat}
\end{figure}

\begin{figure}[!ht]
	\centerline{\includegraphics[width=\textwidth]{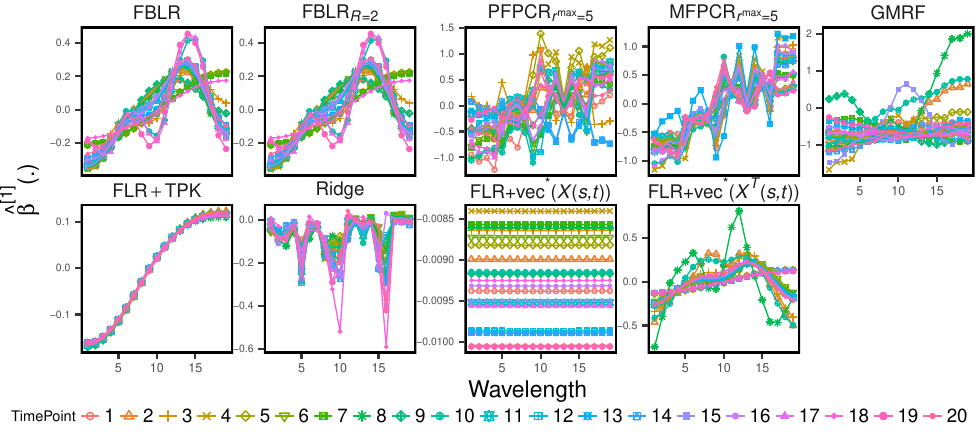}}
	\caption{Plots of the estimated 1D coefficient function $\hat\beta^{[1]}(\cdot)$ that corresponds to the wavelength domain in Model \eqref{eqn:2dflr}. 20 curves in each panel correspond to 20 time points.}
	\label{fig:rd_rightHat}
\end{figure}

Figures \ref{fig:rd_betaHat} - \ref{fig:rd_rightHat} further show the advantage of FBLR over other methods on LIDAR data in terms of interpretation, smoothness, and stableness. 
Figure \ref{fig:rd_betaHat} shows the heat-maps of the estimated 2D coefficient function $\hat\beta(\cdot,\cdot)$ in Model \eqref{eqn:2dflr-other} for these nine methods at time point 2, which is randomly selected (the other time points have similar message). It shows that both FBLR and ${\rm FBLR}_{R=2}$ obtain smoother coefficient function estimations compared with other 2D methods. The small visual difference between FBLR and ${\rm FBLR}_{R=2}$ indicates that the second term in Model \eqref{eqn:2dflr-2} has a small magnitude. Furthermore, although PFPCR and MFPCR are supposed to provide smooth estimations, the resulting estimated coefficient function is not very smooth. 
For the 1D methods, Ridge estimation inherently lacks smoothness, and FLR+$\textrm{vec}^*(X(s,t))$ (stacking columns) and FLR+$\textrm{vec}^*(X^T(s,t))$ (stacking rows) exhibit excessive smoothing because one dimension has nearly no variation.

Figure \ref{fig:r-error-fblr2} shows the estimated 2D coefficient functions $\hat\beta(\cdot,\cdot)$ in Model \eqref{eqn:2dflr-other} by ${\rm FBLR}_{R=2}$ for LIDAR data at 20 time points. Every 2D coefficient function at any time point is smooth and the 2D coefficient functions evolve smoothly over time. 

We next compare the evolvement of the 2D coefficient functions over time of all nine methods. A direct comparison of the 2D surfaces is difficult, so we perform SVD of all 2D coefficient functions. Figures \ref{fig:rd_leftHat} - \ref{fig:rd_rightHat} show the leading left and right singular vectors, which correspond to $\hat\alpha^{[1]}(\cdot)$ and $\hat\beta^{[1]}(\cdot)$ in Model \eqref{eqn:2dflr}, respectively.
Again, the estimated 1D coefficient functions by FBLR and ${\rm FBLR}_{R=2}$ are smooth for each time point and evolve smoothly over time; in contrast, the estimated 1D coefficient functions by 2D methods such as PFPCR, MFPCR, and GMRF are in general not very smooth for each time point and do not evolve quite smoothly over time; the estimated 1D coefficient functions by 1D methods such as FLR+$\textrm{vec}^*(X(s,t))$ and FLR+$\textrm{vec}^*(X^T(s,t))$ tend to be over-smoothed for one of the two domains. This demonstrates the ``stableness'' of the (iterative) FBLR estimations. FLR+TPK is overly ``stable'' since the estimated functions do not evolve over time at all.

\end{appendix}

% \clearpage
% \newpage
% DY to do
% \begin{enumerate}
%     \item cover letter
% %    \item acknowledge section: need others
%     \item cannot theory initialization; Haipeng, write something in the report
% \end{enumerate}

% \clearpage
% \newpage
% \input{./TEX/revision_report.tex}
\end{document}